\documentclass[structabstract]{aa}  
\usepackage{graphicx}
\usepackage{txfonts}
\usepackage{array}
\usepackage{longtable}
\usepackage{natbib}
\usepackage{multirow}
\usepackage{subfigure}

\newcommand{\bibfont}{14pt}

\begin{document}

\def\gsim{\;\lower.6ex\hbox{$\sim$}\kern-7.75pt\raise.65ex\hbox{$>$}\;}

   \title{VLT/FLAMES spectroscopy of Red Giant Branch stars\\ in the Carina dwarf spheroidal galaxy.\thanks{Based on FLAMES observations collected at the European Southern Observatory, proposals numbers 074.B-0415 and 076.B-0146.}}

   \subtitle{}

   \author{B. Lemasle\inst{1}
          \and
          V. Hill\inst{2}
	  \and
          E.Tolstoy\inst{1}
          \and
          K. A. Venn\inst{3}
          \and
	  M. D. Shetrone\inst{4}
	  \and
          M. J. Irwin\inst{5}
          \and
          T. J. L. de Boer\inst{1}
          \and
          E. Starkenburg\inst{1}
          \and
          S. Salvadori\inst{1}
          }

   \institute{Kapteyn Astronomical Institute, University of Groningen, PO Box 800, 9700AV Groningen, The Netherlands\\
              \email{lemasle@astro.rug.nl}
         \and
             Universit\' e de Nice Sophia-Antipolis, CNRS, Observatoire de la C\^ ote d'Azur, Laboratoire Cassiop\' ee, 06304 Nice Cedex 4, France
         \and
	     Department of Physics \& Astronomy, University of Victoria, 3800 Finerty Road, Victoria, BC V8P 1A1, Canada
	 \and
	     McDonald Observatory, University of Texas at Austin, HC75 Box 1337-MCD, Fort Davis, TX 79734
	 \and
	     Institute of Astronomy, University of Cambridge, Madingley Road, Cambridge CB3 0HA, UK
             }

   \date{Received September 21, 2011; accepted November 27, 2011}

 
  \abstract
{   
The ages of individual Red Giant Branch stars can range from 1~Gyr old
to the age of the Universe, and it is believed that the abundances of
most chemical elements in their photospheres remain unchanged with time (those that are not 
affected by the first dredge-up). This means that they trace the interstellar
medium in the galaxy at the time the star formed, and hence the chemical enrichment history of the galaxy.
}
   {
Colour-Magnitude Diagram analysis has shown 
the Carina dwarf spheroidal to have had an unusually episodic star
formation history and this is expected to be reflected in the
abundances of different chemical elements.
}
   {
We use the VLT-FLAMES multi-fibre spectrograph in high-resolution mode
(R$\approx$20000) to measure the abundances of several chemical
elements, including Fe, Mg, Ca and Ba, in a sample of 35 individual Red
Giant Branch stars in the Carina dwarf spheroidal galaxy.  We also combine
these abundances with photometry to derive age estimates for
these stars. This allows us to determine which of two distinct star
formation episodes the stars in our sample belong to, and thus to define
the relationship between star formation and chemical enrichment during
these two episodes.
}
   {
As is expected from the star formation history, Carina contains two
distinct populations of Red Giant Branch stars: one old ($\gtrsim$~10~Gyr), which we have found to be 
metal-poor ([Fe/H]$ < -1.5$), and $\alpha$-rich ([Mg/Fe]$ > 0$); the other
intermediate age ($\approx 2-6$~Gyr), which we have found to have a metallicity 
range ($-1.8 < $[Fe/H]$ < -1.2$) 
with a large spread in [$\alpha$/Fe] abundance, going from extremely
low values ([Mg/Fe]$ < -$0.3) to the same mean values as the older population
($<$[Mg/Fe]$>\sim 0.3$).
}
   {
We show that the chemical enrichment history of the Carina dwarf
spheroidal was different for each star formation episode.  The
earliest was short ($\sim 2-3$~Gyr) and resulted in the rapid chemical
enrichment of the whole galaxy to [Fe/H]$\sim -1.5$ with both SNe II and SNe Ia
contributions.  The subsequent
episode occured after a gap of $\sim 3-4$~Gyr, forming $\sim$~70\% of
the stars in the Carina dSph, 
but it appears to have resulted in relatively little
evolution in either [Fe/H] or [$\alpha$/Fe].
}


   \maketitle
%

\section{Introduction}

{ 
The main advantage of nearby galaxies in the Local Group is
that it is possible to study their resolved stellar populations. This
means that stars of particular ages can be picked out of a
Colour-Magnitude Diagram (CMD) for a more detailed study of specific
epochs in the evolution of a galaxy. An important and bright
population of stars that trace the chemical evolution over almost the
entire star formation history (SFH) of a galaxy are the Red Giant Branch (RGB) 
stars \citep[see][and references therein]{Tolstoy2009}.

In the context of a $\Lambda$-CDM universe dwarf galaxies play a key
role in galaxy formation \citep[e.g.][]{Kauf1993}. The most simple
scenario assumes that the small and unevolved dwarf spheroidal (dSph)
galaxies that we see around the Milky Way should be related to the
type of objects that originally merged to form the Milky Way. This
suggests that the stellar populations at the ages when mergers
occurred should be similar in small dSph galaxies and the
Milky Way.  The comparison of stellar abundances in dSph galaxies
around the Milky Way and the stars in the Milky Way itself have shown
that the merging of these nearby dwarf galaxies to create the Milky
Way could only have happened at very early times, within the first
Gyr of star formation occuring in either system 
\citep[e.g.][]{Una1996,She2001,Tolstoy2003,Venn2004,Tolstoy2011}. After
this time the differences in $\alpha$-element enrichment (the position of the ``knee'')
become too different between large and small systems.

The Carina dSph galaxy has long been known to have a most unusual and
episodic SFH. The presence of RR~Lyrae
variable stars first indicated an ancient stellar
population \citep{Saha1986}, and anomalous Cepheid
variable stars \citep{Mat1998} showed that a young ($<$~1~Gyr)
population is also present. Anomalous Cepheids are believed to be metal-poor classical Cepheids \citep{Bono1997},
but they could also be the result of mass transfer and possibly coalescence in a low-mass binary systems
\citep{Zinn1976}. The Carina dSph is still the only galaxy where such 
distinct main sequence turnoffs can clearly be seen in
deep CMDs \citep[first shown by][]{SH1996}. This is the unequivocal
signature of periods of active star formation separated by similarly
long periods of no star formation at all.  

There have been numerous analyses
of the CMDs of the Carina dSph, 
including the distinct Main Sequence turn-offs, in a variety
of filters, for different fields of view to determine the full SFH 
\citep[e.g.,][]{Migh1997,HK1998,Dolphin2002,Mon2003,Riz2003,Bono2010}.  There
is still quite some discrepancy between the SFHs obtained; specifically precisely when the
different episodes of star formation occurred.  These differences
usually relate to the field of view of the observations modelled and
thus how well different features in the CMD are populated
\citep[see][]{Cig2010} and also the different methods used to determine
the SFH.  However, all studies agree that there were at least two
major episodes of star formation in the Carina dSph which were
separated by a long quiescent era. They also agree that the dominant
stellar population ($\sim 70\%$ of the stars) comes from an 
intermediate age episode of star formation.

Despite its complex SFH the Carina dSph has an extremely narrow RGB.
This is a result of the age-metallicity degeneracy, where the
metal-rich, young stars have almost the same colour as the metal-poor
older stars. The thinness of the RGB is also a consequence of the
dominant intermediate age population which formed over a relatively
short time (a few Gyr) with only a small spread in [Fe/H]. The thin
RGB has made reliable photometric determinations of the metallicity
and its evolution challenging \citep[e.g.,][]{Riz2003, Bono2010}.

The first spectroscopic studies of the metallicities of individual
stars in the Carina dSph came from the empirical calibration of the
strong Ca~II triplet (CaT) lines \citep{Arm1991,DaCos1994,SH1999}.  These studies 
typically reported a mean metallicity between [Fe/H]=$-1.5$ and $-2.0$ dex,
with a small dispersion.  The largest sample was $\sim$~50 RGB
stars. The advent of multi-fibre instruments like FLAMES at the VLT
led to a dramatic increase in the sizes of these CaT samples
\citep[e.g., 437 stars from][]{Koch2006}, leading to a much better
defined and statistically robust metallicity distribution function
\citep{Helmi2006,Stark2010}. The spread in iron \citep[e.g., a FWHM of 0.92 dex derived from CaT by][]{Koch2006}
is much larger than the one suggested by photometric evidence.

The first detailed, high resolution, abundance study of a 
sample of five individual RGB stars in the Carina dSph used VLT/UVES
\citep{She2003}.  These
results were analyzed in terms of galactic evolution with the help of
both photometric ages and detailed spectroscopic abundances
\citep{Tolstoy2003}.  Recently a study of iron and $\alpha$-element
abundances from UVES spectra of a further 10 RGB stars has been made
\citep{Koch2008} and a study of a large range of elements in 9 stars with 
UVES and MIKE on Magellan spectra \citep{Venn2011}. As the total
number of RGB stars studied in the Carina dSph has increased so has
the apparent star to star scatter in the abundance ratios of most
elements, although the dispersion in elemental ratios was already 
apparent from the start \citep{She2003,Koch2008}.

There have been several attempts to self-consistently model the SFH and 
the chemical evolution of the Carina dSph
\citep[e.g.,][]{Lan2003,Lan2004,Lan2006,Revaz2009}. This is challenging
mostly because the Carina dSph is so unusual compared to the other dSph in the
Local Group. It is hard to find a unique explanation for
the clear gaps seen in the SFH \citep[e.g.][]{Pas2011}.  

It is hoped that detailed chemical
abundances of a large sample of individual RGB stars, with age estimates,
in the Carina dSph will
better define the effects of star formation on the chemical evolution.
These large
samples will also better quantify the scatter seen in the abundances of
a variety of chemical elements. 
}

\par This paper is organized as follows: in Sect.~\ref{sample}, we
describe the sample selection, observations and data reduction. In
Sect.~\ref{atmparam_section}, we explain how we derived the
atmospheric parameters (T$_{\rm eff}$, $log~g$, v$_{t}$ and [Fe/H]) and
the individual abundances. The abundances are discussed in detail in 
Sect.~\ref{results} and the ages in Sect.~\ref{ages_sect}. They are
interpreted in terms of the SFH of the Carina dSph in Sect.~\ref{disc}.

\section{Sample selection, observations and data reduction.}
\label{sample}

\begin{table*}[!ht]
\caption{Observing log.}
\label{obslog}
\centering
\begin{tabular}{clcccccc}
\hline\hline
ESO archive observation name & Setting & Exp. time & Airmass & DIMM seeing  & DIMM seeing & Notes \\
                             &         &    sec    &         & at beginning &   at end    &       \\
\hline
GIRAF.2005-01-09T00:38:36.971.fits & H651.5A &  600 & 1.49 &  1.17 &   -   & low S/N \\ 
GIRAF.2005-01-09T00:51:15.578.fits & H651.5A & 3600 & 1.43 &   -   &  1.66 & low S/N \\ 
GIRAF.2005-01-09T01:57:26.664.fits &  H627.3 & 3600 & 1.24 &  1.71 &  1.09 & low S/N \\ 
GIRAF.2005-01-09T03:12:34.973.fits &  H548.8 & 3600 & 1.14 &  1.18 &  0.68 &         \\ 
GIRAF.2005-01-09T04:13:33.996.fits &  H548.8 & 3600 & 1.11 &  0.68 &  0.57 &         \\ 
GIRAF.2005-01-09T05:29:40.059.fits & H651.5A & 3900 & 1.16 &   -   &  0.58 &         \\ 
GIRAF.2005-01-09T06:36:21.407.fits &  H627.3 & 3600 & 1.28 &  0.58 &  0.65 &         \\ 
GIRAF.2005-01-09T07:38:07.042.fits &  H627.3 & 3900 & 1.49 &  0.54 &   -   &         \\ 
GIRAF.2005-01-10T03:40:05.439.fits & H651.5A & 3600 & 1.12 &  1.01 &  0.78 & low S/N \\ 
GIRAF.2005-01-10T04:41:45.604.fits &  H627.3 & 4400 & 1.12 &  0.75 &  0.88 & low S/N \\ 
GIRAF.2005-01-10T06:15:26.119.fits & H651.5A & 4200 & 1.24 &   -   &  1.07 & low S/N \\ 
GIRAF.2005-01-10T07:27:05.588.fits &  H627.3 & 4133 & 1.46 &  1.24 &  1.16 & low S/N \\ 
GIRAF.2005-01-11T01:06:52.044.fits & H651.5A & 3600 & 1.35 &  0.71 &  0.76 &         \\ 
GIRAF.2005-01-11T02:08:30.061.fits &  H627.3 & 3600 & 1.21 &  0.74 &  1.69 &         \\ 
GIRAF.2005-01-11T03:23:59.489.fits &  H548.8 & 3600 & 1.12 &  1.09 &  1.69 & low S/N \\ 
GIRAF.2005-01-11T04:24:59.522.fits &  H548.8 & 3600 & 1.12 &  1.57 &  3.20 & low S/N \\ 
GIRAF.2005-01-11T05:27:34.650.fits &  H548.8 & 3600 & 1.17 &  2.55 &  2.84 & low S/N \\ 
GIRAF.2005-12-06T03:26:47.857.fits &  H627.3 & 3900 & 1.36 &  1.25 &  1.24 &         \\ 
GIRAF.2005-12-06T04:46:21.726.fits &  H548.8 & 3600 & 1.18 &  1.12 &  1.40 &         \\ 
GIRAF.2005-12-06T05:59:52.728.fits &  H627.3 & 3900 & 1.12 &  1.26 &  1.40 & low S/N \\ 
GIRAF.2005-12-07T05:39:28.586.fits &  H548.8 & 3600 & 1.13 &  0.79 &  0.78 &         \\ 
GIRAF.2005-12-08T03:13:50.991.fits & H651.5A & 3600 & 1.38 &  0.84 &  0.56 & low S/N \\ 
GIRAF.2005-12-08T04:24:14.821.fits &  H627.3 & 3900 & 1.20 &  0.62 &  0.54 &         \\ 
GIRAF.2005-12-08T05:37:44.005.fits & H651.5A & 3600 & 1.12 &  0.52 &  0.52 & low S/N \\ 
GIRAF.2006-01-06T04:50:17.596.fits &  H548.8 & 3600 & 1.12 &  0.69 &  0.68 &         \\ 
\hline
\end{tabular}
\end{table*}

Our program consists of high resolution spectroscopy of individual RGB stars in the Carina dSph galaxy. We used FLAMES \citep{Pas2002} to obtain spectra simultaneously with both GIRAFFE (in high resolution mode) and UVES spectrographs (programs 074.B-0415 and 076.B-0146). This paper deals with the abundances determined from the FLAMES/GIRAFFE observations and the FLAMES/UVES results are described in a separate paper \citep{Venn2011}.

\subsection{Sample selection and observations.}

Our targets consist of RGB stars located within a 25 arcmin diameter field of view close to the center of the Carina dSph. They were selected partly from ESO low resolution spectra \citep[][]{Helmi2006} and partly from a CMD. We cannot exclude that our sample contains some contamination from AGB stars.

Observations were carried out between 9 and 11 January 2005, 6 and 8 December 2005, and on 6 January, 2006. 
Due to bad weather conditions, only 12 out of 25 frames had a sufficient S/N to be useful. The observing log is listed in Table~\ref{obslog}.

\begin{figure}[!htb]
   \centering
   \includegraphics[angle=-90,width=0.90\columnwidth]{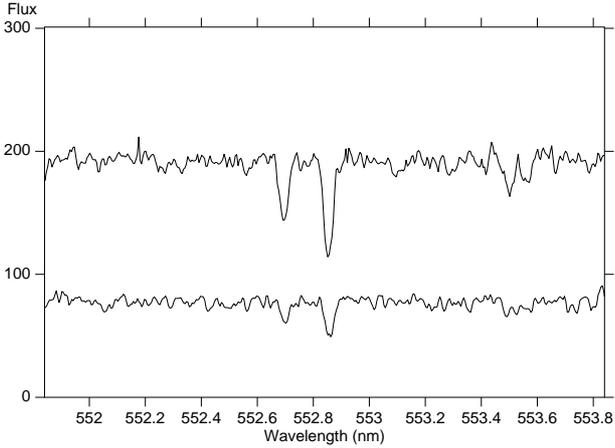}
      \caption{Representative spectra of two stars of our sample, centered on the Mg line at 552.841 nm. {\it(top)} MKV0900: S/N=44, Vmag=17.79; {\it(bottom)} MKV0614: S/N=22, Vmag=18.72. A ScII line ($\lambda$=552.679 nm) is located on the left of the Mg line.}
         \label{Spectra}
   \end{figure}

We used GIRAFFE in the Medusa mode with 98 out of 132 fibres placed on targets over the 25 arcmin diameter field of view. The remaining 34 fibres were put on blank sky positions to provide for sky subtraction of the target spectra. Observations must provide not only a sufficient wavelength coverage for a canonical analysis (i.e., sufficient neutral and ionized iron lines), but also a good number of useful $\alpha$-element and heavy element lines. To achieve this we observed three different wavelength settings with three different gratings, namely HR10, HR13 and HR14 (see Table~\ref{grat}). To increase the S/N, several exposures were taken with each HR setting. Due to bad weather conditions, only five HR10 frames, five HR13 frames and two HR14 frames could be used, resulting in the total exposure times given in Table~\ref{grat}. The large number of low S/N observations, which had to be rejected, obviously limited the total number of stars that could be analysed from the initial selection. In order to illustrate the data quality of our sample, we show in Fig.~\ref{Spectra} the spectra of two stars (MKV0614, S/N=22; MKV0900, S/N=44) centered on the Mg line at 552.8 nm (HR10 grating).

%


\begin{table}[!hb]
\caption{FLAMES-GIRAFFE gratings used}
\label{grat}
\centering
\begin{tabular}{cccc}
\hline\hline
Grating & HR10 & HR13 & HR14 \\
\hline
$\lambda_{min}$ ($\AA$) &  5339 &  6120   &  6308   \\
$\lambda_{max}$ ($\AA$) &  5619 &  6405   &  6701   \\
Resolution at center    & 19800 & 22500   & 17740   \\
exposure time           & 5h    & 4h15min & 2h05min \\
\hline
\end{tabular}
\end{table}

\subsection{Data reduction.}

We used the ESO pipeline to perform the basic data reduction, as well as the extraction and wavelength calibration of the spectra. 
For sky subtraction, we used a routine (from M. Irwin) that produces an average sky spectrum from the sky-dedicated fibers, which is then subtracted from each object spectrum after being rescaled to match the sky features in each fibre.\\
As spectra were taken at several different epochs, they need to be registered to the same rest frame. We computed the barycentric correction to radial velocity with the {\it dopcor} task and coadded individual spectra with the {\it scombine} task in IRAF. We used a flux weighted average, with median sigma clipping to remove cosmic rays.

\subsection{Membership.}

\begin{figure}[!ht]
   \centering
   \includegraphics[width=\columnwidth]{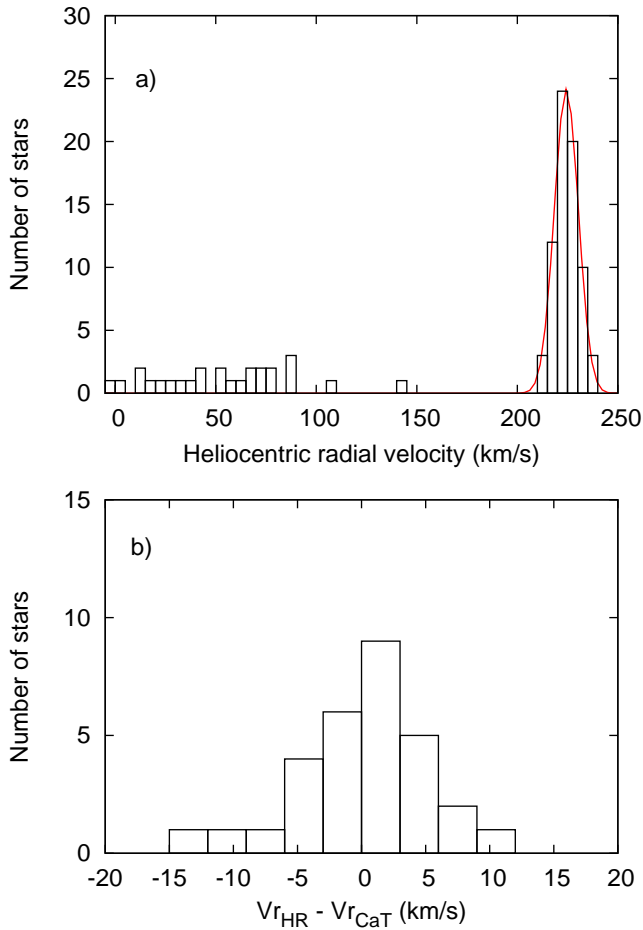}
      \caption{(a) The distribution of heliocentric radial velocities for our whole FLAMES/GIRAFFE sample, including a gaussian fit which gives V$_{c}$=224.4 km s$^{-1}$ $\pm$5.95 km s$^{-1}$. (b) Comparison between radial velocities derived from our data (HR) compared to CaT \citep{Helmi2006}.}
         \label{radvel}
   \end{figure}

We derived radial velocities from the reduced spectra with DAOSPEC \citep{StetPan2008}, which cross-correlates all the lines detected by the software with an input line list. The accuracy is in general better than $\pm$2 km s$^{-1}$. The radial velocity distribution is shown in Fig. \ref{radvel}a, where the systemic radial velocity peak of the Carina dSph stands out from the Galactic foreground contamination at 224.4 km s$^{-1}$, which is consistent with previous determinations \citep{Mat1998,Maj2005,Koch2006,Wal2007,Fab2011}. We also compared our velocites with those obtained in our low resolution CaT sample \citep{Helmi2006}, and this is shown in Fig.\ref{radvel}b. There is no systematic bias between the CaT and HR radial velocities and values agree within their error bars. 

\begin{figure}[!hbt]
   \centering
   \includegraphics[angle=-90,width=\columnwidth]{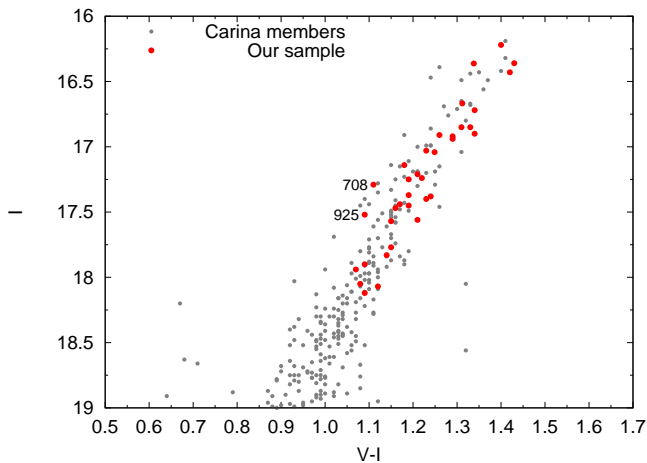}
      \caption{I vs. (V-I) CMD: our FLAMES/GIRAFFE sample is shown in red dots and other Carina members (from CaT) are shown in grey dots.}
         \label{CMD}
   \end{figure}

We considered as likely members those stars with a radial velocity within 2$\sigma$ of the mean radial velocity, which means between 212 and 236 km s$^{-1}$. We found that 26 out of our 98 targets have velocities that are inconsistent with membership of the Carina dSph. All of these are stars for which no radial velocity from CaT spectra was available before selection. An additional 37 stars had to be discarded because their spectra were too low S/N to perform an accurate abundance analysis. This depends on both the setting and the intrinsic metallicity of the star, so no S/N limit is specified. Our sample thus reduced to 35 stars which are likely members of the Carina dSph with sufficient S/N for an abundance analysis. They are shown in the I vs. (V-I) CMD in Fig.~\ref{CMD}. Target coordinates, photometry and radial velocities of these stars are listed in Table~\ref{targets}.

%

\addtocounter{table}{1} 

\section{Determination of stellar parameters and abundances.}
\label{atmparam_section}

We measured the equivalent widths (EWs) of the absorption lines with DAOSPEC \citep{StetPan2008}, which is an automatic tool optimised for the characteristics of GIRAFFE HR spectra. Lines are fitted by saturated gaussians because at the resolution of our data, the line profile is dominated by the instrumental effects and not astrophysical processes. All lines with EW$>$200~m\AA~are discarded as they will most likely depart from a Gaussian profile. The EWs determined by DAOSPEC for our sample are listed in Table~\ref{EW}\\

\addtocounter{table}{1} 

We used the same line list as in the other DART \citep[\textbf{D}warf \textbf{A}bundances and \textbf{R}adial velocity \textbf{T}eam,][]{Tolstoy2006}  papers based on GIRAFFE medium resolution spectroscopy \citep{Let2010,Hill2011}. It is based on the line list of \cite{She2003} with additional lines from \cite{Pomp2008}.\\

We used a grid of OSMARCS atmosphere models in spherical symmetry  
\citep{Gus2008} valid for T$_{\rm eff}$ = [4000~\textendash~5500] K, $log~g$ = [0.0~\textendash~3.5] dex, [Fe/H] = [-3.0~\textendash~+1.0] dex and [$\alpha$/Fe] following the trend of increasing as [Fe/H] decreases. Abundances are derived with Calrai, a LTE spectrum synthesis code originally developped by \cite{Spite1967} and continuously improved since then. Abundances were first computed for each individual line and the uncertainties on the EW measurements given by DAOSPEC are propagated into uncertainties on individual abundances. They are subsequently propagated into error estimates on the abundances for each element by weighting each line by 1/($\sigma^{2}$).\\

Stellar temperatures T$_{\rm eff}$ and surface gravities $log~g$ were determined from photometric data in the VIJHK bands. We have V and I bands from ESO-WFI for the whole sample and ESO-SOFI JHK magnitudes for $\approx$60\% of our sample (M. Gullieuszik, private communication). The photometry is listed in Table~\ref{targets}.\\

When available, we averaged T$_{\rm eff}$ given by the four different colours (V-I), (V-J), (V-H) and (V-K) to determine the photometric T$_{\rm eff}$, following the calibrations for giants from \cite{Rami2005}. For those stars lacking IR photometry, we used the temperature determined from the (V-I) colour. In most cases, the temperatures from different colors are in very good agreement (see Table~\ref{atmparam}). We adopted the reddening law A$_{v}$/E(B-V)=3.24 with an extinction of 0.06 mag \citep{Schle1998}. A first evaluation of T$_{\rm eff}$ was computed with the metallicities derived from the CaT, and when no CaT value was available the mean value [Fe/H] = $-$1.5 dex was assumed. These values were later updated with the final metallicities of the stars.\\ 

Using our temperature estimates together with a distance modulus of
$\mu_{0}$=20.06$\pm$0.12 mag \citep{Mat1998} and assuming an average
stellar mass of 0.8 M$_{\sun}$, the photometric surface gravities
where computed using the bolometric correction from
\cite{Alonso1999}. Adopting a 1.0 M$_{\sun}$ for the intermediate-age stars in our sample would increase $log~g$ by 0.1 dex with no effect on T$_{\rm eff}$ and, in turn, have a very small effect on the chemical abundances. The distance modulus from \cite{Mat1998} was derived from the Period-Luminosity relation of the dwarf Cepheids found in the Carina dSph. It is in good agreement with more recent values also based on primary distance indicators (RR~Lyrae, anomalous Cepheids, tip of the RGB), for example from \cite{Dall2003}: $\mu_{v}$=20.19$\pm$0.12 mag or \cite{Piet2009}: 20.09$\pm$0.03$\pm$0.12 mag in the J band, 20.14$\pm$0.04$\pm$0.14 mag in the K band. Adopting a lower extinction value (for example, \cite{Mon2003} proposed E(B-V)=0.03 mag) leads to lower T$_{\rm eff}$ by 40-60K but has a negligible influence on $log~g$.\\

\begin{figure}
   \centering
   \includegraphics[angle=-90,width=\columnwidth]{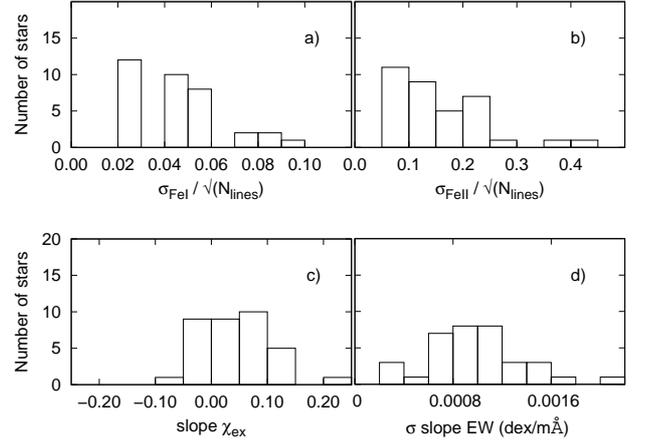}
      \caption{The distribution of errors on the stellar atmospheric diagnostics for the stars in our sample: (a) The error on the mean [FeI/H] and (b) [FeII/H] which are measured from the dispersion around the mean divided by the square root of the number of lines measured. (c) The slope of [FeI/H] versus the excitation potential $\chi_{ex}$. (d) The propagated error on the measurement of the slope of [FeI/H] versus the line strength.
              }
         \label{disp}
   \end{figure}

As our [FeII/H] could only be computed from a handful of weak lines, we were not able to accurately determine the ionization balance. From Fig. \ref{disp}b, it can be seen that [FeII/H] is rather uncertain with average errors up to $\approx$ $\pm$0.25 dex. In comparison, [FeI/H] errors (Fig. \ref{disp}a) do not exceed $\pm$0.09 dex. This uncertainty on [FeII/H] translates to an uncertainty on $log~g$ $\gtrsim$0.3 dex. The same outcome applies for the TiI/TiII comparison, where both [TiI/H] and [TiII/H] are derived from only between 1 and 5 lines. We therefore decided to keep a common photometric $log~g$ scale rather than using the uncertain ionization balance of iron or titanium lines to determine individual spectroscopic gravities. Surface gravity has a minor effect on our abundances derived from neutral ions: a variation of 0.5 dex in $log~g$ results in a variation of $\approx$0.1 dex in [FeI/H] and $\approx$0.25 dex in [FeII/H].\\

The photometric T$_{\rm eff}$ values were checked by ensuring that [FeI/H] does
 not depend on the excitation potential $\chi_{ex}$ and
indeed the slopes of [FeI/H] versus $\chi_{ex}$ are small
(see Fig. \ref{disp}c). However, the HR analyses of \cite{Venn2011}
and \cite{Koch2008} had to increase their photometric temperatures to
reach the excitation balance, and Fig. \ref{disp}c shows a systematic
offset of $\approx$0.05 which is in the same sense. This suggests a
systematic bias between the photometric temperature and the excitation
temperature. Following \cite{Venn2011}, we tried to increase our
  T$_{\rm eff}$ by 200K and consequently adjusted the other
  atmospheric parameters. As a result, [Fe/H] increases by $\simeq$0.2-0.3 dex
  and [Mg/H] increases by $\simeq$0.05-0.15 dex. The FeI/FeII ionization
  balance is better satisfied but our [Fe/H] show a systematic offset
  of $\approx$+0.3 dex with those derived from CaT. We decided not to apply 
any offset to our photometric parameters because there was no compelling reason to do so.

The relatively high T$_{\rm eff}$ uncertainty quoted in Table~\ref{atm_errors} reflects a conservative error bar associated with this possible bias in the T$_{\rm eff}$ scale.\\

The microturbulent velocity, v$_{t}$, is determined with an iterative process, which imposes the requirement that [FeI/H] does not depend on EW, that is to say that the slope between [FeI/H] and EW is zero. Measuring the value of the slope around the correct value of v$_{t}$ shows that v$_{t}$ varies linearly and symmetrically as a function of the slope and it is therefore possible to use this relation to estimate the uncertainty on the slope between [FeI/H] and EW. Given that the error on the slope is lower than $\pm$0.0016 dex/m\AA~(see Fig. \ref{disp}), the uncertainty on microturbulent velocity does not exceed 0.3 km/s. We have used the theoretical EWs (computed from the atomic parameters of the line and the atmospheric parameters of the star) rather than the measured values in order to avoid systematic biases on v$_{t}$ caused by random errors on the EW measurements \citep{Mag1984}. The stellar parameters can be found in Table~\ref{atmparam}.\\

\begin{table*}[!ht]
\centering
\caption{The atmospheric parameters (derived from photometry) for stars in our sample. Columns 2 to 5 gather the temperatures calculated from different colours, following the calibrations for giants from \cite{Rami2005}. Column 6 lists the bolometric correction, and column 7 the bolometric magnitude. Column 8 lists the mean temperature assumed for each star. Columns 9, 10 and 11 list respectively $log~g$, V$_{t}$ and [Fe/H].}
\begin{tabular}{cccccccclllccc}
\hline\hline
 Target & T$_{(V-I)}$ & T$_{(V-J)}$ & T$_{(V-H)}$ & T$_{(V-K)}$ & $BC_{V}$ & M$_{bol}$ & T$_{\rm eff}$ & {\it log g} & V$_{t}$ & [Fe/H] & S/N (H10) & S/N (H13) & S/N (H14) \\ 
        &       K     &       K     &       K     &       K     &    mag   &    mag    &       K       &     dex     &   dex   &   dex  &           &           &           \\ 
 \hline
 MKV0397 & 4494 &   -  &   -  &   -  & -0.50 & -2.19 & 4490 & 1.1  & 1.9 & -2.0  &  31   &  22   &  27   \\ 
 MKV0458 & 4518 &   -  &   -  &   -  & -0.48 & -2.18 & 4520 & 1.1  & 1.8 & -1.6  &  31   &  25   &  28   \\ 
 MKV0514 & 4454 & 4453 & 4407 & 4417 & -0.54 & -2.51 & 4430 & 0.9  & 2.5 & -2.3  &  40   &  20   &  41   \\ 
 MKV0556 & 4636 & 4683 & 4668 & 4675 & -0.41 & -1.36 & 4670 & 1.5  & 2.2 & -1.6  &  13   &   9   &  14   \\ 
 MKV0577 & 4623 & 4699 & 4616 & 4689 & -0.43 & -1.66 & 4660 & 1.4  & 1.4 & -1.6  &  20   &  14   &  20   \\ 
 MKV0596 & 4313 & 6397 & 5755 & 5714 & -0.11 & -2.20 & 4660 & 1.5  & 1.7 & -1.5  &  33   &  34   &  51   \\ 
 MKV0614 & 4587 &   -  &   -  &   -  & -0.45 & -1.87 & 4590 & 1.3  & 1.8 & -1.6  &  22   &  17   &  20   \\ 
 MKV0628 & 4381 &   -  &   -  &   -  & -0.56 & -2.50 & 4380 & 1.0  & 1.7 & -1.65 &  17   &  32   &  32   \\ 
 MKV0640 & 4425 &   -  &   -  &   -  & -0.53 & -2.39 & 4420 & 1.0  & 1.8 & -1.7  &  24   &  30   &  34   \\ 
 MKV0652 & 4540 & 4615 & 4572 & 4549 & -0.48 & -2.06 & 4570 & 1.2  & 1.8 & -2.3  &  21   &  13   &  25   \\ 
 MKV0677 & 4306 &   -  &   -  &   -  & -0.61 & -3.06 & 4300 & 0.7  & 1.7 & -1.75 &  38   &  48   &  40   \\ 
 MKV0698 & 4218 & 4114 & 4119 & 4116 & -0.74 & -3.27 & 4150 & 0.55 & 2.0 & -1.5  &  47   &  42   &  51   \\ 
 MKV0708 & 4668 &   -  &   -  &   -  & -0.42 & -2.16 & 4670 & 1.2  & 1.8 & -1.6  &  20   &  21   &  26   \\ 
 MKV0729 & 4551 &   -  &   -  &   -  & -0.46 & -1.97 & 4550 & 1.2  & 2.1 & -1.35 &  18   &  18   &  25   \\ 
 MKV0733 & 4491 &   -  &   -  &   -  & -0.50 & -1.87 & 4490 & 1.25 & 2.0 & -1.7  &  21   &  20   &  28   \\ 
 MKV0740 & 4585 &   -  &   -  &   -  & -0.44 & -1.61 & 4585 & 1.3  & 2.2 & -1.2  &  19   &  16   &  18   \\ 
 MKV0743 & 4694 &   -  &   -  &   -  & -0.40 & -1.33 & 4690 & 1.55 & 1.5 & -1.2  &  13   &   8   &  15   \\ 
 MKV0770 & 4535 & 4490 & 4508 & 4492 & -0.49 & -2.31 & 4500 & 1.0  & 1.7 & -1.6  &  25   &  30   &  33   \\ 
 MKV0780 & 4556 & 4539 & 4538 & 4565 & -0.47 & -2.00 & 4550 & 1.2  & 2.3 & -1.8  &  23   &  17   &  24   \\ 
 MKV0812 & 4709 & 4765 & 4712 & 4715 & -0.38 & -1.39 & 4720 & 1.5  & 2.0 & -1.35 &  12   &  10   &  15   \\ 
 MKV0825 & 4345 & 4335 & 4339 & 4340 & -0.59 & -2.57 & 4340 & 0.9  & 1.7 & -1.4  &  35   &  34   &  41   \\ 
 MKV0840 & 4378 & 4388 & 4390 & 4391 & -0.56 & -2.47 & 4380 & 0.9  & 1.5 & -1.1  &  26   &  27   &  35   \\ 
 MKV0842 & 4718 &   -  &   -  &   -  & -0.38 & -1.51 & 4720 & 1.4  & 2.3 & -1.45 &  13   &  13   &  16   \\ 
 MKV0880 & 4212 & 4171 & 4186 & 4181 & -0.70 & -3.00 & 4190 & 0.6  & 1.9 & -1.5  &  42   &  36   &  39   \\ 
 MKV0900 & 4207 & 4166 & 4174 & 4181 & -0.71 & -3.06 & 4180 & 0.6  & 1.7 & -1.7  &  44   &  28   &  49   \\ 
 MKV0902 & 4360 &   -  &   -  &   -  & -0.58 & -2.54 & 4360 & 0.9  & 1.9 & -1.9  &  27   &  22   &  41   \\ 
 MKV0914 & 4365 &   -  &   -  &   -  & -0.58 & -2.48 & 4360 & 0.9  & 2.4 & -2.5  &  29   &  21   &  31   \\ 
 MKV0916 & 4454 & 4411 & 4421 & 4464 & -0.53 & -2.04 & 4440 & 1.1  & 2.1 & -1.5  &  13   &  13   &  20   \\ 
 MKV0925 & 4698 &   -  &   -  &   -  & -0.40 & -1.93 & 4700 & 1.3  & 2.2 & -1.5  &  17   &  17   &  21   \\ 
 MKV0948 & 4339 &   -  &   -  &   -  & -0.59 & -2.75 & 4340 & 0.8  & 1.6 & -2.05 &  29   &  25   &  44   \\ 
 MKV0976 & 4692 & 4731 & 4706 & 4748 & -0.38 & -1.54 & 4720 & 1.4  & 1.7 & -1.3  &  15   &  11   &  16   \\ 
 MKV1007 & 4443 & 4277 & 4261 & 4279 & -0.60 & -2.49 & 4315 & 0.9  & 1.4 & -1.4  &  32   &  27   &  34   \\ 
 MKV1009 & 4522 & 4479 & 4483 & 4472 & -0.50 & -2.00 & 4490 & 1.2  & 2.0 & -1.7  &  22   &  16   &  21   \\ 
 MKV1012 & 4488 & 4415 & 4419 & 4430 & -0.53 & -2.25 & 4440 & 1.0  & 1.8 & -1.6  &  22   &  27   &  31   \\ 
 MKV1061 & 4453 & 4534 & 4456 & 4489 & -0.50 & -2.03 & 4480 & 1.1  & 1.7 & -1.5  &  19   &  17   &  27   \\ 
\hline
\end{tabular}
\label{atmparam}
\end{table*}

DAOSPEC returns an error estimate on the EW measurements that was propagated throughout the abundance determination. We also used the abundance dispersion $\sigma(X)$ around the mean value, to take into account not only the uncertainties on the EW measurements but also the uncertainties on the atomic parameters of the lines. This dispersion is defined for a given element X as $\frac{\sigma(X)}{\sqrt{N_X}}$ where N$_{X}$ is the number of lines measured for this element. It can happen that the abundance dispersion is underestimated due to small number statistics when abundances are determined from a limited number of lines. We therefore do not allow that any abundance can be measured more accurately than [FeI/H] and set the [FeI/H] dispersion as a lower limit, leading to a lower limit for the error estimate of $\frac{\sigma(FeI)}{\sqrt{N_{X}}}$. The maximum of these three values was adopted as the final error on [X/H]. It includes all the errors due to measurements and will be used throughout the rest of the paper. The errors on the abundance ratios [X/Fe] were subsequently computed as the quadratic sum of the errors on [X/H] and [Fe/H]. Averaged over the whole sample, they lead to representative error bars of 0.04 dex for [FeI/H], 0.19 dex for [Ba/Fe], 0.16 dex for [Ca/Fe] and 0.21 dex for [Mg/Fe].\\

\indent The uncertainties resulting from the choice of the atmospheric parameters have been estimated by varying one by one these parameters by their uncertainties ($\Delta$T$_{\rm eff}$=$\pm$200K, $\Delta$$log~g$=$\pm$0.5 dex, $\Delta$v$_{t}$=$\pm$0.3 km s$^{-1}$) for the stars MKV0770 and MKV0900, and determining how this modification affects the nominal abundances. The overall uncertainties due to atmospheric parameters are then calculated as the quadratic sum of the individual uncertainties due to T$_{\rm eff}$, $log~g$ and v$_{t}$. By construction, this method ignores covariances between the atmospheric parameters and thus overestimates the total error \citep{McWil1995}. The uncertainties due to the determination of atmospheric parameters are given in Table~\ref{atm_errors}.\\

\begin{table}[!h]
\centering
\caption{Errors on abundances due to sensitivity on the stellar parameters, computed for MKV0770 and MKV0900.}
\begin{tabular}{crrcc}
\hline\hline
Element & $\Delta$T$_{\rm eff}$= & $\Delta$$log g$= & $\Delta$v$_{t}$=     & Quadratic \\
        &   $\pm$200 K       &   $\pm$0.5 dex   & $\pm$0.3 km s$^{-1}$ &   sum     \\
\hline
$\Delta$[BaII/H] & -0.04 &  0.17 & 0.28 & 0.33 \\{}
$\Delta$[CaI/H]  & -0.27 &  0.04 & 0.07 & 0.28 \\{}
$\Delta$[CoI/H]  & -0.14 &  0.10 & 0.12 & 0.21 \\{}
$\Delta$[CrI/H]  & -0.36 &  0.03 & 0.32 & 0.48 \\{}
$\Delta$[FeI/H]  & -0.12 &  0.09 & 0.18 & 0.23 \\{}
$\Delta$[FeII/H] &  0.34 &  0.23 & 0.11 & 0.42 \\{}
$\Delta$[MgI/H]  & -0.10 & -0.08 & 0.15 & 0.20 \\{}
$\Delta$[MnI/H]  & -0.27 &  0.06 & 0.07 & 0.29 \\{}
$\Delta$[NdII/H] &  0.00 &  0.19 & 0.03 & 0.19 \\{}
$\Delta$[NiI/H]  & -0.01 &  0.06 & 0.03 & 0.07 \\{}
$\Delta$[TiI/H]  & -0.38 &  0.06 & 0.07 & 0.39 \\{}
$\Delta$[TiII/H] &  0.08 &  0.20 & 0.13 & 0.25 \\
\hline
\end{tabular}
\label{atm_errors}
\end{table}

\indent The hyperfine structure (HFS) of a line tends to desaturate it, which can lead to an overestimate of the abundance of an element. Following \cite{Let2010}, we applied a line-by-line HFS correction to La and Eu lines in the few cases where we managed to detect these elements. This correction only depends on the EW of the lines.\\

\newpage
\section{Results.}
\label{results}

We have determined the atmospheric parameters and detailed abundances for 35 RGB stars in the Carina dSph. These abundances are listed in Tables~\ref{abund1}, \ref{abund2} and \ref{fewdetect} and the abundance ratios can be found in Table~\ref{abratio}. The solar abundances of \cite{And1989} have been adopted in this study, except for Ti, Fe and La for which the values of \cite{Gre1998} have been chosen.

\begin{table*}[!h]
\caption{The abundances and errors of iron and $\alpha$-elements in our sample of RGB stars in the Carina dSph. The number of lines used to determine each value is given in brackets.}
\begin{tabular}{ccccccc}
\hline
\hline
star    &       [Fe1/H]       &       [Fe2/H]      &       [Mg/H]       &      [Ca/H]       &       [Ti1/H]      &       [Ti2/H]      \\
        &         dex         &         dex        &         dex        &        dex        &         dex        &         dex        \\
\hline											    					     		     
MKV0397 & -1.99$\pm$0.05 (14) & -1.85$\pm$0.08 (5) &          -         & -1.60$\pm$0.17 (1)&          -         & -1.84$\pm$0.12 (2) \\
MKV0458 & -1.60$\pm$0.03 (22) & -1.09$\pm$0.06 (5) & -1.53$\pm$0.14 (1) & -1.18$\pm$0.25 (2)& -1.36$\pm$0.08 (3) & -1.23$\pm$0.10 (2) \\
MKV0514 & -2.32$\pm$0.04 (28) & -1.80$\pm$0.16 (3) & -2.10$\pm$0.23 (1) & -2.16$\pm$0.10 (5)&          -         & -2.19$\pm$0.16 (2) \\
MKV0556 & -1.57$\pm$0.04 (13) & -0.45$\pm$0.23 (3) & -1.09$\pm$0.16 (1) & -1.42$\pm$0.09 (4)&          -         & -0.80$\pm$0.26 (1) \\
MKV0577 & -1.57$\pm$0.05 (24) & -0.78$\pm$0.26 (1) & -1.72$\pm$0.26 (1) &          -        & -1.15$\pm$0.26 (1) & -1.34$\pm$0.26 (1) \\
MKV0596 & -1.54$\pm$0.04 (31) & -1.37$\pm$0.08 (7) & -1.72$\pm$0.21 (1) & -1.76$\pm$0.09 (5)& -1.22$\pm$0.21 (1) & -1.57$\pm$0.15 (3) \\
MKV0614 & -1.57$\pm$0.04 (22) & -1.16$\pm$0.20 (3) & -1.54$\pm$0.18 (1) &          -        & -1.37$\pm$0.18 (1) & -1.07$\pm$0.14 (2) \\
MKV0628 & -1.67$\pm$0.03 (28) & -1.15$\pm$0.09 (4) & -1.26$\pm$0.17 (1) & -1.56$\pm$0.07 (6)& -2.01$\pm$0.17 (1) & -1.08$\pm$0.12 (2) \\
MKV0640 & -1.73$\pm$0.04 (31) & -1.42$\pm$0.09 (7) & -1.42$\pm$0.24 (1) & -1.58$\pm$0.08 (8)& -1.65$\pm$0.17 (2) & -1.05$\pm$0.17 (2) \\
MKV0652 & -2.29$\pm$0.08 (21) & -2.05$\pm$0.36 (1) & -2.25$\pm$0.36 (1) & -2.61$\pm$0.36 (1)& -1.24$\pm$0.36 (1) & -2.50$\pm$0.36 (1) \\
MKV0677 & -1.75$\pm$0.03 (27) & -1.36$\pm$0.07 (8) & -1.40$\pm$0.14 (1) & -1.59$\pm$0.06 (8)& -1.93$\pm$0.08 (3) & -1.09$\pm$0.14 (2) \\
MKV0698 & -1.48$\pm$0.02 (19) & -0.96$\pm$0.07 (5) & -1.33$\pm$0.08 (1) & -1.42$\pm$0.04 (8)& -1.62$\pm$0.04 (4) & -1.41$\pm$0.06 (2) \\
MKV0708 & -1.57$\pm$0.07 (14) & -1.34$\pm$0.18 (2) & -1.29$\pm$0.25 (1) &          -        &          -         & -0.90$\pm$0.25 (1) \\
MKV0729 & -1.39$\pm$0.04 (21) & -1.13$\pm$0.15 (4) & -1.00$\pm$0.19 (1) & -1.25$\pm$0.10 (4)& -1.35$\pm$0.13 (3) & -1.36$\pm$0.15 (2) \\
MKV0733 & -1.64$\pm$0.05 (19) & -1.10$\pm$0.14 (4) & -1.89$\pm$0.21 (1) & -1.74$\pm$0.18 (3)&          -         & -1.55$\pm$0.21 (1) \\
MKV0740 & -1.20$\pm$0.09 (20) & -0.56$\pm$0.21 (4) & -1.93$\pm$0.40 (1) & -1.57$\pm$0.40 (1)& -0.79$\pm$0.40 (1) & -1.16$\pm$0.23 (3) \\
MKV0743 & -1.21$\pm$0.08 (17) & -1.22$\pm$0.18 (3) & -2.16$\pm$0.32 (1) & -1.00$\pm$0.32 (1)& -0.67$\pm$0.23 (2) & -0.90$\pm$0.23 (2) \\
MKV0770 & -1.63$\pm$0.03 (23) & -1.26$\pm$0.06 (7) & -1.38$\pm$0.14 (1) & -1.50$\pm$0.07 (4)&          -         & -1.19$\pm$0.10 (2) \\
MKV0780 & -1.78$\pm$0.05 (32) & -1.39$\pm$0.11 (7) & -1.98$\pm$0.30 (1) & -1.78$\pm$0.21 (2)&          -         & -1.55$\pm$0.25 (2) \\
MKV0812 & -1.34$\pm$0.03 (10) & -0.93$\pm$0.40 (2) &          -         & -1.15$\pm$0.13 (1)&          -         & -0.25$\pm$0.47 (2) \\
MKV0825 & -1.43$\pm$0.03 (29) & -1.04$\pm$0.08 (8) & -1.26$\pm$0.18 (1) & -1.38$\pm$0.09 (4)& -1.56$\pm$0.09 (4) & -1.03$\pm$0.13 (2) \\
MKV0840 & -1.18$\pm$0.03 (24) & -0.89$\pm$0.09 (6) & -0.95$\pm$0.15 (1) & -1.21$\pm$0.09 (6)& -1.97$\pm$0.15 (1) & -1.09$\pm$0.28 (3) \\
MKV0842 & -1.47$\pm$0.02 (14) & -1.06$\pm$0.10 (4) & -1.20$\pm$0.10 (1) &          -        &          -         & -1.26$\pm$0.06 (3) \\
MKV0880 & -1.58$\pm$0.03 (36) & -1.15$\pm$0.08 (5) & -1.37$\pm$0.18 (1) & -1.55$\pm$0.07 (6)& -1.84$\pm$0.08 (5) & -1.33$\pm$0.12 (3) \\
MKV0900 & -1.72$\pm$0.02 (31) & -1.06$\pm$0.22 (3) & -1.45$\pm$0.22 (1) & -1.72$\pm$0.05 (7)& -2.05$\pm$0.09 (3) & -1.38$\pm$0.10 (3) \\
MKV0902 & -1.99$\pm$0.02 (22) & -1.58$\pm$0.14 (4) & -2.42$\pm$0.11 (1) & -2.12$\pm$0.08 (2)&          -         & -1.84$\pm$0.06 (3) \\
MKV0914 & -2.51$\pm$0.07 (17) & -2.15$\pm$0.20 (2) & -2.10$\pm$0.28 (1) & -2.33$\pm$0.21 (2)&          -         & -2.29$\pm$0.28 (1) \\
MKV0916 & -1.51$\pm$0.05 (16) & -0.89$\pm$0.24 (4) & -1.42$\pm$0.21 (1) &          -        &          -         & -0.49$\pm$0.47 (2) \\
MKV0925 & -1.55$\pm$0.04 (20) & -1.44$\pm$0.10 (3) & -1.14$\pm$0.17 (1) &          -        &          -         & -0.95$\pm$0.17 (1) \\
MKV0948 & -2.04$\pm$0.04 (20) & -1.82$\pm$0.18 (2) &          -         & -2.09$\pm$0.16 (1)&          -         & -1.52$\pm$0.16 (1) \\
MKV0976 & -1.24$\pm$0.06 (16) & -0.98$\pm$0.23 (1) & -1.22$\pm$0.23 (1) & -1.28$\pm$0.23 (1)& -0.77$\pm$0.23 (1) &          -         \\
MKV1007 & -1.39$\pm$0.06 (21) & -1.16$\pm$0.16 (3) & -1.11$\pm$0.27 (1) & -1.14$\pm$0.19 (2)& -1.38$\pm$0.12 (5) & -1.13$\pm$0.19 (2) \\
MKV1009 & -1.75$\pm$0.04 (32) & -1.22$\pm$0.12 (3) & -1.64$\pm$0.20 (1) & -1.41$\pm$0.08 (6)&          -         & -1.45$\pm$0.20 (1) \\
MKV1012 & -1.60$\pm$0.04 (26) & -1.18$\pm$0.11 (5) & -1.30$\pm$0.20 (1) & -1.46$\pm$0.09 (5)& -1.68$\pm$0.20 (1) & -1.79$\pm$0.14 (2) \\
MKV1061 & -1.50$\pm$0.05 (27) & -1.12$\pm$0.11 (6) & -0.97$\pm$0.24 (1) & -1.18$\pm$0.24 (1)& -1.43$\pm$0.17 (2) & -1.02$\pm$0.17 (2) \\
\hline
\end{tabular}
\label{abund1}
\end{table*}

\begin{table*}[!h]
\caption{The abundances and errors of iron-peak and heavy elements in our sample of RGB stars in the Carina dSph. The number of lines used to determine each value is given in brackets.}
\begin{tabular}{ccccccc}
\hline
\hline
star    &       [Fe1/H]       &       [Cr/H]        &       [Co/H]       &       [Ni/H]       &       [Ba/H]       \\   
        &         dex         &         dex         &        dex         &         dex        &         dex        \\	
\hline	 		       			   			  		       			  	 		       
MKV0397 & -1.99$\pm$0.05 (14) & -2.80$\pm$0.17 (1)  &          -         &          -         & -2.86$\pm$0.17 (1) \\ 	
MKV0458 & -1.60$\pm$0.03 (22) & -1.77$\pm$0.14 (1)  &          -         & -1.07$\pm$0.14 (1) & -1.16$\pm$0.10 (2) \\ 	
MKV0514 & -2.32$\pm$0.04 (28) & -2.66$\pm$0.23 (1)  &          -         & -2.28$\pm$0.23 (1) & -3.31$\pm$0.16 (2) \\ 	
MKV0556 & -1.57$\pm$0.04 (13) & -1.36$\pm$0.11 (2)  &          -         &          -         & -1.17$\pm$0.21 (1) \\ 	
MKV0577 & -1.57$\pm$0.05 (24) & -1.51$\pm$0.26 (1)  &          -         &          -         & -1.23$\pm$0.26 (1) \\ 	
MKV0596 & -1.54$\pm$0.04 (31) & -1.39$\pm$0.21 (1)  & -1.58$\pm$0.21 (1) & -1.06$\pm$0.21 (1) & -1.45$\pm$0.15 (2) \\ 	
MKV0614 & -1.57$\pm$0.04 (22) & -2.00$\pm$0.18 (1)  &          -         & -1.56$\pm$0.18 (1) & -1.17$\pm$0.18 (2) \\ 	
MKV0628 & -1.67$\pm$0.03 (28) & -1.69$\pm$0.12 (2)  & -1.60$\pm$0.17 (1) &          -         & -1.30$\pm$0.17 (1) \\ 	
MKV0640 & -1.73$\pm$0.04 (31) & -1.66$\pm$0.17 (2)  & -1.59$\pm$0.24 (1) & -1.12$\pm$0.24 (1) & -1.00$\pm$0.24 (1) \\ 	
MKV0652 & -2.29$\pm$0.08 (21) & -3.03$\pm$0.36 (1)  &          -         &          -         &          -         \\ 	
MKV0677 & -1.75$\pm$0.03 (27) & -1.45$\pm$0.14 (1)  & -1.86$\pm$0.14 (1) &          -         & -1.48$\pm$0.10 (2) \\ 	
MKV0698 & -1.48$\pm$0.02 (19) & -1.73$\pm$0.08 (2)  &          -         &          -         &          -         \\ 	
MKV0708 & -1.57$\pm$0.07 (14) & -1.60$\pm$0.25 (1)  &          -         & -1.11$\pm$0.25 (1) &          -         \\ 	
MKV0729 & -1.39$\pm$0.04 (21) & -1.09$\pm$0.19 (1)  & -1.28$\pm$0.19 (1) & -0.84$\pm$0.19 (1) & -1.41$\pm$0.19 (2) \\ 	
MKV0733 & -1.64$\pm$0.05 (19) & -2.29$\pm$0.21 (1)  &          -         & -1.49$\pm$0.21 (1) & -1.87$\pm$0.21 (1) \\ 	
MKV0740 & -1.20$\pm$0.09 (20) & -1.72$\pm$0.40 (1)  & -1.46$\pm$0.40 (1) &          -         &          -         \\ 	
MKV0743 & -1.21$\pm$0.08 (17) & -1.55$\pm$0.32 (1)  &          -         & -0.90$\pm$0.32 (1) & -1.30$\pm$0.32 (1) \\ 	
MKV0770 & -1.63$\pm$0.03 (23) & -1.65$\pm$0.14 (1)  & -1.76$\pm$0.14 (1) & -1.22$\pm$0.10 (2) &          -         \\ 	
MKV0780 & -1.78$\pm$0.05 (32) & -2.50$\pm$0.30 (1)  &          -         & -1.99$\pm$0.30 (1) & -1.81$\pm$0.30 (1) \\ 	
MKV0812 & -1.34$\pm$0.03 (10) & -1.35$\pm$0.15 (1)  &          -         &          -         & -1.48$\pm$0.15 (1) \\ 	
MKV0825 & -1.43$\pm$0.03 (29) & -1.33$\pm$0.18 (1)  & -0.96$\pm$0.18 (1) & -1.40$\pm$0.18 (1) & -1.17$\pm$0.18 (1) \\ 	
MKV0840 & -1.18$\pm$0.03 (24) & -1.24$\pm$0.15 (1)  &          -         & -0.96$\pm$0.16 (1) & -0.93$\pm$0.11 (2) \\ 	
MKV0842 & -1.47$\pm$0.02 (14) & -1.76$\pm$0.16 (1)  & -1.24$\pm$0.11 (1) & -1.21$\pm$0.42 (1) & -2.10$\pm$0.16 (1) \\ 	
MKV0880 & -1.58$\pm$0.03 (36) & -2.01$\pm$0.18 (1)  & -1.42$\pm$0.18 (1) & -1.40$\pm$0.14 (2) & -1.56$\pm$0.18 (1) \\ 	
MKV0900 & -1.72$\pm$0.02 (31) & -1.84$\pm$0.12 (1)  & -1.80$\pm$0.12 (1) & -1.66$\pm$0.12 (1) & -1.31$\pm$0.12 (1) \\ 	
MKV0902 & -1.99$\pm$0.02 (22) &          -          &          -         & -2.71$\pm$0.18 (1) & -2.43$\pm$0.14 (1) \\ 	
MKV0914 & -2.51$\pm$0.07 (17) & -3.09$\pm$0.28 (1)  &          -         & -2.74$\pm$0.28 (1) & -3.06$\pm$0.28 (1) \\ 	
MKV0916 & -1.51$\pm$0.05 (16) & -1.64$\pm$0.21 (1)  &          -         & -1.33$\pm$0.21 (1) &          -         \\ 	
MKV0925 & -1.55$\pm$0.04 (20) & -2.09$\pm$0.17 (1)  &          -         & -1.12$\pm$0.27 (1) & -1.87$\pm$0.12 (2) \\ 	
MKV0948 & -2.04$\pm$0.04 (20) & -2.39$\pm$0.16 (1)  &          -         &          -         & -1.17$\pm$0.11 (2) \\ 	
MKV0976 & -1.24$\pm$0.06 (16) & -0.83$\pm$0.23 (1)  &          -         & -1.55$\pm$0.23 (1) & -0.92$\pm$0.23 (1) \\ 	
MKV1007 & -1.39$\pm$0.06 (21) & -1.15$\pm$0.27 (1)  & -0.89$\pm$0.27 (1) &          -         & -0.82$\pm$0.27 (1) \\ 	
MKV1009 & -1.75$\pm$0.04 (32) & -1.49$\pm$0.20 (1)  &          -         &          -         & -1.30$\pm$0.22 (1) \\ 	
MKV1012 & -1.60$\pm$0.04 (26) & -1.79$\pm$0.20 (1)  & -1.62$\pm$0.20 (1) & -1.61$\pm$0.14 (2) & -1.43$\pm$0.14 (2) \\ 	
MKV1061 & -1.50$\pm$0.05 (27) & -1.53$\pm$0.24 (1)  & -1.54$\pm$0.24 (1) & -1.01$\pm$0.24 (1) &          -         \\	
\hline
\end{tabular} 
\label{abund2}
\end{table*}



\subsection{Iron and iron-peak elements.}
\label{iron}

{\it The metallicity distribution of our sample:}\\

The iron abundances of our sample are determined from between 15 and 36 FeI lines (see Table \ref{abund1}) which are all in good agreement. Our sample spans a metallicity range from [Fe/H] $\approx$ $-$1.2 dex to [Fe/H] $\approx$ $-$2.5 dex. It is clearly biased towards higher metallicities, with 24 stars more metal-rich than the mean [Fe/H] of $\approx$ $-$1.70 dex and only 6 stars with [Fe/H] $\leq$ $-$2.0 dex. An obvious explanation for this bias is that at lower S/N, the analysis of metal-rich stars remains possible while the weaker lines of the more metal-poor stars are drowned in the noise. \\

{\it A comparison with the Calcium Triplet MDF:}\\
\label{CaT}

   \begin{figure}[!ht]
   \centering
   \includegraphics[width=\columnwidth]{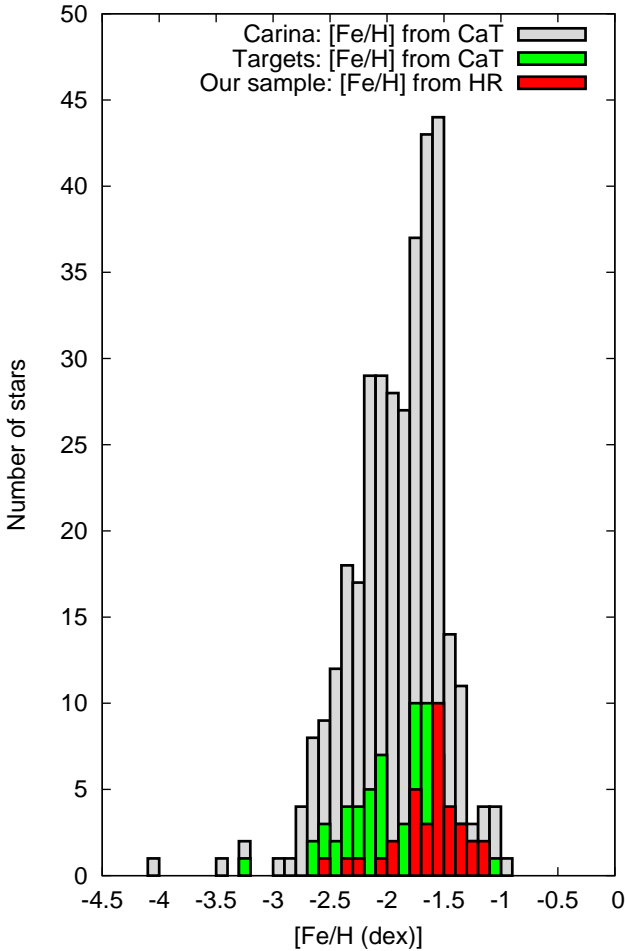}
      \caption{The [Fe/H] distribution of our sample (in red) and the full HR sample showing the stronger effect of low S/N on more metal poor stars (in green). The larger set of CaT measurements \citep{Stark2010} is in gray for comparison.}
         \label{MDF}
   \end{figure}

Figure \ref{MDF} shows the metallicity distribution of our sample, compared to the larger sample with low-resolution spectroscopy in the CaT region \citep[from][]{Stark2010}. The effect of our metallicity bias can clearly be seen. We also show our full sample and how metal-poor stars are preferentially excluded due to the weak lines being more sensitive to S/N.\\

\begin{figure}[!htb]
   \centering
   \includegraphics[angle=-90,width=\columnwidth]{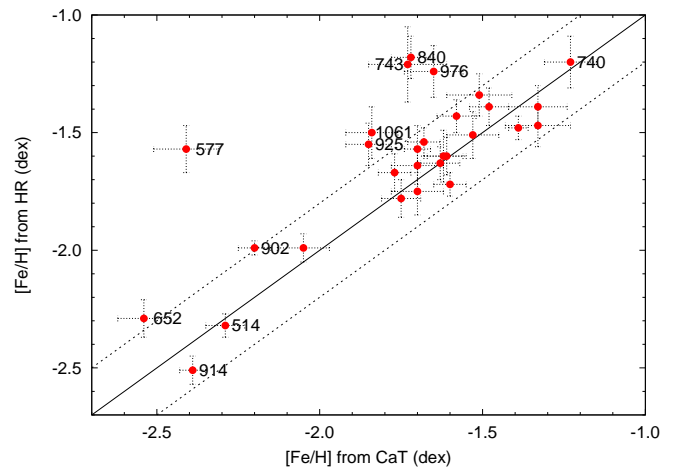}
      \caption{For our sample of RGB stars in the Carina dSph, we show the comparison between [Fe/H] measured from our high resolution spectroscopy and from the CaT \citep{Stark2010}. The solid line indicates where [Fe/H]$_{CaT}$=[Fe/H]$_{HR}$ while the dashed lines indicate an uncertainty of $\pm$0.2 dex.}
         \label{CaT-HR}
   \end{figure}

For 29 out of the 35 stars in our sample, both the CaT metallicity and our high resolution determination of [Fe/H] are available. In general, they agree well, despite the presence of outliers (see Fig. \ref{CaT-HR}). Some of these outliers are stars with large errors on [Fe/H], but we have carefully investigated all of them and found no obvious error in the high resolution analysis. 
To try to understand the origin of the differences we first consider [Ca/Fe], which was found to be the dominant factor driving the EW of the CaT lines \citep{Stark2010}. Unfortunately we couldn't measure Ca in the most distant outlier, MKV0577, that is also Mg-depleted, with [Mg/Fe]$\approx$ $-$0.15 dex. As Mg is one of the main electron donors, a depletion in Mg lowers the H$^{-}$ concentration and hence affects the continuous opacity and the temperature structure of the star. Several other outliers also have extreme [Ca/Fe] and/or [Mg/Fe] ratios, that sometimes fall out of the range (+0.0 $<$ [$\alpha$/Fe] $<$ +0.4 dex) of the stellar models used to calibrate the CaT vs [Fe/H] relation e.g. MKV0743, MKV0925, MKV1061 (see Table~\ref{abratio}). However, this cannot be considered as a general pattern as two stars (MKV0840, MKV0976) with a rather common chemical composition also don't match very well and conversely the agreement between both methods is very good for two other stars (MKV0740, MKV0902) which are both strongly depleted in Mg and/or Ca.

\cite{Venn2011} report 2 stars with large under-abundances for all elements measured, except the iron-peak and proposed that these stars are iron enriched. Both of them also show discrepancies of the order of 0.5 dex between metallicities obtained from a classical analysis and those derived from the CaT. \cite{She2009} already suggested that the CaT calibration is sensitive to changes in the continuous opacity and hence related to the global budget of free electrons (mostly released by metals) rather than to any particular element. As a consequence, the CaT feature might not be such a good proxy for the iron abundance for stars with a peculiar chemical composition. For large samples of stars with more standard abundance patterns, the metallicity estimates from the CaT agree very well in the mean (within 0.1-0.2 dex over the range $-$2.5$<$[Fe/H]$<$ $-$0.5) with high resolution spectroscopic determinations as shown in \cite{Batt2008}.\\ 

{\it Other iron peak elements: Sc, Cr, Mn, Co, Ni:}\\

The iron peak elements are synthesized in small quantities during SNe II events, but mostly come from SNe Ia. We were able to measure Sc, Cr, Mn, Co and Ni for many stars in our sample (see Tables~\ref{abund2} and \ref{fewdetect}). Cr I (measured from 2 lines) is, as expected, underabundant for metal-poor stars and the Cr abundance increases with [Fe/H], with a handful of stars remaining underabundant at intermediate metallicities. Mn (one line) and Co (up to four lines) show the same behaviour (see Table~\ref{abund2}). The situation is more complex in the case of Ni, where the abundance is derived from only 1-2 lines (see Table~\ref{abund2}). We find a global increase of [Ni/Fe] with [Fe/H], and the persistence of Ni-poor stars at intermediate metallicities. Our sample also harbours a group of Ni-rich ($>$+0.5 dex) stars. Sc, measured from one line, was detected in only two stars (see Table~\ref{fewdetect}).\\

\subsection{$\alpha$-elements.}

The $\alpha$-elements are named after their formation process, the capture of He nuclei during different stages of stellar evolution. They are mainly formed and released when high mass ($>$8 M$_{\sun}$), short lived stars explode as type II supernovae (SNe II).

The change in slope of measurements in a [Mg/Fe] versus [Fe/H] plot is often called a ``knee'' \citep[e.g.][]{Matt1990,Matt2003} whose position indicates the level of Fe-enrichment reached by a system when SNe Ia begin to influence the chemical evolution. We know from CMD analysis that Local Group dSphs experienced a range of different SFHs, and so we expect that the ``knee'' will occur at different [Fe/H] in different galaxies. At present, the only dSph for which the knee position is clear (at [Fe/H]=$-$1.8 dex) is Sculptor \citep{Tolstoy2009,Hill2011}. In the case of the Fornax dSph, current data suggests that the knee should occur below [Fe/H] $< -$1.5 dex, while in the case of Sagittarius it could be at [Fe/H] $< -$1.0 dex. It might be expected that the Carina dSph will have more than one ``knee'' given the multiple star formation episodes separated in time \citep[e.g.,][]{Gil1991}.

We focus on Mg and Ca results, where we also add to our measurements literature samples \citep{She2003,Koch2008,Venn2011}. It consists of 5 stars for \cite{She2003}, 10 stars \cite{Koch2008} and 9 stars for \cite{Venn2011}, of which 4 are overlapping with \cite{Koch2008}. We also have 1 star in common with \cite{Koch2008}.\\

\begin{figure*}[!htb]
   \centering
   \includegraphics[angle=-90,width=0.80\textwidth]{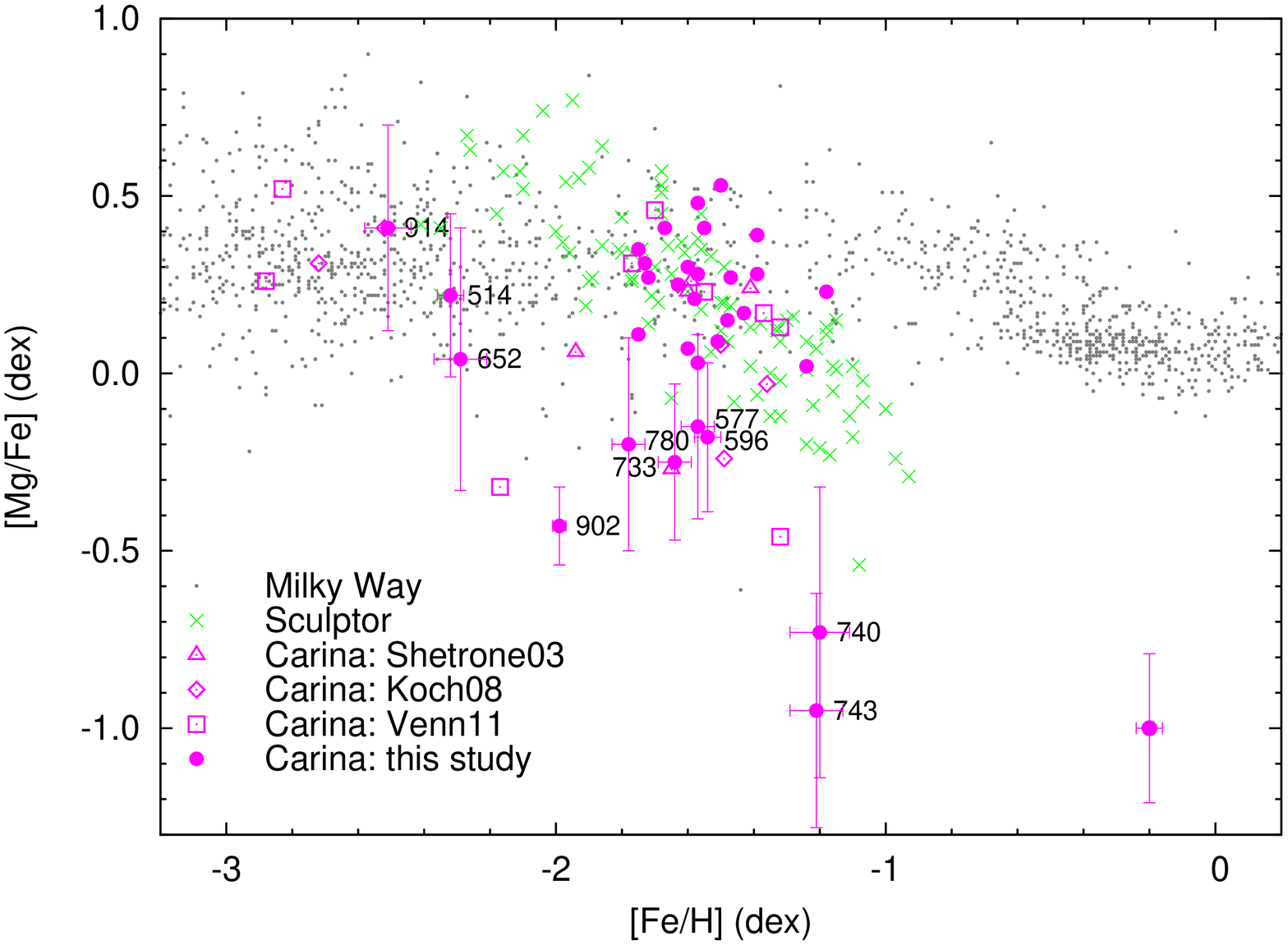}
      \caption{The distribution of [Mg/Fe] for our sample of RGB stars in the Carina dSph as pink filled circles. Also included in the plot are the 5 RGB stars in the Carina dSph from \cite{She2003}: pink open triangles; the 9 RGB stars from \cite{Venn2011}: pink open squares and 6 of the RGB stars from \cite{Koch2008}: pink open diamonds. Stars in Sculptor are in green crosses \citep{She2003,Gei2005,Hill2011}. Milky Way halo stars are in small grey dots (from \cite{Venn2004} and references therein). Individual error bars are given for some peculiar stars and a representative error bar for the rest is given in bottom right hand corner.}
         \label{MgFe}
   \end{figure*}

{\it Magnesium:} We can only determine [Mg/H] using a single, well defined line at $\lambda$=5528.41$\AA$ (see Table~\ref{abund1}). Three stars lack Mg abundances because of an EW in excess of 200~m\AA. Most of the stars in our sample are Mg-rich, with [Mg/Fe] $\approx$ 0.3 dex (see Fig. \ref{MgFe}). However, seven stars have low or extremely low [Mg/Fe]. MKV0740 and MKV0743 have among the lowest [Mg/Fe] ever reported, although the errors on the abundances of these stars are fairly large. As can be seen in Fig.~\ref{MgFe}, the many RGB stars in the Carina dSph have [Mg/Fe] similar to Galactic halo stars at same [Fe/H]. The same feature can also be observed for similar metallicity stars in the Sculptor dSph, with about the same [Fe/H] limit \citep{Tolstoy2009,Hill2011}.\\ 

There does appear to be a general downward trend of the mean [Mg/Fe] with increasing [Fe/H] for the Carina dSph stars in Fig.~\ref{MgFe}, but we need ages of the stars to disentangle the different star formation episodes. It is clear that the Carina dSph has not followed the same chemical enrichment path as the Sculptor dSph. The sparse numbers of low [Fe/H] stars measured in the Carina dSph have lower [Mg/Fe] than in the Sculptor dSph. There is an overlap of the majority of stars for the more [Fe/H] rich stars in the Carina dSph with those in Sculptor, but these stars have a much higher scatter in Carina, going down to extremely underabundant [Mg/Fe].\\

%

   \begin{figure*}[!hbt!]
   \centering
   \includegraphics[angle=-90,width=0.80\textwidth]{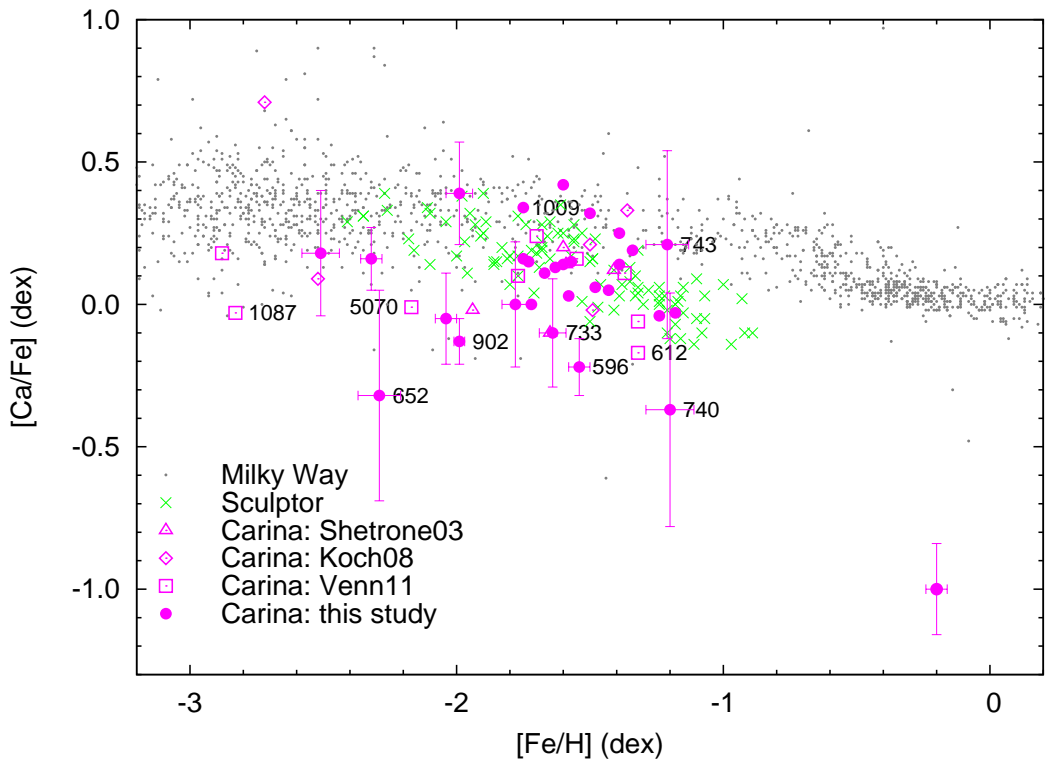}
      \caption{The distribution of [Ca/Fe] for our sample of RGB stars in the Carina dSph. Additional data, definition of symbols and references are as in Fig.~\ref{MgFe}}
      \label{CaFe}
   \end{figure*}

{\it Calcium:} For our sample, as many as eight lines can be used to determine [Ca/Fe] (see Table~\ref{abund1}). As with Mg, our sample is mostly Ca-rich (see Fig. \ref{CaFe}), but some stars have low [Ca/Fe]. The samples of \cite{She2003}, \cite{Koch2008} and \cite{Venn2011} also include stars with similarly low [Ca/Fe]. Both the spread and the extremes are much less pronounced than for [Mg/Fe]. Magnesium (and Oxygen) are produced during the hydrostatic He burning in massive stars, and therefore their yields are not predicted to be affected by the SNe II explosion. Conversely, Ca (and Si, Ti) are formed during the SNe II explosion itself and thus their yields depend sensitively on the energy of the SNe. However, explosion energies cover only a small energy range which leads to a small scatter in [Ca/Fe].\\

In Fig.~\ref{CaFe}, we see that [Ca/Fe] for the Carina dSph stars overlaps less markedly the halo stars of the Milky Way than [Mg/Fe] (in Fig.~\ref{MgFe}). This is also true for the Sculptor dSph measurements. The [Ca/Fe], like [Mg/Fe], for the Carina dSph stars does overlap the measurements of Sculptor stars, but as for [Mg/Fe] the scatter in the Carina measurements is larger, with a number of outliers with particularly low [Ca/Fe]. In most cases, stars with a low [Mg/Fe] also have a low [Ca/Fe]

The [Ca/Fe] for the Carina dSph (Fig.~\ref{CaFe}) shows no obvious knee, but rather a shallow downward trend, with an increased scatter at intermediate metallicities. This could be due to the overlapping influence of both SNe Ia and SNe II from distinct episodes of star formation.\\

{\it Titanium:} We could measure both TiI (5 lines) and TiII (1-5 lines) but the abundances obtained do not agree well with each other (see Table~\ref{abund1}). TiI is systematically lower than TiII. We expect this effect to be mainly due to NLTE effects, with a departure from ionization equilibrium affecting TiI. However, errors in the determination of T$_{\rm eff}$ could also shift the ionization equilibrium in the same way. This is the same discrepancy already reported by \cite{Let2010}, for the Fornax dSph, observed with the same FLAMES setup.
\cite{Let2010} warns us that CaI could also be affected by NLTE effects in the
same way as TiI, without any possibility to check since we cannot measure CaII
lines in these stars. \\

{\it Silicon:} The available Si lines are very weak and could only be measured in three stars (see Table~\ref{fewdetect}). Like the other $\alpha$-elements, Si is over-abundant.

\begin{table*}[!b]
\centering
\caption{The abundances and errors for elements that could be measured from single lines in only a handful of stars.}
\begin{tabular}{ccccccccc}
\hline
\hline
star    &       [Na/H]       &       [Si/H]       &       [Sc/H]       &       [Y/H]        &       [La/H]       &       [Nd/H]       &      [Eu/H]        \\  
        &         dex        &         dex        &         dex        &         dex        &         dex        &         dex        &        dex         \\  
\hline	 			       			 	       		 		     	      						      		       
MKV0514 &          -         & -0.90$\pm$0.23 (1) &          -         &          -         &          -         &          -         &          -         \\  
MKV0640 &          -         &          -         & -1.22$\pm$0.24 (1) & -1.66$\pm$0.24 (1) & -0.83$\pm$0.24 (1) &          -         &          -         \\  
MKV0825 &          -         &          -         &          -         &          -         & -0.86$\pm$0.18 (1) &          -         &          -         \\  
MKV0840 &          -         &          -         & -1.13$\pm$0.15 (1) &          -         & -0.81$\pm$0.15 (1) &          -         &          -         \\  
MKV0880 &          -         & -1.04$\pm$0.18 (1) &          -         & -1.53$\pm$0.18 (1) &          -         &          -         & -1.17$\pm$0.18 (1) \\  
MKV0900 &          -         &          -         &          -         &          -         &          -         & -1.20$\pm$0.12 (1) &          -         \\  
MKV1007 &          -         &          -         &          -         &          -         & -0.87$\pm$0.27 (1) &          -         &          -         \\  
MKV1061 & -1.14$\pm$0.24 (1) & -0.42$\pm$0.24 (1) &          -         &          -         &          -         &          -         &          -         \\  
\hline
\end{tabular}
\label{fewdetect}
\end{table*}

\subsection{Neutron-capture elements.}

Elements with atomic number Z$>$30 are neutron-capture elements. The two main ways to produce them are the slow {\it s}-process and the rapid {\it r}-process. For the {\it s}-process a longer period (with respect to the timescale for $\beta^{-}$ decay) elapses between successive neutron capture events. This process is thought to occur in thermally pulsating AGB stars, which are long-lived low or intermediate mass stars. The {\it r}-process in contrast, is where successive neutron captures occur on a very short timescale. The {\it r}-process divides into two different branches: the weak {\it r}-process, which forms the lighter n-capture elements (e.g. Sr, Y, Zr) and the main {\it r}-process, which forms the heavier n-capture elements. There remains significant uncertainty about exactly what is the origin of the r-process elements \citep[e.g.,][]{Sne2008}.

 We can measure [Ba/Fe] for the majority of stars in our sample. A few stars have Y, La, Eu and Nd measurements as well.\\

%

{\it Barium:}
Our analysis is based on 1-2 strong lines (see Table~\ref{abund2}).
The most metal-poor RGB stars in the Carina dSph have low [Ba/Fe] with a lot of scatter. Studies which could measure both [Ba/Fe] and [Eu/Fe] \citep{She2003,Venn2011} suggest that the Ba in very metal-poor stars has an {\it r}-process-only origin. This scatter is also seen in the RGB stars measured in the Sculptor dSph \citep{Hill2011}. Our sample, combined with \cite{She2003} and \cite{Venn2011} measurements, shows an increase of [Ba/Fe] with [Fe/H], once again similar to what is seen in Sculptor \citep{Hill2011}. This can be interpreted as an increasing contribution of the {\it s}-process from AGB stars \citep{She2001}. A couple of the stars in the Carina dSph reach high Ba abundances at rather low [Fe/H], more similar to Fornax \citep{Let2010}.\\

{\it Y, La, Eu, Nd:}
Only a handful of measurements could be made for these elements in our sample, and always in stars with an enhanced content in neutron-capture elements. [La/Fe] is consistent
with [Ba/Fe], where stars with high [Ba/Fe] also have high [La/Fe] (see Tables~\ref{abund2} and \ref{fewdetect}). 

\subsection{Comparison with previous work.}

We have no stars in common with \cite{She2003} or \cite{Venn2011}, and only one star in common with the UVES study of \cite{Koch2008}, our MKV0948 is his LG04d\_006628. Making a careful comparison we can see that the photometric T$_{\rm eff}$ values agree, however \cite{Koch2008} chose to use the spectroscopic T$_{\rm eff}$ in their study and so they had to increase the photometric estimate by $\approx$ 450 K to reach the excitation equilibrium for this star. As a consequence, our $log~g$ and v$_{t}$ are also different, though only slightly for v$_{t}$. In addition, for the few lines in common in both linelists, our EWs are systematically lower (by $\simeq$15 m$\AA$) than those of \cite{Koch2008}. Finding out why is not straightforward and seems to be more complicated than due only to a difference in continuum placement. Despite these differences, our abundances are in reasonably good agreement: [FeII/H] matches almost exactly and [FeI/H] differs by $-$0.25 dex (which is nearly within the error bars). [TiII/H] is also in good agreement, however [Ca/H] is strongly discrepant by $\sim$ +0.4 dex.

\section{Ages.}
\label{ages_sect}

Due to its distinct episodes of star formation, the age-metallicity
relation for stars in the Carina dSph is not expected to be as
straightforward as in more simple systems like the Sculptor dSph 
\citep[e.g.][]{DeBoer2011}. Accurate ages would allow us to determine the 
speed of chemical enrichment of different elements and hence the
relative importance of SNe II and SNe Ia at any given time.

\subsection{Age determination.}

Combining measurements of [Fe/H] and [Mg/Fe] with accurate photometry we can estimate the ages
of individual RGB stars using simple isochrone fitting. The age is determined by finding
the isochrone with appropriate [Fe/H] and [$\alpha$/Fe] abundances
that best matches the position of the star in the CMD.

\begin{figure}[!hbt]
   \centering
   \includegraphics[angle=-90,width=\columnwidth]{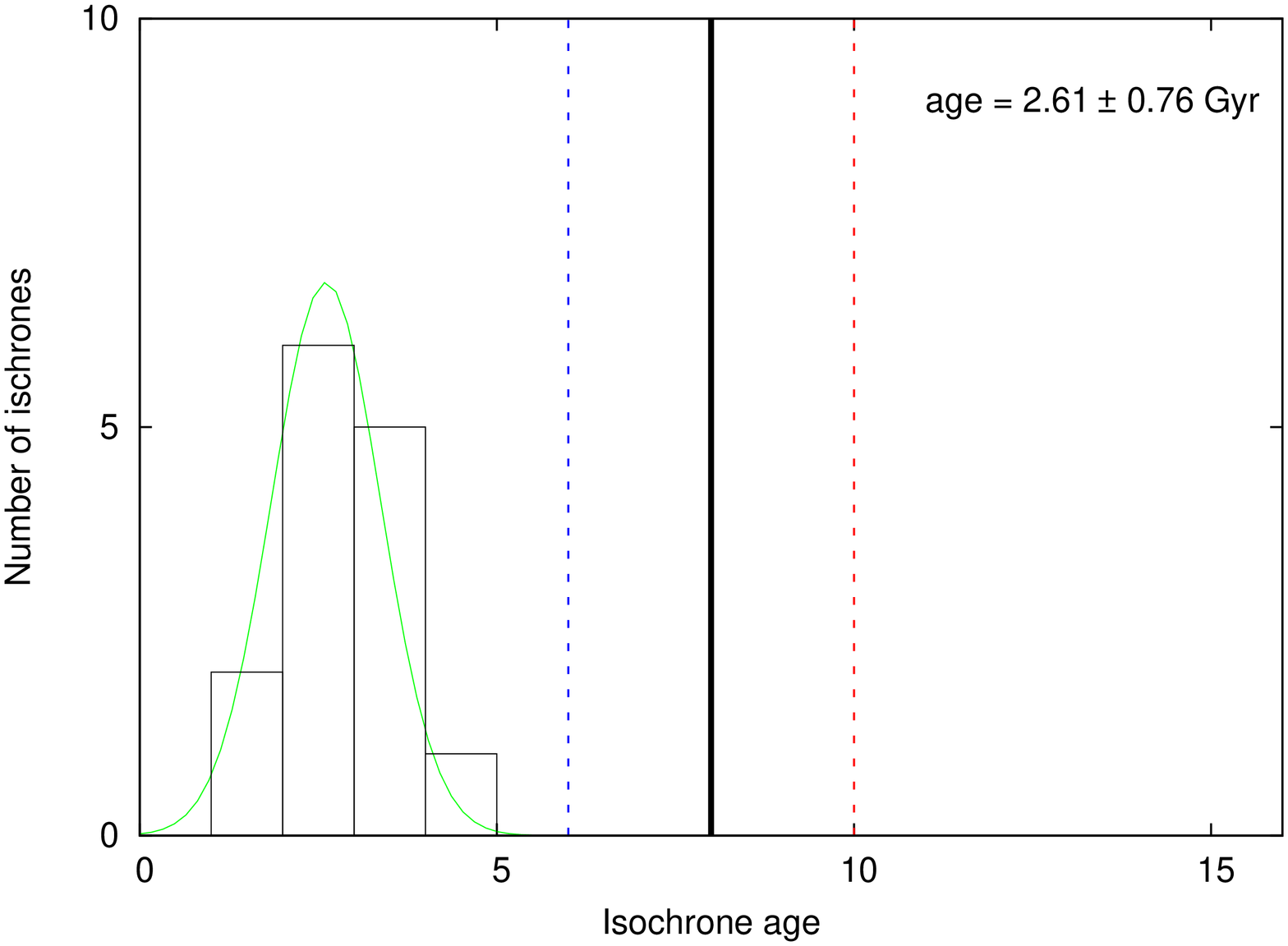}
   \includegraphics[angle=-90,width=\columnwidth]{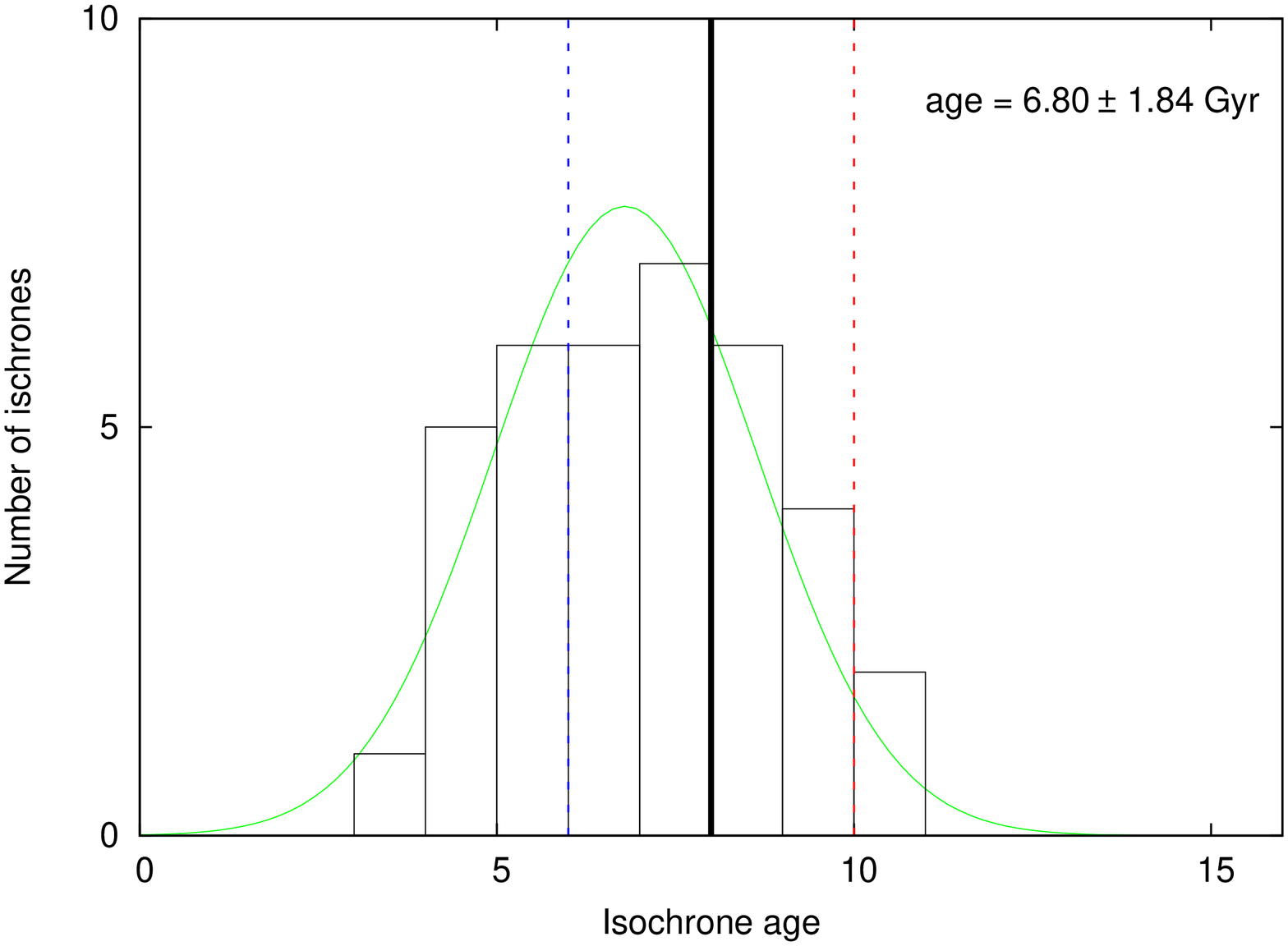}
      \caption{Age probability distributions built up from isochrone fitting and their Gaussian fits. We show here two examples: (a) MKV0976, age~=~2.61$\pm$0.76~Gyr and (b) MKV0614, age~=~6.80$\pm$1.84~Gyr.}
         \label{agedistrib}
   \end{figure}

We first built a grid of metallicities by dividing the possible [Fe/H] range~(given by the error-bars) into bins of 0.05 dex.
We used [Mg/Fe] as a proxy for [$\alpha$/Fe] (and [Ca/Fe] when [Mg/Fe] was not available). The isochrones do not sample the full [$\alpha$/Fe] range, so we used the lowest value available to compute the ages of the $\alpha$-poor stars. As a consequence, 
the accuracy of our ages will be reduced for these stars.\\
For each point on the metallicity grid, we generated isochrones using the Teramo/BaSTI stellar evolutionary models \citep{Pietri2004,Pietri2006} for ages between 1 and 15 Gyr, with a spacing of 0.5 Gyr. Those isochrones which are consistent with 
the observed colours and magnitudes within the photometric error-bars were then used to build up 
a probability distribution function for ages. Fig.~\ref{agedistrib}a shows the age probability distribution for an intermediate age 
star (MKV0976, age = 2.61$\pm$0.76 Gyr) while Fig.~\ref{agedistrib}b shows the age probability distribution for a star (MKV0614) with a more uncertain age (6.80$\pm$1.84 Gyr) but that probably still belongs to the intermediate age population. 
The age distribution is fit by a Gaussian where the mean is the likely age of the star and the standard 
deviation the error bar on the age. According to their colours, a few stars (7)
are found to be older than 14 Gyr and hence inconsistent with the
age of the Universe. It is likely that these are old stars, and so we
set their age to 15 Gyr (without error-bar) to identify the larger
than normal uncertainty in their true age.
\par We applied the same method to the RGB stars in the samples of \cite{She2003}, \cite{Koch2008} and \cite{Venn2011}, using homogeneous V and I magnitudes from ESO-WFI. The ages and their uncertainties of all the stars in the Carina dSph with HR abundances can be found in Table~\ref{abratio}.\\

\subsection{The age range: two populations.}

\begin{figure*}[h!]
   \centering
   \resizebox{\hsize}{!}{\includegraphics[angle=-90]{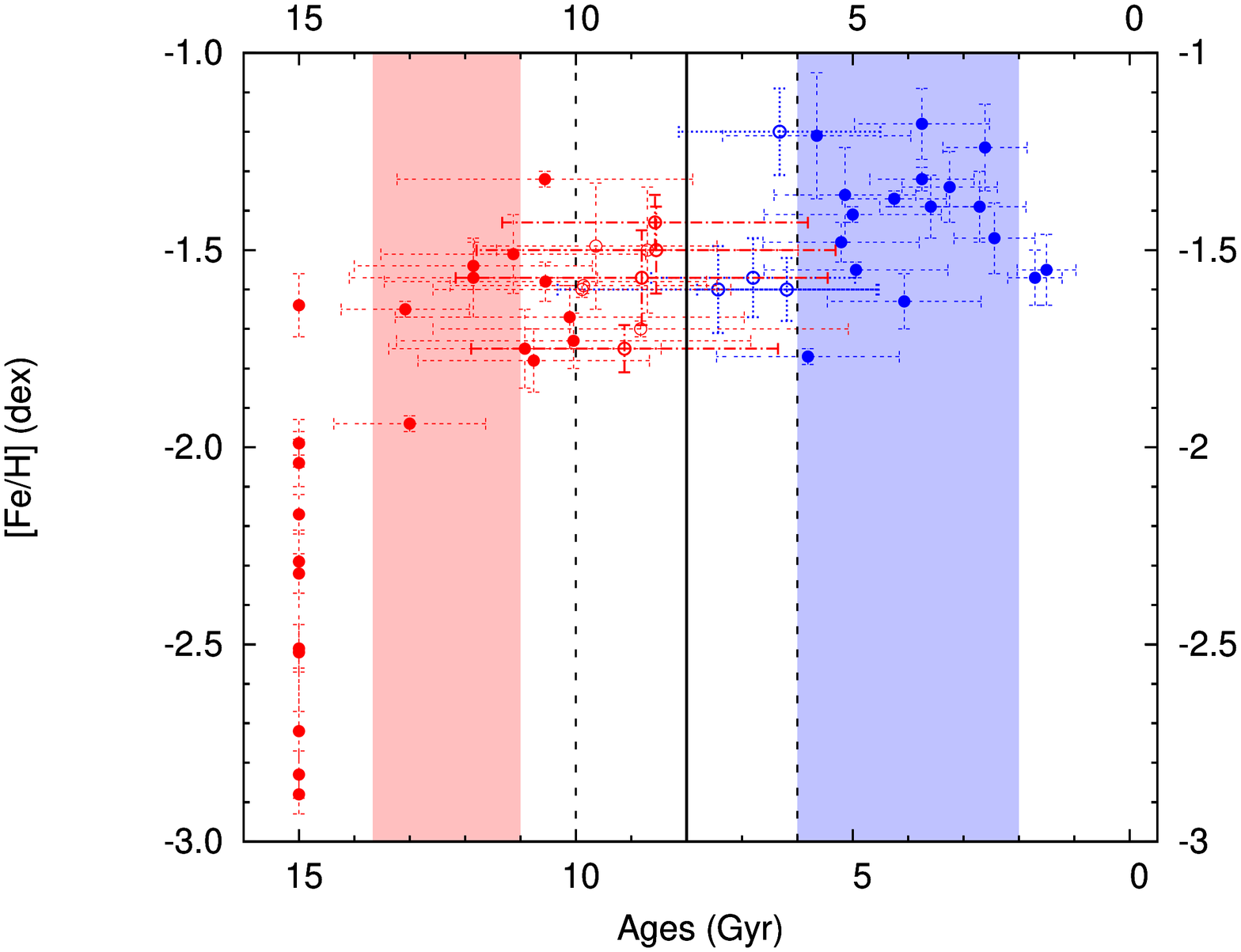}}
      \caption{ The ages (in Gyr) as a function of [Fe/H] (dex) for the RGB stars with high resolution abundances in the Carina dSph (see also Table~\ref{abratio}). The shaded areas represent the expected age ranges of star formation as indicated by photometric studies.}
         \label{ages}
   \end{figure*}

We noticed that the ages of the RGB stars in the Carina dSph
(Table~\ref{abratio}) fall broadly into two groups, as might be
expected from the SFH \citep{SH1996,Mon2003,Riz2003,Bono2010}. Because our ages have large error bars, our aim is simply to split our sample into an old population and an intermediate age population. Our abundances are of RGB stars, so we cannot observe stars from the very young population ($<$1 Gyr old). We chose 8 Gyr as the border between the old and intermediate age populations. Most of the photometric studies place the intermediate population in the 2~\textendash~6 Gyr range, while the old population should be older than 10 Gyr. The majority of our stars fall naturally into these age groups, and only a few stars lie near the threshold value (see Fig. \ref{ages}). Indeed, only 13 (out of 54) stars have ages between 6 and 10 Gyr, the rest are clearly younger or older than 8 Gyr. However it is still possible to attribute these 13 stars to either age group given their age probability distribution. The case of MKV0614 (see Fig. \ref{agedistrib}b) shows for example that it is probably an intermediate-age star, as most of the possible isochrones indicate an age $<$8~Gyr, though we cannot exclude an older age for this star.\\

\par To quantify the distribution of stars in both age groups, we computed the median and median absolute deviation (mad). The mad method is a robust estimate of the dispersion, much less sensitive to outliers than the standard deviation. This was done for two different samples: the whole sample of RGB stars in the Carina dSph with detailed abundances available; and a smaller sample in which all the stars whose age is between 6 and 10 Gyr (i. e., within $\pm$ 2 Gyr from the border) have been removed. The results shown in Table~\ref{mad} indicate that a more restricted sample has a negligible effect on the median and the mad for the intermediate age population. This is because they already had a narrower age distribution but also because only 4 stars have been removed. In the case of the older population, reducing the sample to those stars older than 10 Gyr only has a noticeable effect on the median age, that is shifted toward higher values by $\approx$2 Gyr. This effect is not surprising as 9 stars are removed and they are preferentially among the youngest of the old sample. 

\begin{table*}[!ht]
\centering
\caption{Ages and abundance ratios for all the stars in the Carina dSph with detailed abundance measurements. The first part of the table displays our new results and the second part shows abundances taken from the literature together with our age determinations. (1): \cite{She2003},(2): \cite{Venn2011}, (3):\cite{Koch2008}.}
\begin{tabular}{rrrrrr}
\hline\hline
Star~~~~&    [Fe/H]    &    [Ba/Fe]    &     [Ca/Fe]    &     [Mg/Fe]    &      Age~~~~~~ \\
        &       dex~~~ &       dex~~~~ &        dex~~~~ &        dex~~~~ &      Gyr~~~~~~ \\
\hline
MKV0397 & -1.99$\pm$0.05 & -0.87$\pm$0.18 &  0.39$\pm$0.18 &      -~~~~~~~~ & 15.00~~~~~~~~~~ \\ 
MKV0458 & -1.60$\pm$0.03 &  0.44$\pm$0.10 &  0.42$\pm$0.25 &  0.07$\pm$0.14 &  6.19$\pm$1.62 \\ 
MKV0514 & -2.32$\pm$0.04 & -0.99$\pm$0.16 &  0.16$\pm$0.11 &  0.22$\pm$0.23 & 15.00~~~~~~~~~~ \\ 
MKV0556 & -1.57$\pm$0.04 &  0.40$\pm$0.21 &  0.15$\pm$0.10 &  0.48$\pm$0.16 &  8.81$\pm$3.36 \\ 
MKV0577 & -1.57$\pm$0.05 &  0.34$\pm$0.26 &      -~~~~~~~~ & -0.15$\pm$0.26 & 11.85$\pm$2.24 \\ 
MKV0596 & -1.54$\pm$0.04 &  0.09$\pm$0.16 & -0.22$\pm$0.10 & -0.18$\pm$0.21 & 11.85$\pm$2.15 \\ 
MKV0614 & -1.57$\pm$0.04 &  0.40$\pm$0.18 &      -~~~~~~~~ &  0.03$\pm$0.18 &  6.80$\pm$1.84 \\ 
MKV0628 & -1.67$\pm$0.03 &  0.37$\pm$0.17 &  0.11$\pm$0.08 &  0.41$\pm$0.17 & 10.11$\pm$3.15 \\ 
MKV0640 & -1.73$\pm$0.04 &  0.73$\pm$0.24 &  0.15$\pm$0.09 &  0.31$\pm$0.24 & 10.04$\pm$3.20 \\ 
MKV0652 & -2.29$\pm$0.08 &      -~~~~~~~~ & -0.32$\pm$0.37 &  0.04$\pm$0.37 & 15.00~~~~~~~~~~ \\ 
MKV0677 & -1.75$\pm$0.03 &  0.27$\pm$0.10 &  0.16$\pm$0.07 &  0.35$\pm$0.14 &  9.12$\pm$2.77 \\ 
MKV0698 & -1.48$\pm$0.02 &      -~~~~~~~~ &  0.06$\pm$0.04 &  0.15$\pm$0.08 &  5.21$\pm$1.41 \\ 
MKV0708 & -1.57$\pm$0.07 &      -~~~~~~~~ &      -~~~~~~~~ &  0.28$\pm$0.26 &  1.71$\pm$0.49 \\ 
MKV0729 & -1.39$\pm$0.04 & -0.02$\pm$0.19 &  0.14$\pm$0.11 &  0.39$\pm$0.19 &  2.71$\pm$0.84 \\ 
MKV0733 & -1.64$\pm$0.05 & -0.23$\pm$0.22 & -0.10$\pm$0.19 & -0.25$\pm$0.22 & 15.00~~~~~~~~~~ \\ 
MKV0740 & -1.20$\pm$0.09 &      -~~~~~~~~ & -0.37$\pm$0.41 & -0.73$\pm$0.41 &  6.32$\pm$1.82 \\ 
MKV0743 & -1.21$\pm$0.08 & -0.09$\pm$0.33 &  0.21$\pm$0.33 & -0.95$\pm$0.33 &  5.65$\pm$1.70 \\ 
MKV0770 & -1.63$\pm$0.03 &      -~~~~~~~~ &  0.13$\pm$0.08 &  0.25$\pm$0.14 &  4.07$\pm$1.39 \\ 
MKV0780 & -1.78$\pm$0.05 & -0.03$\pm$0.30 &  0.00$\pm$0.22 & -0.20$\pm$0.30 & 10.76$\pm$2.09 \\ 
MKV0812 & -1.34$\pm$0.03 & -0.14$\pm$0.15 &  0.19$\pm$0.13 &      -~~~~~~~~ &  3.25$\pm$0.86 \\ 
MKV0825 & -1.43$\pm$0.03 &  0.26$\pm$0.18 &  0.05$\pm$0.09 &  0.17$\pm$0.18 &  8.57$\pm$2.76 \\ 
MKV0840 & -1.18$\pm$0.03 &  0.25$\pm$0.11 & -0.03$\pm$0.09 &  0.23$\pm$0.15 &  3.75$\pm$1.22 \\ 
MKV0842 & -1.47$\pm$0.02 & -0.63$\pm$0.16 &      -~~~~~~~~ &  0.27$\pm$0.10 &  2.44$\pm$0.73 \\ 
MKV0880 & -1.58$\pm$0.03 &  0.02$\pm$0.18 &  0.03$\pm$0.08 &  0.21$\pm$0.18 & 10.55$\pm$2.91 \\ 
MKV0900 & -1.72$\pm$0.02 &  0.41$\pm$0.12 &  0.00$\pm$0.05 &  0.27$\pm$0.12 & 10.01$\pm$2.18 \\ 
MKV0902 & -1.99$\pm$0.02 & -0.44$\pm$0.14 & -0.13$\pm$0.08 & -0.43$\pm$0.11 & 15.00~~~~~~~~~~ \\ 
MKV0914 & -2.51$\pm$0.07 & -0.55$\pm$0.29 &  0.18$\pm$0.22 &  0.41$\pm$0.29 & 15.00~~~~~~~~~~ \\ 
MKV0916 & -1.51$\pm$0.05 &      -~~~~~~~~ &      -~~~~~~~~ &  0.09$\pm$0.22 & 11.13$\pm$2.39 \\ 
MKV0925 & -1.55$\pm$0.04 & -0.32$\pm$0.13 &      -~~~~~~~~ &  0.41$\pm$0.17 &  1.50$\pm$0.53 \\ 
MKV0948 & -2.04$\pm$0.04 &  0.87$\pm$0.12 & -0.05$\pm$0.16 &      -~~~~~~~~ & 15.00~~~~~~~~~~ \\ 
MKV0976 & -1.24$\pm$0.06 &  0.32$\pm$0.24 & -0.04$\pm$0.24 &  0.02$\pm$0.24 &  2.61$\pm$0.76 \\ 
MKV1007 & -1.39$\pm$0.06 &  0.57$\pm$0.28 &  0.25$\pm$0.20 &  0.28$\pm$0.28 &  3.59$\pm$0.92 \\ 
MKV1009 & -1.75$\pm$0.04 &  0.45$\pm$0.22 &  0.34$\pm$0.09 &  0.11$\pm$0.20 & 10.92$\pm$2.46 \\ 
MKV1012 & -1.60$\pm$0.04 &  0.17$\pm$0.15 &  0.14$\pm$0.10 &  0.30$\pm$0.20 &  7.43$\pm$2.90 \\ 
MKV1061 & -1.50$\pm$0.05 &      -~~~~~~~~ &  0.32$\pm$0.25 &  0.53$\pm$0.25 &  8.55$\pm$3.24 \\ 
\hline
\hline
S03Ca10 & -1.94$\pm$0.02 &  0.25$\pm$0.09 & -0.02$\pm$0.05 &  0.06$\pm$0.11 & 13.00$\pm$1.37$^{1}$ \\
S03Ca12 & -1.41$\pm$0.02 &  0.11$\pm$0.08 &  0.12$\pm$0.05 &  0.24$\pm$0.10 &  5.00$\pm$1.60$^{1}$ \\
 S03Ca2 & -1.60$\pm$0.02 &  0.11$\pm$0.08 &  0.20$\pm$0.05 &  0.23$\pm$0.10 &  9.89$\pm$2.69$^{1}$ \\
 S03Ca3 & -1.65$\pm$0.02 &  0.20$\pm$0.10 & -0.10$\pm$0.06 & -0.27$\pm$0.12 & 13.08$\pm$1.16$^{1}$ \\
 S03Ca4 & -1.59$\pm$0.02 &  0.02$\pm$0.08 &  0.14$\pm$0.04 &  0.26$\pm$0.09 &  9.86$\pm$2.41$^{1}$ \\

UKV0484 & -1.55$\pm$0.02 &  0.23$\pm$0.26 &  0.16$\pm$0.04 &  0.23$\pm$0.15 &  4.94$\pm$1.66$^{2}$ \\
UKV0524 & -1.77$\pm$0.02 &  0.16$\pm$0.21 &  0.10$\pm$0.04 &  0.31$\pm$0.15 &  5.81$\pm$1.65$^{2}$ \\
UKV0612 & -1.32$\pm$0.02 & -0.57$\pm$0.16 & -0.17$\pm$0.04 & -0.46$\pm$0.16 & 10.56$\pm$2.67$^{2}$ \\
UKV0705 & -1.37$\pm$0.02 & -0.54$\pm$0.14 &  0.11$\pm$0.04 &  0.17$\pm$0.18 &  4.25$\pm$0.94$^{2}$ \\
UKV0769 & -1.70$\pm$0.02 & -0.10$\pm$0.14 &  0.24$\pm$0.06 &  0.46$\pm$0.18 &  8.83$\pm$3.75$^{2}$ \\
UKV1013 & -1.32$\pm$0.03 & -0.08$\pm$0.26 & -0.06$\pm$0.05 &  0.13$\pm$0.26 &  3.75$\pm$0.94$^{2}$ \\
UKV1087 & -2.83$\pm$0.06 & -1.00$\pm$0.34 & -0.03$\pm$0.20 &  0.52$\pm$0.34 &  9.50$\pm$0.50$^{2}$ \\
UKV5070 & -2.17$\pm$0.05 & -1.06$\pm$0.27 & -0.01$\pm$0.10 & -0.32$\pm$0.38 & 15.00~~~~~~~~~$^{2}$ \\
UKV7002 & -2.88$\pm$0.05 & -0.91$\pm$0.18 &  0.18$\pm$0.10 &  0.26$\pm$0.23 & 15.00~~~~~~~~~$^{2}$ \\

K000377 & -1.50$\pm$0.16 &      -~~~~~~~~ &  0.21$\pm$0.04 &  0.08$\pm$0.19 &  8.71$\pm$2.72$^{3}$ \\
K000626 & -2.50$\pm$0.15 &      -~~~~~~~~ &  0.09$\pm$0.11 &  0.41$\pm$0.09 & 15.00~~~~~~~~~~$^{3}$ \\
K000777 & -1.36$\pm$0.12 &      -~~~~~~~~ &  0.33$\pm$0.06 & -0.03$\pm$0.20 &  5.14$\pm$1.28$^{3}$ \\
K000951 & -2.72$\pm$0.16 &      -~~~~~~~~ &  0.71$\pm$0.05 &  0.31$\pm$0.12 & 15.00~~~~~~~~~~$^{3}$ \\
K004260 & -1.49$\pm$0.16 &      -~~~~~~~~ & -0.02$\pm$0.07 & -0.24$\pm$0.13 &  9.64$\pm$2.19$^{3}$ \\
K006628 & -1.79$\pm$0.15 &      -~~~~~~~~ &  0.13$\pm$0.10 & -0.04$\pm$0.13 & 11.86$\pm$1.67$^{3}$ \\
\hline
\end{tabular}
\label{abratio}	
\end{table*}

\begin{table*}[!hb]
\caption{The median and median absolute deviation (mad) computed for 2 age groups (intermediate-age stars and old stars), either for the whole sample or for a more restricted sample in which all the stars whose age is betwen 6 and 10 Gyr have been removed.}
\centering
\begin{tabular}{ccccc}
\hline
\hline
                        & \multicolumn{2}{c}{        old stars        } & \multicolumn{2}{c} { intermediate age stars }\\
                                       &  whole sample & restricted sample &  whole sample & restricted sample \\
\hline
\multirow{1}{*}{median} &   11.025   & 13.08 & 4.16  & 3.75 \\
               { mad  } &   2.205    &  1.92 & 1.47  & 1.22 \\
\hline
\end{tabular}
\label{mad}
\end{table*}

%
 
\section{Discussion.}
\label{disc}

The Carina dSph has a particularly complex SFH, with at least two, and possibly
three or four distinct bursts of star formation over its history as shown by
multiple MSTOs in its CMD \citep[e.g.,][]{SH1996,HK1998}. 
The extremely narrow RGB shows the
strong presence of an age-metallicity degeneracy. 
Spectroscopic abundances of stars with age estimates are 
required to determine how the chemical
evolution progressed with the SFH.
Our FLAMES study (35 stars) significantly enlarges the sample of RGB stars in the Carina dSph
with accurate abundances, which was 15 stars \citep[][focusing only on $\alpha$-elements]{She2003,Koch2008}. 
The \cite{Venn2011} sample looks at a smaller
number of stars (9) but with the accuracy and wealth of chemical elements of UVES, whereas our
sample contains more stars, but with fewer elements and larger
uncertainties.\\

We would like to briefly mention that none of the values discussed ([Fe/H], [Mg/Fe], ages...])
correlates with the distance from the center of Carina. The GIRAFFE field of view (25') does not cover the whole galaxy, but from CaT measurements that do \citep{Koch2006}, no obvious radial metallicity gradient was found. However, the shape of the MDF varies from the inner to the outer region, where the more metal-rich stars are more centrally concentrated than the metal-poor.

\subsection{SFH and the [$\alpha$/Fe] ``clock''.}

One of the most obvious signs of chemical evolution in galaxies is the
[$\alpha$/Fe] ``clock'' \citep{Tin1979,Gil1991,Matt2003}. The traditional
interpretation is that the $\alpha$-elements are predominantly released into
the interstellar medium (ISM) by SNe II explosions. As the progenitors of SNe
II are high-mass stars with short lifetimes, the $\alpha$-enrichment of the ISM
starts soon after the beginning of star formation. Some Fe is also formed in
SNe II, but it is mainly produced by explosive nucleosynthesis in SNe Ia. The
progenitors of SNe Ia are long lived low-mass stars. The first SNe Ia
explosions are believed to occur $\approx$1 Gyr after the first SNe II and to be much less concentrated in time \citep{Matt1990,Matt2003}. Their explosions cause a decrease in [$\alpha$/Fe] if the SFR is constant with time. The variation of [$\alpha$/Fe] allows us to trace the relative importance of SNe II and SNe Ia. 
For this purpose the $\alpha$-element Mg is more suitable than Ca or Ti, 
as it is only produced in very small quantities by SNe Ia \citep{Tsu1995}.\\

\begin{figure*}[!htb]
   \centering
   \includegraphics[angle=-90,width=0.80\textwidth]{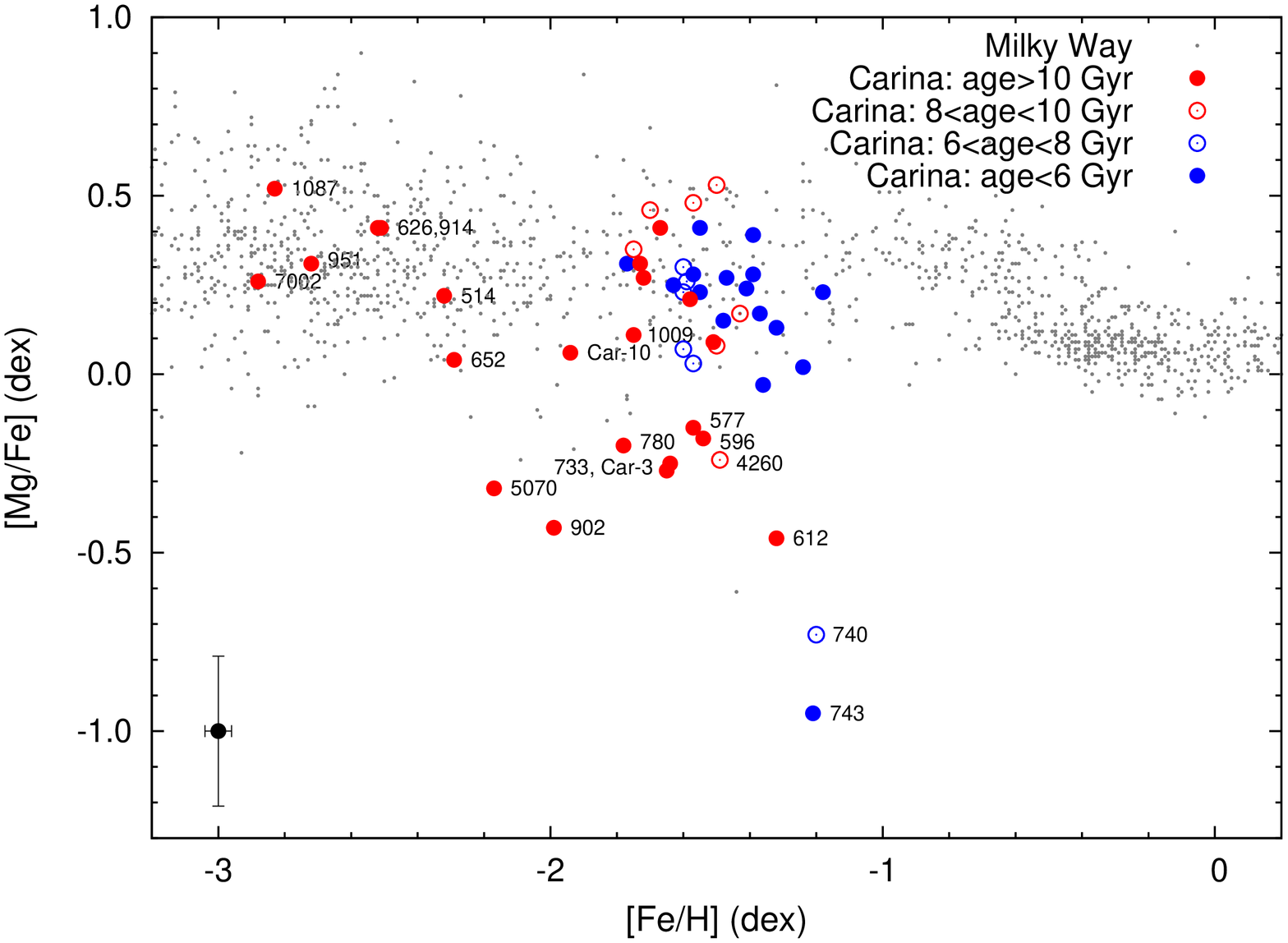}
      \caption{The [Mg/Fe] distribution for the RGB stars from the Carina dSph separated into two ages groups: stars $>$10 Gyr old are represented by red filled circles, stars $<$10 Gyr old are represented by blue filled circles. Stars in the age range 6-10 Gyr are represented either by red open circles (8$<$age$<$10 Gyr) or by blue open circles (6$<$age$<$8 Gyr). Milky Way halo stars are shown as small grey dots.}
         \label{MgFe_ages}
   \end{figure*}

\begin{figure*}[!bht]
   \centering
   \includegraphics[angle=-90,width=0.80\textwidth]{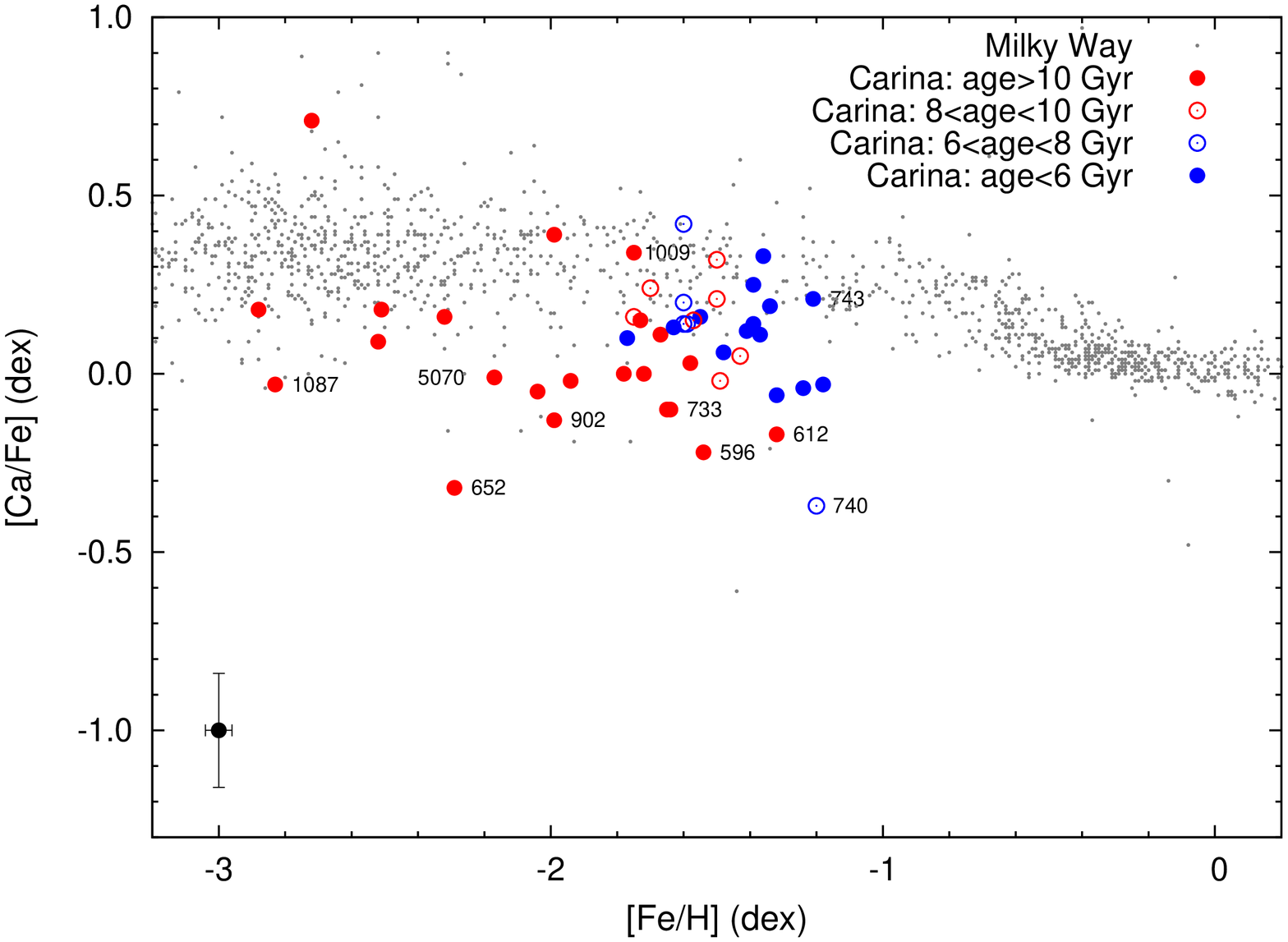}
      \caption{The [Ca/Fe] distribution with the RGB stars from the Carina dSph separated into two ages groups as described in Fig.~\ref{MgFe_ages}.}
         \label{CaFe_ages}
   \end{figure*}

Fig.~\ref{MgFe_ages} displays [Mg/Fe] against [Fe/H] for the RGB stars observed in the Carina dSph, with symbols identifying the two age groups and those stars with error bars that overlap the boundaries of the different age groups. 
The interpretation of both the 
large spread in [Mg/Fe] and the existence of an intermediate age, Mg-rich 
population, is extremely puzzling. [Ca/Fe] (see Fig.~\ref{CaFe_ages}) shows a similar behaviour but with a smaller dispersion.\\ 

%

\subsection{The Old Population ($>$8 Gyr old).}
We can use the [Fe/H] values provided in Fig.~\ref{ages} to divide the oldest 
stellar population observed in Carina in two groups: very metal-poor, 
[Fe/H]$<-2.2$ dex, and metal-poor, $-2.2<$[Fe/H]$<-1.0$, stars.
\subsubsection{Very metal-poor old stars.} 

Our age estimates indicate that most of the low-metallicity RGB stars in the
Carina dSph are very old ($>$13 Gyr) (see also Table \ref{abratio} and
Figs~\ref{MgFe_ages}, \ref{CaFe_ages}). These stars were probably formed at a
time when the ISM was only enriched by $\alpha$-elements from SNe II, which
means during the first Gyr after star formation started in the Carina dSph. The
low values and large scatter of [Ba/Fe] measurements (see Fig.~\ref{BaFe_ages})
also support this. At the oldest ages Ba is expected to form, from 
pre-existing Fe-peak seeds, mainly through the r-process whose production 
sites are still debated but probably related to neutron-rich regions 
surrounding SNe \citep[e.g.,][]{Sne2008} 
Very metal-poor stars in our sample have low values of [Ba/Fe]$<-0.5$, which 
gradually decrease with decreasing [Fe/H] reaching the extreme value of 
[Ba/Fe]$=-1.1$ in the most metal-poor star.

In Fig.~\ref{MgFe_ages} we can see that [Mg/Fe] in very metal-poor stars in 
the Carina dSphs are consistent with those measured in [Fe/H]$<-2.2$ Milky Way 
halo stars. This result may imply that at very early times the conditions for 
the chemical enrichment were the same. In particular, as the 
$\alpha$-elements are mostly produced by massive stars exploding as SNe II, 
a similar [$\alpha$/Fe] tells us that the IMF was most likely similar. Note 
that even though these stars are Ca-enhanced, they do not reach the high 
values found in very metal-poor stars in Sculptor or the Milky Way halo (see 
Fig.~\ref{CaFe}). This is usually attributed to a lower star formation 
rate \citep{Matt2003}.\\ 

It is hard to establish if there is a ``knee'' associated with the first episode of star formation 
and what may be its position. Instead, one could also imagine that the low [$\alpha$/Fe] ratios 
seen in Carina stars among the oldest stars is merely a large scatter, that might be due 
to inhomogeneous mixing \citep[See Sect.~\ref{MPO} and][]{Venn2011}.  
However, Fig.~\ref{MgFe_ages} suggests that there may be a knee 
somewhere between [Fe/H]$\sim (-2.7,-2.3)$. A larger sample in this 
metallicity range is needed to confirm the reality and exact location of the knee. A low [Fe/H] 
value for the position of the knee is consistent with the Carina dSph only experiencing 
a modest chemical evolution before the onset of SNe Ia.

\subsubsection{Metal-poor old stars.} 
\label{MPO}

\par In Fig.~\ref{MgFe_ages} we can see that our sample includes a number of old 
metal-poor stars with a very large scatter in [Mg/Fe]. Although
these stars are typically younger ($\approx$3$\pm$1 Gyr) than the Mg-rich, 
very metal-poor population (see Table~\ref{abratio}), the interpretation of 
their abundances is not straightforward.
\par The most massive among the stars formed during the early episodes of 
star formation exploded as SNe II and their ejecta polluted the ISM. The 
subsequent generation of stars are then formed from an $\alpha$- and Fe- 
enriched ISM. Hence, these new stars should have a similar [$\alpha$/Fe] 
to those that formed in the first burst, but at higher [Fe/H]. Approximately 
1~Gyr after the first SNe II explosion, SNe Ia are expected to contribute to the 
ISM enrichment, thus causing a decrease in [Mg/Fe] and an increase in [Fe/H]. 

\par In our sample of old metal-poor stars we find the presence of both Mg-
rich and Mg-poor objects. Interestingly, 
three of the stars that seem to be {\it old} (MKV0612) or {\it very old} 
(car-5070 and MKV0902) have very low [Mg/Fe], [Ca/Fe] and [Ba/Fe], suggesting 
that they formed from SNe Ia-contaminated material, while SNe II seem to have 
played only a minor role for Fe. The presence of these stars, along with the existence of 
Mg-rich stars up to [Fe/H]$\sim -1.4$, is extremely puzzling, and could be 
consistent with inhomogeneous mixing \citep[see][for more details]{Venn2011}. 

\subsection{Intermediate age stars ($<8$ Gyr old).}

Stars with intermediate ages appear to span only a restricted range of [Fe/H] 
(from $-1.8$ to $-1.2$ dex), concentrated at [Fe/H]$\sim -1.5$, 
(See Fig.~\ref{MgFe_ages} and ~\ref{CaFe_ages}). Unexpectedly, only 1-2 stars 
are depleted in [Mg/Fe], most of the others being Mg-rich stars 
which overlap with the old and metal-poor population described in Sec.~6.2.2.
Therefore, most of the $<8$~Gyr old stars in Carina, are presumably formed from 
gas that was apparently not strongly polluted by SNe Ia compared to SNe II. In
Fig.~\ref{MgFe_ages} we can see that a couple of $\approx (5-6)$ Gyr old stars
have extremely low [Mg/Fe]$<-0.5$ dex. Although these outliers have quite high
error bars, it is not clear how to explain their properties.

There is no obvious second ``knee'' associated with the dominant intermediate 
age star formation episode. This may be partly because there is a
large scatter in the abundances at higher [Fe/H]. It is also possible
that due to significant errors in both ages and abundances stars from
different star formation episodes may be misidentified, confusing the
picture. As these results stand there is no knee and very little sign
of chemical evolution in the intermediate age group of RGB stars in the currently observed stars in the Carina dSph.
 
\par
\begin{figure*}[!htb]
   \centering
   \includegraphics[angle=-90,width=0.80\textwidth]{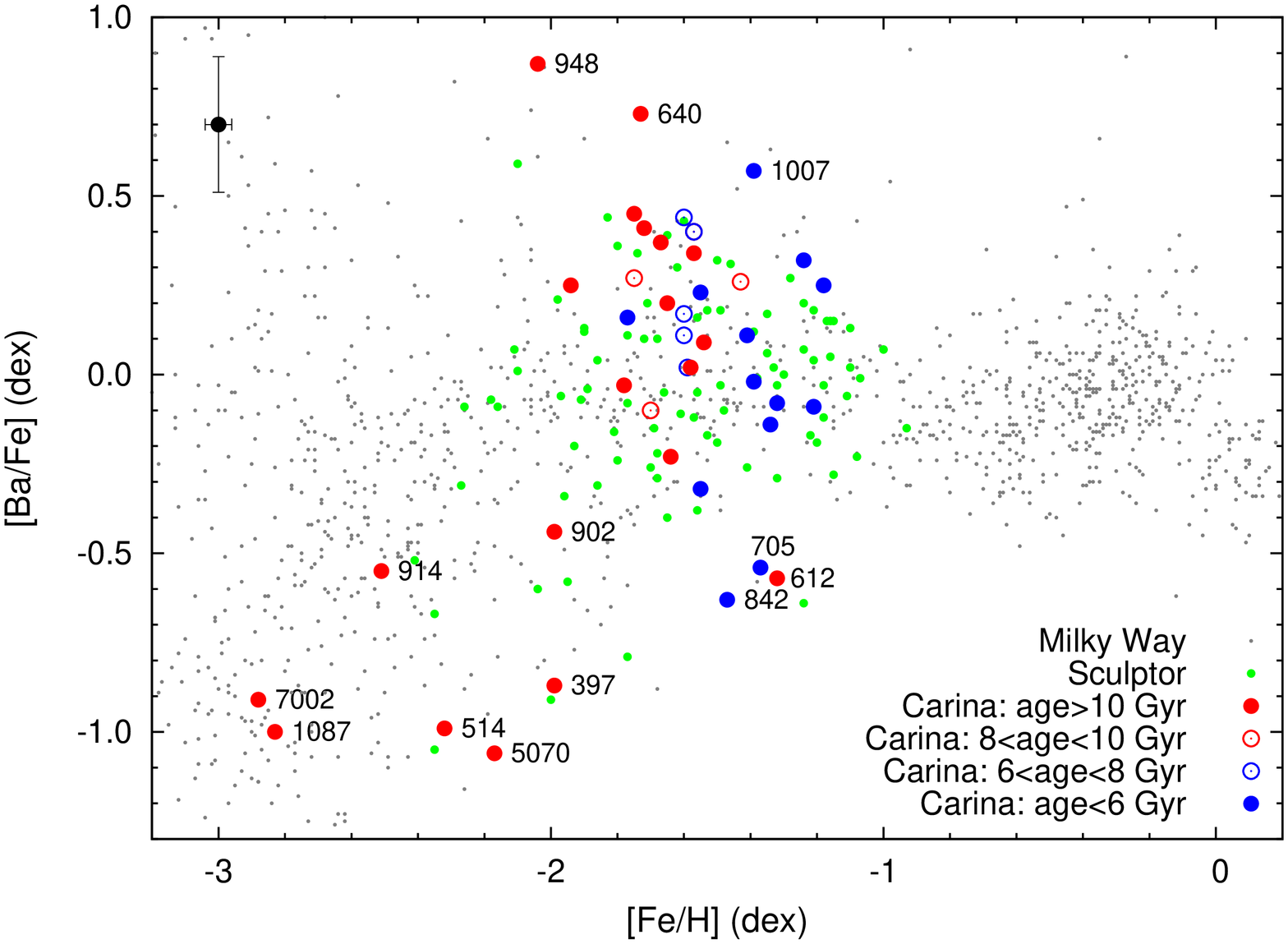}
      \caption{The [Ba/Fe] distribution for the RGB stars from the Carina dSph separated into two ages groups: stars $>$10 Gyr old are represented by red filled circles, stars $<$10 Gyr old are represented by blue filled circles. Stars in the age range 6-10 Gyr are represented either by red open circles (8$<$age$<$10 Gyr) or by blue open circles (6$<$age$<$8 Gyr). Sculptor stars \citep{She2003,Gei2005,Hill2011} are shown as small green dots and Milky Way halo stars as small grey dots.}
         \label{BaFe_ages}
   \end{figure*}

 The large spread in [Ba/Fe] at all ages seems to extend toward lower [Ba/Fe] than in the Milky Way halo stars, 
similar to what is observed in the Sculptor dSph (see Fig.~\ref{BaFe_ages}). Unfortunately, we could measure [Eu/Fe] 
in only one of our stars and only upper limits could be determined for most of the high resolution sample of \cite{Venn2011}, 
preventing us from commenting on relative weights of {\it r}- and {\it s}- processes for these stars. 

\subsection{The chemical evolution of the Carina dSph.}

Our observations, combined with additional abundances from other high
resolution spectroscopic studies, confirm that the Carina dSph has had
a complex chemical evolution. From the segregation of stars in two age
groups we can try to disentangle the chemical properties of the gas
driving these two star formation episodes. It can be seen that the
oldest star formation episode ($> 10$~Gyr old), suggests a fairly
``normal'' chemical evolution, from [Fe/H]$\sim -3$ to $-1.5$, with
[Mg/Fe] decreasing with time after a ``knee'' at [Fe/H]$\sim -2.3$, as
SNe Ia start to contribute to enrichment of the ISM. The second
episode of star formation ($< 6$~Gyr old) does not appear to conform
to the same pattern. There is a relatively small range in
[Fe/H], with most of the stars at [Fe/H]$\sim -1.5\pm0.2$ dex. This
episode also appears to predominantly contain stars with high [Mg/Fe] 
($\sim +0.3$ dex). There does not seem to be a
significant contribution by SNe Ia to the chemical evolution during
this episode of star formation. 

There is always the chance, in such a complex situation, that the
sample of stars which have been observed, which we know is biased to
high metallicity stars, has missed the low metallicity tail of this
intermediate age sample. Stars of low metallicity show very little
spread in colour for different ages, so these stars may also be hard
to identify with straight forward photometric age determinations. But
even if we are missing the low metallicity tail, we still have to
explain the apparent lack of a ``knee'' at any metallicity for this
age group. It seems that, in the mean, the stars managed to form at
roughly the same [Fe/H] ($\sim -$1.5 dex) and [Mg/Fe] ($\sim$~+0.3
dex) for the entire star formation episode. It is not clear what
happened to the SNe Ia products, which should have had time to
contribute, if the length of the star formation episode determined
from CMD analysis is to be believed. It may also be possible that this
star formation episode was much shorter (and also much more intense)
than current CMD analysis suggests, or consists of discontinuous star formation
episodes. Recently \cite{Stet2011} suggested a restricted age range of 2 Gyr for the 
intermediate-age (4-6 Gyr old) population. Short star formation episodes 
($<$1~Gyr) could explain the lack of SNe Ia enrichment. Indeed, if SNe II are able to remove the
ISM of a galaxy before the onset of SNe Ia, then the effectiveness of
SNe Ia in enriching the ISM may be limited.

It seems that the chemical evolution does not run smoothly from the
first star formation episode to the second. The [Fe/H] range overlaps between [Fe/H]$\sim -2$ and $-1.4$ dex. It is likely that we are more incomplete for metal poor stars in our sample (see Sect.~\ref{iron}) than the metal rich, which is only likely to increase the overlap. This overlap means that the most metal poor stars in the intermediate age group, are at lower [Fe/H] than the most metal rich stars in the old group. This requires that either star formation in this second episode started from scratch, with a new reservoir of gas (at [Fe/H]$\lesssim -2$, [Mg/Fe]$\sim +0.3$, [Ba/Fe]$\sim 0$) or additional more metal poor gas was added to gas that was enriched by the first star formation episode to be able to start forming stars at lower [Fe/H]. It is hard to imagine a scenario, in such a small system, where explaining the properties of these two very distinct episodes of star formation does not involve the addition of external gas.

\subsection{Physical interpretation} 

We briefly consider what are the physical processes that may influence the 
evolution of a small dwarf galaxy like Carina and thus offer some hints as 
how to interpret the puzzling results we found namely: (i) the presence of 
numerous Mg-rich stars in the intermediate age population and (ii) the 
absence of a knee associated to the second burst of star formation. 

The cycle of forming stars, heating the ISM with supernovae explosions, and
partially removing it thus decreasing the star formation, is generally called
``feedback'' \citep{Larson1974}. This physical process mainly 
depends on the supernovae rate and on the binding energy of the galaxy.
Therefore, it is likely to be a critically important factor driving the 
evolution of low mass galaxies \citep[e.g.,][]{Dekel1986,MacLow1999,Ferr2000,Salva2008}
like the Carina dSph, with $M_{tot}=6.1\pm2.3 \times 10^6M_{\odot}$ within the 
half radius $241\pm23$kpc \citep{Wal2009}. If the bursts of star formation 
observed in Carina are part of an ongoing cycle of heating and cooling of an 
initial gas reservoir \citep[e.g.,][]{Stin2007,Val2008}, there should be a clear 
evolutionary path in the observed abundances, showing the influence of SNe Ia 
enrichment at increasing [Fe/H] values. On the other hand, if the ISM is 
removed after each burst of star formation then new gas is required to power 
the subsequent star formation activity. This gas will have to originate 
from external processes such as merging, or accretion from the intergalactic medium (IGM),
and will thus most likely have chemically distinct properties. A high abundance of Mg 
over a large time range is perhaps a sign of this.

The reduced mixing of SN ejecta within the ISM of gas-poor galaxies may also 
play a role in the chemical enrichment of a small galaxy like Carina. 
Metal-enhanced winds \citep{Vader1986,Martin2002} are more efficient in 
actively star forming galaxies \citep{Fuji2004}, and may strongly reduce the 
effectiveness of both SNe II and SNe Ia in enriching the ISM. This may be especially 
apparent during the second burst of star formation, which is the most intense. An 
extremely slow chemical enrichment due to high wind losses may be able to explain 
the narrow [Fe/H] range of intermediate age stars along with the lack of a second knee for 
these stellar population. Of course, full calculations have to be made and 
complete models made to try to explain the complex age-metallicity relation of the relic stars observed.

The chemical evolution of the Carina dSph has been modelled 
several times \citep[e.g.][]{Lan2006,Revaz2009} and compared to the (then) 
observed abundance ratios \citep{She2003,Koch2008}. Interestingly, the 
N-body/Tree-SPH simulations \citep{Revaz2009} have successfully predicted 
the large scatter in $\alpha$-abundances seen in our observations, 
including the very low [Mg/Fe] stars. 
However, the SFH is not a good match, with gaps in star formation predicted by the model shorter than
those observed in Carina. This implies that the age-metallicity relation may be different to our observations. 
This discrepancy may be related to the fact that the model assumes that the dwarf galaxy was isolated 
with a fixed gas reservoir. In contrast the complicated star formation 
history of Carina is probably driven by external physical processes \citep[e.g.,][]{Pas2011}.
Carina has currently a very small galactocentric velocity (v$_{galacto}$ = 7 km/s) which 
suggests that it is at apocentre, and has been orbiting the Milky Way environment 
for a considerable time \citep[e.g.,][]{Pia2003}. Updated models are clearly needed, 
but Carina has always proved a complicated challenge.

\section{Summary and conclusions.}

We determined detailed high resolution abundances for a variety of
elements, including Mg, Ca and Ba, for most of our 35 RGB stars in the
Carina dSph. This is a significant increase in the number of RGB stars
analysed from previous studies. We confirm that the abundances of
individual RGB stars, at a given [Fe/H] in the Carina dSph are
typically more scattered than in the Milky Way or other dSph, as has already been noted by \citet{Koch2008}. This scatter exists at all [Fe/H] and is clearly representative of the global properties of the Carina dSph. 

We also determined ages for our sample of RGB stars and for $\sim$20 additional RGB stars in the Carina dSph with detailed abundances from the literature. This allows us for the first time to distinguish significant numbers of stars from different episodes of star formation in the Carina dSph. The ages we determined are broadly consistent with the two episodes of star formation expected from CMD analysis. Our ages are not accurate enough to do more than place our stars in one star formation episode or another. 

As in other dSphs, Carina contains old metal-poor, alpha-rich stars, and among these stars there appears to be a general trend of decreasing [Mg/Fe] with increasing [Fe/H], which is presumably due to the onset of SNe Ia enrichment (although a stochastic star formation could also be invoked). This population of stars is associated to the oldest episode of star formation in the Carina dSph. We have also identified stars associated to a second episode of star formation. Their enrichment history is not so easy to interpret. The most metal poor stars in this later, longer lasting and more intense episode of star formation are somewhat more metal poor than the most metal rich stars in the oldest sample. This suggests that there has not been a simple chemical evolution path following from one episode to the next in a fixed reservoir of gas.

An important step would be to carry out a combined model of the detailed SFH and the abundances, as has been done for Sculptor (de Boer et al. 2011, in prep), and of course to better populate the entire [Fe/H] range of Carina with accurate and detailed stellar abundances. 

Despite a large increase in the number of RGB stars with accurate abundances in the Carina dSph, we still have not been able to fully disentangle the chemical evolution history.

\begin{acknowledgements}
The authors thank the anonymous referee for her/his comments and suggestions that helped to improve the paper. The authors thank ISSI (Bern) for support of the team ``Defining the Life-Cycle of Dwarf Galaxy Evolution: the Local Universe as a Template'' and M. Gullieuszik for sending the IR photometry for stars in our sample. B.L., E.T., T.d.B.,and E.S. have been funded by the Netherlands Foundation for Scientific Research (NWO) through a VICI grant and S.S. through a NOVA fellowship. B.L. thanks the Leids Kerkhoven-Bosscha Fonds (LKBF) for financial support to attend the ``A Universe of dwarf galaxies'' conference, June 2010, Lyon and E. Tiesinga for his help in installing DAOSPEC on a 64-bit machine.
\end{acknowledgements}

\longtab{3}{
\begin{center}
\begin{longtable}[t]{ccccccccrc}
\label{targets}\\
\caption{Photometric information for our FLAMES targets, including ESO-SOFI J, H and K magnitudes (M. Gullieuszik, private communication). We also provide our measured radial velocities.}\\
\hline
\hline
 Target &   RA  &  Dec  &  V  &  I  &  J  &  H  &  K  &     Vr~~~~~ & Membership \\ 
        & h m s & d m s & mag & mag & mag & mag & mag & km.s$^{-1}$ & \\
\hline
MKV0397 & 06:41:58:27 & -50:46:41.40 & 18.460 & 17.240 &        &        &        & 231.4$\pm$1.5 &            \\ 
MKV0446 & 06:42:09.17 & -50:48:42.00 & 19.350 & 18.320 &        &        &        & 223.4$\pm$1.2 &            \\ 
MKV0458 & 06:41:58.19 & -50:48:58.00 & 18.440 & 17.250 &        &        &        & 227.1$\pm$1.2 &            \\ 
MKV0482 & 06:41:58.31 & -50:49:50.00 & 19.140 & 18.090 & 17.351 & 16.814 & 16.752 & 236.8$\pm$3.6 &            \\ 
MKV0499 & 06:40:42.29 & -50:50:19.70 & 19.310 & 17.899 &        &        &        &  65.8$\pm$3.8 & non-member \\ 
MKV0511 & 06:41:02.64 & -50:50:48.40 & 18.980 & 17.840 & 16.952 & 16.398 & 16.324 & 233.7$\pm$2.9 &            \\ 
MKV0512 & 06:40:46.55 & -50:50:50.60 & 19.180 & 18.140 & 17.531 & 16.959 & 16.845 & 229.8$\pm$1.9 &            \\ 
MKV0514 & 06:41:46.00 & -50:51:00.50 & 18.170 & 16.910 & 16.032 & 15.402 & 15.266 & 221.9$\pm$1.3 &            \\ 
MKV0516 & 06:40:45.44 & -50:51:01.60 & 19.180 & 18.160 & 17.308 & 16.725 & 16.660 & 228.1$\pm$0.4 &            \\ 
MKV0523 & 06:42:35.73 & -50:51:07.90 & 19.280 & 17.958 &        &        &        &  71.6$\pm$2.4 & non-member \\ 
MKV0533 & 06:42:16.11 & -50:51:17.30 & 17.690 & 16.570 & 15.833 & 15.311 & 15.184 &  87.7$\pm$1.9 & non-member \\ 
MKV0535 & 06:41:18.58 & -50:51:19.70 & 17.780 & 16.690 & 15.932 & 15.386 & 15.225 &  54.6$\pm$1.5 & non-member \\ 
MKV0537 & 06:40:42.18 & -50:51:22.40 & 17.780 & 16.521 &        &        &        & 142.2$\pm$1.6 & non-member \\ 
MKV0542 & 06:41:59.48 & -50:51:42.90 & 19.240 & 18.200 & 17.371 & 16.774 & 16.738 & 218.5$\pm$1.2 &            \\ 
MKV0550 & 06:42:08.99 & -50:52:00.20 & 19.330 & 18.290 & 17.484 & 16.838 & 16.874 & 228.9$\pm$1.7 &            \\ 
MKV0555 & 06:41:03.94 & -50:52:05.50 & 19.340 & 18.340 & 17.475 & 16.942 & 16.759 & 229.7$\pm$0.7 &            \\ 
MKV0556 & 06:42:22.26 & -50:52:09.50 & 19.190 & 18.070 & 17.301 & 16.741 & 16.631 & 220.1$\pm$2.2 &            \\ 
MKV0558 & 06:40:54.75 & -50:52:11.60 & 17.980 & 16.570 & 15.673 & 15.059 & 14.921 &  63.0$\pm$1.7 & non-member \\ 
MKV0560 & 06:42:25.71 & -50:52:12.30 & 18.820 & 17.720 &        &        &        &  38.2$\pm$2.6 & non-member \\ 
MKV0563 & 06:40:40.23 & -50:52:20.50 & 18.890 & 17.665 &        &        &        & 210.0$\pm$1.3 &            \\ 
MKV0569 & 06:41:40.16 & -50:52:33.40 & 19.120 & 18.010 & 17.160 & 16.569 & 16.495 & 231.9$\pm$0.3 &            \\ 
MKV0571 & 06:40:35.78 & -50:52:37.60 & 18.130 & 17.041 &        &        &        &  11.5$\pm$4.1 & non-member \\ 
MKV0576 & 06:41:25.13 & -50:52:47.90 & 18.740 & 17.580 & 16.766 & 16.213 & 16.016 & 226.8$\pm$0.8 &            \\ 
MKV0577 & 06:41:45.03 & -50:52:49.00 & 18.920 & 17.770 & 17.005 & 16.402 & 16.350 & 217.9$\pm$0.7 &            \\ 
MKV0584 & 06:41:05.69 & -50:52:59.60 & 19.210 & 18.230 & 17.446 & 16.934 & 16.864 & 228.6$\pm$0.8 &            \\ 
MKV0587 & 06:41:19.97 & -50:53:02.30 & 19.120 & 17.790 & 16.931 & 16.327 & 16.193 & 108.6$\pm$2.3 & non-member \\ 
MKV0592 & 06:40:29.55 & -50:53:09.70 & 18.350 & 16.780 &        &        &        &  73.1$\pm$1.7 & non-member \\ 
MKV0596 & 06:41:18.04 & -50:53:11.40 & 18.060 & 16.720 & 17.009 & 16.464 & 16.366 & 227.0$\pm$0.5 &            \\ 
MKV0598 & 06:42:01.12 & -50:53:18.10 & 19.020 & 17.870 & 17.077 & 16.484 & 16.379 & 220.0$\pm$1.9 &            \\ 
MKV0605 & 06:41:47.66 & -50:53:28.20 & 16.720 & 15.560 & 14.812 & 14.259 & 14.180 &   3.3$\pm$2.7 & non-member \\ 
MKV0606 & 06:41:05.46 & -50:53:28.60 & 18.990 & 17.800 & 17.067 & 16.633 & 16.505 & 222.9$\pm$0.2 &            \\ 
MKV0608 & 06:42:26.10 & -50:53:28.80 & 17.940 & 16.620 &        &        &        &  -0.5$\pm$1.0 & non-member \\ 
MKV0614 & 06:42:37.95 & -50:53:37.60 & 18.720 & 17.570 &        &        &        & 229.0$\pm$1.7 &            \\ 
MKV0628 & 06:40:35.37 & -50:54:07.60 & 18.210 & 16.920 &        &        &        & 220.4$\pm$1.7 &            \\ 
MKV0630 & 06:42:26.76 & -50:54:09.90 & 18.240 & 17.158 &        &        &        &  43.3$\pm$1.2 & non-member \\ 
MKV0635 & 06:41:12.83 & -50:54:20.00 & 19.390 & 18.320 & 17.573 & 17.085 & 16.851 & 217.4$\pm$1.1 &            \\ 
MKV0640 & 06:40:40.82 & -50:54:29.20 & 18.290 & 17.041 &        &        &        & 226.2$\pm$1.7 &            \\ 
MKV0641 & 06:42:38.89 & -50:54:29.30 & 18.200 & 16.640 &        &        &        &  52.5$\pm$2.1 & non-member \\ 
MKV0647 & 06:40:11.30 & -50:54:39.10 & 18.920 & 17.380 &        &        &        &  78.9$\pm$5.0 & non-member \\ 
MKV0652 & 06:41:40.82 & -50:54:45.80 & 18.560 & 17.370 & 16.599 & 16.012 & 15.848 & 233.8$\pm$1.7 &            \\ 
MKV0668 & 06:42:09.36 & -50:55:14.20 & 19.100 & 18.080 & 17.285 & 16.697 & 16.552 & 222.9$\pm$0.8 &            \\ 
MKV0675 & 06:41:26.91 & -50:55:23.20 & 18.650 & 17.130 & 16.136 & 15.519 & 15.368 &  67.1$\pm$3.2 & non-member \\ 
MKV0677 & 06:40:31.12 & -50:55:24.50 & 17.700 & 16.362 &        &        &        & 215.0$\pm$4.2 &            \\ 
MKV0692 & 06:41:18.32 & -50:55:39.10 & 19.110 & 18.130 & 17.401 & 16.935 & 16.764 & 226.8$\pm$1.0 &            \\ 
MKV0698 & 06:41:26.09 & -50:55:43.70 & 17.620 & 16.220 & 15.203 & 14.500 & 14.348 & 217.1$\pm$7.4 &            \\ 
MKV0701 & 06:41:42.51 & -50:55:50.20 & 17.970 & 16.740 & 15.942 & 15.363 & 15.224 &  76.0$\pm$1.4 & non-member \\ 
MKV0708 & 06:42:38.52 & -50:56:00.80 & 18.400 & 17.290 &        &        &        & 211.6$\pm$1.5 &            \\ 
MKV0713 & 06:41:45.20 & -50:56:04.70 & 19.060 & 17.940 & 17.212 & 16.691 & 16.574 & 236.6$\pm$2.3 &            \\ 
MKV0716 & 06:42:00.50 & -50:56:08.50 & 19.280 & 18.250 &        &        &        & 218.3$\pm$0.5 &            \\ 
MKV0728 & 06:40:37.63 & -50:56:25.40 & 18.980 & 17.770 &        &        &        &  11.7$\pm$1.9 & non-member \\ 
MKV0729 & 06:42:17.45 & -50:56:26.30 & 18.630 & 17.470 &        &        &        & 227.7$\pm$1.7 &            \\ 
MKV0733 & 06:42:30.38 & -50:56:33.30 & 18.770 & 17.560 &        &        &        & 233.2$\pm$1.5 &            \\ 
MKV0735 & 06:40:32.97 & -50:56:34.50 & 17.220 & 16.004 &        &        &        &  24.4$\pm$2.6 & non-member \\ 
MKV0740 & 06:41:29.08 & -50:56:46.50 & 18.970 & 17.830 &        &        &        & 227.8$\pm$2.2 &            \\ 
MKV0743 & 06:41:45.91 & -50:56:54.30 & 19.210 & 18.120 &        &        &        & 218.2$\pm$0.9 &            \\ 
MKV0770 & 06:41:12.28 & -50:57:25.70 & 18.320 & 17.140 & 16.262 & 15.688 & 15.538 & 220.0$\pm$1.6 &            \\ 
MKV0780 & 06:40:57.07 & -50:57:44.10 & 18.610 & 17.440 & 16.593 & 16.014 & 15.918 & 228.8$\pm$1.8 &            \\ 
\hline
\newpage
\caption{continued.} \\
\hline
\hline
 Target &   RA  &  Dec  &  V  &  I  &  J  &  H  &  K  &     Vr~~~~~ & Membership \\ 
        & h m s & d m s & mag & mag & mag & mag & mag & km.s$^{-1}$ & \\
\hline
MKV0785 & 06:42:24.13 & -50:57:54.50 & 17.360 & 16.200 & 15.506 & 14.966 & 14.784 &  16.9$\pm$2.7 & non-member \\ 
MKV0797 & 06:41:18.32 & -50:58:07.60 & 19.370 & 18.410 & 17.540 & 17.064 & 17.157 & 222.1$\pm$1.8 &            \\ 
MKV0798 & 06:40:24.32 & -50:58:08.50 & 18.790 & 17.450 &        &        &        &  86.7$\pm$2.5 & non-member \\ 
MKV0812 & 06:40:51.62 & -50:58:21.70 & 19.130 & 18.050 & 17.306 & 16.728 & 16.617 & 212.5$\pm$4.2 &            \\ 
MKV0814 & 06:42:29.33 & -50:58:24.00 & 17.540 & 16.448 &        &        &        &  44.4$\pm$1.8 & non-member \\ 
MKV0825 & 06:40:56.96 & -50:58:38.30 & 18.160 & 16.850 & 15.945 & 15.309 & 15.167 & 221.0$\pm$4.5 &            \\ 
MKV0832 & 06:42:24.63 & -50:58:44.90 & 19.090 & 17.718 &        &        &        &  59.1$\pm$1.5 & non-member \\ 
MKV0840 & 06:42:02.81 & -50:58:59.40 & 18.230 & 16.940 & 16.066 & 15.448 & 15.309 & 223.5$\pm$1.4 &            \\ 
MKV0842 & 06:40:22.51 & -50:59:01.90 & 19.010 & 17.940 &        &        &        & 229.7$\pm$0.8 &            \\ 
MKV0843 & 06:41:30.01 & -50:59:02.40 & 18.950 & 17.900 & 17.099 & 16.543 & 16.456 & 219.5$\pm$4.5 &            \\ 
MKV0844 & 06:40:31.35 & -50:59:02.90 & 19.070 & 17.970 &        &        &        & 221.1$\pm$6.4 &            \\ 
MKV0846 & 06:42:15.99 & -50:59:04.50 & 19.080 & 17.960 & 17.134 & 16.616 & 16.467 & 215.0$\pm$2.8 &            \\ 
MKV0847 & 06:42:39.77 & -50:59:05.20 & 17.780 & 16.737 &        &        &        &  85.0$\pm$1.7 & non-member \\ 
MKV0850 & 06:41:37.54 & -50:59:12.80 & 19.380 & 18.270 &        &        &        & 224.0$\pm$1.4 &            \\ 
MKV0860 & 06:40:41.77 & -50:59:19.10 & 19.280 & 17.890 &        &        &        &  30.3$\pm$2.9 & non-member \\ 
MKV0866 & 06:42:17.10 & -50:59:28.40 & 18.460 & 17.290 & 16.473 & 15.922 & 15.841 &  25.6$\pm$0.8 & non-member \\ 
MKV0875 & 06:41:48.71 & -50:59:42.40 & 19.240 & 18.160 & 17.439 & 16.943 & 16.818 & 215.0$\pm$2.8 &            \\ 
MKV0877 & 06:41:12.61 & -50:59:44.10 & 19.150 & 18.070 & 17.268 & 16.706 & 16.526 & 218.6$\pm$0.6 &            \\ 
MKV0880 & 06:41:15.66 & -50:59:47.90 & 17.850 & 16.430 & 15.441 & 14.771 & 14.603 & 227.0$\pm$2.2 &            \\ 
MKV0898 & 06:41:17.47 & -51:00:17.50 & 19.010 & 17.840 & 16.995 & 16.444 & 16.282 & 218.7$\pm$2.2 &            \\ 
MKV0900 & 06:41:27.15 & -51:00:18.30 & 17.790 & 16.360 & 15.367 & 14.691 & 14.543 & 233.0$\pm$2.5 &            \\ 
MKV0901 & 06:42:06.47 & -51:00:18.70 & 19.080 & 17.900 & 17.126 & 16.553 & 16.423 & 220.7$\pm$0.5 &            \\ 
MKV0902 & 06:41:16.31 & -51:00:18.70 & 18.180 & 16.850 &        &        &        & 223.3$\pm$1.4 &            \\ 
MKV0914 & 06:40:42.49 & -51:00:42.70 & 18.240 & 16.900 &        &        &        & 234.1$\pm$0.6 &            \\ 
MKV0916 & 06:42:08.97 & -51:00:48.00 & 18.630 & 17.400 & 16.498 & 15.889 & 15.812 & 228.4$\pm$2.8 &            \\ 
MKV0925 & 06:40:43.17 & -51:01:06.60 & 18.610 & 17.520 &        &        &        & 221.8$\pm$1.5 &            \\ 
MKV0939 & 06:41:49.67 & -51:01:31.30 & 19.280 & 18.300 & 17.390 & 16.892 & 16.745 & 228.3$\pm$4.4 &            \\ 
MKV0948 & 06:41:37.65 & -51:01:43.60 & 17.980 & 16.668 &        &        &        & 231.5$\pm$1.5 &            \\ 
MKV0959 & 06:42:15.67 & -51:01:59.90 & 18.680 & 17.560 & 16.682 & 16.106 & 16.024 & 230.1$\pm$2.1 &            \\ 
MKV0976 & 06:40:57.68 & -51:02:40.70 & 18.990 & 17.900 & 17.138 & 16.581 & 16.511 & 224.3$\pm$1.5 &            \\ 
MKV0981 & 06:41:54.36 & -51:02:46.30 & 18.410 & 17.150 & 16.196 & 15.470 & 15.488 & 225.9$\pm$4.5 &            \\ 
MKV0987 & 06:40:53.55 & -51:02:53.00 & 19.450 & 18.430 & 17.679 & 17.131 & 16.910 & 226.0$\pm$0.8 &            \\ 
MKV0990 & 06:42:04.96 & -51:02:59.00 & 18.720 & 17.610 & 16.843 & 16.232 & 16.124 & 223.5$\pm$0.4 &            \\ 
MKV0998 & 06:41:12.10 & -51:03:12.40 & 19.320 & 18.300 & 17.551 & 16.952 & 16.899 & 222.6$\pm$1.4 &            \\ 
MKV1007 & 06:41:44.65 & -51:03:31.40 & 18.260 & 17.030 & 16.034 & 15.352 & 15.246 & 236.4$\pm$2.0 &            \\ 
MKV1009 & 06:41:51.55 & -51:03:36.00 & 18.640 & 17.450 & 16.569 & 15.977 & 15.830 & 220.6$\pm$1.5 &            \\ 
MKV1011 & 06:41:04.04 & -51:03:39.60 & 19.020 & 17.910 & 17.110 & 16.552 & 16.423 & 230.8$\pm$1.6 &            \\ 
MKV1012 & 06:41:00.31 & -51:03:43.20 & 18.420 & 17.210 & 16.288 & 15.676 & 15.555 & 224.7$\pm$1.7 &            \\ 
MKV1030 & 06:41:14.81 & -51:04:15.30 & 19.380 & 18.350 & 17.593 & 17.011 & 17.108 & 224.9$\pm$1.0 &            \\ 
MKV1061 & 06:41:29.15 & -51:05:22.30 & 18.620 & 17.380 & 16.594 & 15.922 & 15.829 & 222.7$\pm$0.9 &            \\ 
MKV1070 & 06:41:41.33 & -51:05:39.70 & 19.310 & 18.220 & 17.289 & 16.737 & 16.669 & 222.9$\pm$0.9 &            \\ 
\hline
\end{longtable}
\end{center}
}

\longtab{4}{
\begin{longtable}{crrrrrrrrrrrrrrr}
\label{EW} \\
\caption{List of all spectral lines, their atomic parameters and their measured EWs (in m$\AA$) for all the RGB stars in the Carina dSph. Part 1/3.}\\
\hline
\hline
$\lambda$&Elem&$\chi_{ex}$&$log~gf$& 397 & 458 & 514 & 556 & 577 & 596 & 614 & 628 & 640 & 652 & 677 & 698 \\
  $\AA$  &    &           &        &     &     &     &     &     &     &     &     &     &     &     &     \\
\hline											    					     		     
5339.93 & Fe1 & 3.270 & -0.680 &   - & 145 & 108 &   - &  90 & 120 &   - &   - & 117 &  52 &   - & 177 \\ 
5364.86 & Fe1 & 4.450 &  0.220 &  54 &  89 &  63 &   - &   - &  76 & 112 & 101 &   - &  57 & 105 & 122 \\
5367.48 & Fe1 & 4.420 &  0.550 &  63 & 111 &  67 & 157 & 104 &  90 &  60 & 118 & 102 &  70 & 113 &   - \\
5369.96 & Fe1 & 4.370 &  0.540 &  80 & 117 &  70 & 148 &  82 &  96 &  88 & 144 &   - &  65 & 110 & 125 \\
5371.50 & Fe1 & 0.960 & -1.644 &   - &   - &   - &   - & 193 &   - &   - &   - &   - &   - &   - &   - \\
5381.01 & Ti2 & 1.570 & -1.780 &  79 & 108 &  84 &   - &   - &  95 & 102 & 110 & 121 &  41 & 127 & 108 \\
5383.37 & Fe1 & 4.310 &  0.500 &  95 & 131 &  84 & 135 &  74 &  93 & 141 & 135 & 126 &   - &   - & 177 \\
5389.48 & Fe1 & 4.420 & -0.400 &   - &  87 &   - & 121 &   - &  48 &  97 &  63 &  61 &  46 &   - &  97 \\
5393.17 & Fe1 & 3.240 & -0.920 & 108 & 121 &   - &   - & 112 & 120 & 126 & 133 & 128 &   - & 142 & 168 \\
5397.14 & Fe1 & 0.910 & -1.992 &   - &   - &   - &   - & 192 &   - &   - &   - &   - & 164 &   - &   - \\
5400.51 & Fe1 & 4.370 & -0.150 &  54 &  66 &  30 &  75 &   - &  64 &  76 &  91 &  91 &   - &  81 & 131 \\
5402.78 &  Y2 & 1.840 & -0.510 &   - &   - &   - &   - &   - &   - &   - &   - &  18 &   - &   - &   - \\
5405.79 & Fe1 & 0.990 & -1.852 &   - &   - & 197 &   - & 184 &   - &   - &   - &   - &   - &   - &   - \\
5407.42 & Mn1 & 2.140 & -1.743 &   - &   - &   - &   - &   - &  21 &   - &   - &  33 &   - &  30 &  91 \\
5409.80 & Cr1 & 1.030 & -0.720 &  86 & 141 & 113 & 159 & 121 & 141 & 120 & 159 & 161 &  56 & 179 &   - \\
5415.19 & Fe1 & 4.390 &  0.510 &   - & 104 &  78 & 103 & 100 &  97 &  93 & 134 & 110 &  52 & 134 & 143 \\
5418.77 & Ti2 & 1.580 & -2.110 &  64 &  88 &  47 & 116 &  69 &  71 &   - &  96 &  99 &   - &  96 &  93 \\
5420.36 & Mn1 & 2.140 & -1.460 &   - &   - &   - &   - &   - &  31 &   - &   - &  26 &   - &  34 & 115 \\
5424.07 & Fe1 & 4.320 &  0.520 &  74 & 148 &  92 & 134 &  95 & 105 & 122 & 136 & 133 &  72 & 137 & 166 \\
5425.25 & Fe2 & 3.200 & -3.360 &  27 &  46 &  31 &   - &  61 &  41 &   - &  52 &  27 &   - &  30 &  51 \\
5432.55 & Mn1 & 0.000 & -3.795 &   - &  59 &  22 &   - &   - &  53 &   - &   - &  65 &   - &  90 &   - \\
5476.92 & Ni1 & 1.830 & -0.890 &   - &   - & 155 &   - &   - & 173 & 154 &   - & 199 &   - &   - &   - \\
5483.34 & Co1 & 1.710 & -1.488 &   - &   - &   - &   - &   - &  40 &   - &  65 &  64 &   - &  58 &   - \\
5485.71 & Nd2 & 1.260 & -0.120 &   - &   - &   - &   - &   - &   - &   - &   - &   - &   - &   - &   - \\
5490.16 & Ti1 & 1.460 & -0.933 &   - &   - &   - &   - &   - &   - &   - &   - &   - &   - &  25 &  62 \\
5501.48 & Fe1 & 0.960 & -3.050 & 157 & 166 & 146 & 144 & 164 & 168 & 144 &   - & 182 &   - & 189 &   - \\
5506.79 & Fe1 & 0.990 & -2.790 & 164 &   - & 148 & 172 & 128 &   - & 176 & 183 & 179 & 113 &   - &   - \\
5516.77 & Mn1 & 2.180 & -1.847 &   - &   - &   - &   - &   - &   - &   - &   - &   - &   - &   - &  62 \\
5528.41 & Mg1 & 4.350 & -0.357 &   - & 147 & 124 & 188 & 117 & 126 & 143 & 171 & 159 &  93 & 162 & 185 \\
5534.85 & Fe2 & 3.240 & -2.920 &  32 &  78 &  40 &  91 &   - &  63 &   - &  69 &  69 &  31 &  70 &  82 \\
6120.26 & Fe1 & 0.910 & -5.940 &   - &   - &   - &   - &   - &   - &   - &   - &   - &   - &   - &   - \\
6122.23 & Ca1 & 1.890 & -0.320 &   - & 187 & 109 & 138 & 153 & 127 &   - & 149 & 161 &  67 & 168 & 195 \\
6126.22 & Ti1 & 1.070 & -1.425 &   - &  30 &   - &   - &   - &   - &   - &   - &   - &  35 &   - &  75 \\
6136.62 & Fe1 & 2.450 & -1.500 &  50 &   - & 118 &   - &   - &   - & 185 &   - &   - &  90 &   - &   - \\
6137.70 & Fe1 & 2.590 & -1.366 &   - & 155 & 115 & 128 & 104 & 126 & 138 & 158 & 143 & 100 & 143 & 199 \\
6141.73 & Ba2 & 0.700 & -0.077 &   - & 166 &  84 & 181 & 134 & 138 & 157 & 164 &   - &   - & 165 &   - \\
6149.25 & Fe2 & 3.890 & -2.720 &   - &   - &   - &   - &   - &   - &   - &   - &   - &   - &  36 &   - \\
6151.62 & Fe1 & 2.180 & -3.370 &   - &   - &  47 &   - &   - &  40 &   - &  52 &   - &  30 &  67 & 103 \\
6157.75 & Fe1 & 4.070 & -1.260 &   - &   - &   - &   - &  46 &   - &  48 &   - &  74 &   - &   - &  86 \\
6160.75 & Na1 & 2.100 & -1.260 &   - &   - &   - &   - &   - &   - &   - &   - &   - &   - &   - &   - \\
6161.30 & Ca1 & 2.520 & -1.270 &   - &   - &   - &   - &   - &   - &   - &  61 &  40 &   - &  35 &  89 \\
6165.36 & Fe1 & 4.140 & -1.470 &   - &   - &   - &   - &   - &  29 &   - &   - &   - &   - &  21 &   - \\
6166.44 & Ca1 & 2.520 & -1.140 &   - &  46 &   - &   - &   - &   - &   - &  44 &  33 &   - &  33 &  75 \\
6169.04 & Ca1 & 2.520 & -0.800 &   60&   - &  48 &  81 &   - &  37 &   - &  71 &  69 &   - &  83 & 117 \\
6169.56 & Ca1 & 2.520 & -0.480 &   - &   - &  44 &  98 &   - &  56 &   - &  76 &  99 &   - &  99 & 133 \\
6173.34 & Fe1 & 2.220 & -2.850 &   - &  96 &   - & 132 &  76 &  71 &  69 &  98 &  85 &  42 & 108 & 132 \\
6176.82 & Ni1 & 4.090 & -0.430 &   - &  55 &   - &   - &   - &   - &   - &   - &   - &   - &   - &   - \\
6177.25 & Ni1 & 1.830 & -3.500 &   - &   - &   - &   - &   - &   - &   - &   - &   - &   - &   - &   - \\
6187.99 & Fe1 & 3.940 & -1.580 &   - &  50 &   - &   - &   - &   - &   - &  30 &   - &   - &  39 &  57 \\
6191.57 & Fe1 & 2.430 & -1.416 &   - & 156 & 134 & 160 &   - & 130 & 161 & 153 & 159 &  93 &   - & 197 \\
6213.43 & Fe1 & 2.220 & -2.660 &   - &   - &  78 & 104 &  68 & 116 & 110 & 110 & 131 &  61 & 124 & 152 \\
6219.29 & Fe1 & 2.200 & -2.438 &   - & 106 &  94 &  79 &  90 &  95 & 113 & 122 & 124 &  73 & 119 & 158 \\
6229.23 & Fe1 & 2.840 & -2.900 &   - &   - &   - &   - &   - &   - &   - &  38 &   - &   - &  39 &  78 \\
6230.74 & Fe1 & 2.560 & -1.276 &   - & 124 & 124 & 136 & 105 & 144 &   - & 157 & 152 &   - & 169 &   - \\
6238.38 & Fe2 & 3.890 & -2.480 &   - &   - &   - &   - &   - &  24 &   - &   - &  36 &   - &  32 &   - \\
6240.66 & Fe1 & 2.220 & -3.230 &   - &  61 &   - &  54 &  54 &  45 &  40 &  63 &  49 &   - &  56 & 110 \\
6243.82 & Si1 & 5.610 & -1.270 &   - &   - &  34 &   - &   - &   - &   - &   - &   - &   - &   - &   - \\
\hline
\newpage
\caption{Continued.}\\
\hline
\hline
$\lambda$&Elem&$\chi_{ex}$&$log~gf$& 397 & 458 & 514 & 556 & 577 & 596 & 614 & 628 & 640 & 652 & 677 & 698 \\
  $\AA$  &    &           &        &     &     &     &     &     &     &     &     &     &     &     &     \\
\hline	
6244.48 & Si1 & 5.610 & -1.270 &   - &   - &   - &   - &   - &   - &   - &   - &   - &   - &   - &   - \\
6247.56 & Fe2 & 3.890 & -2.360 &   - &  34 &   - & 116 &   - &  38 &  75 &   - &  44 &   - &  36 &  66 \\
6252.57 & Fe1 & 2.400 & -1.757 &   - & 127 & 126 &  80 & 120 &   - & 136 & 138 & 111 &  85 & 134 & 183 \\
6258.10 & Ti1 & 1.443 & -0.355 &   - &   - &   - &   - &   - &   - &  70 &  50 &  52 &   - &  62 & 129 \\
6290.97 & Fe1 & 4.730 & -0.760 &   - &   - &   - &   - &   - &   - &   - &   - &   - &   - &   - &   - \\
6297.80 & Fe1 & 2.220 & -2.740 &   - &   - &   - &  77 &   - &   - &  96 &  99 &  75 &   - &   - &   - \\
6301.50 & Fe1 & 3.650 & -0.720 &   - &  92 &  82 &   - &  89 &   - & 100 &  87 & 102 &   - & 108 & 144 \\
6302.49 & Fe1 & 3.690 & -1.150 &   - &  58 &   - &   - &   - &  50 &   - &   - &  46 &   - &   - &   - \\
6309.90 & Sc2 & 1.500 & -1.520 &   - &   - &   - &   - &   - &   - &   - &   - &  48 &   - &   - &   - \\
6311.51 & Fe1 & 2.830 & -3.220 &   - &   - &   - &   - &   - &   - &   - &   - &   - & 109 &   - &   - \\
6320.43 & La2 & 0.170 & -1.562 &   - &   - &   - &   - &   - &   - &   - &   - &   - &   - &   - &   - \\
6330.09 & Cr1 & 0.940 & -2.920 &   - &   - &   - &  33 &   - &   - &   - &  40 &  32 &   - &   - &  94 \\
6355.04 & Fe1 & 2.840 & -2.290 &  51 & 112 &  57 & 124 &  70 &  83 &  82 &  86 &  91 &   - &  96 & 160 \\
6369.46 & Fe2 & 2.890 & -4.250 &   - &   - &   - &   - &   - &   - &   - &   - &   - &   - &   - &   - \\
6380.75 & Fe1 & 4.190 & -1.500 &   - &   - &  48 &   - &   - &   - &   - &  65 &   - &   - &   - &  72 \\
6390.46 & La2 & 0.320 & -1.400 &   - &   - &   - &   - &   - &   - &   - &   - &  52 &   - &   - &   - \\
6392.54 & Fe1 & 2.280 & -3.950 &   - &  60 &   - &   - &  97 &  26 &   - &   - &  38 &   - &  35 &  83 \\
6393.61 & Fe1 & 2.430 & -1.630 & 146 & 145 & 119 &   - & 149 &   - & 102 & 193 & 139 & 141 &   - &   - \\
6416.93 & Fe2 & 3.890 & -2.790 &   - &  52 &   - &   - &   - &   - &   - &   - &   - &   - &   - &   - \\
6419.96 & Fe1 & 4.730 & -0.240 &   - &   - &   - &   - &   - &   - &   - &   - &   - &   - &  34 &  73 \\
6421.36 & Fe1 & 2.280 & -2.014 &  84 &   - &  83 & 185 & 118 & 138 & 112 & 159 & 118 &  60 & 141 &   - \\
6430.86 & Fe1 & 2.180 & -1.946 & 116 & 135 & 126 & 189 &   - & 142 &   - & 140 & 154 &  60 & 154 &   - \\
6432.68 & Fe2 & 2.890 & -3.710 &  39 &   - &  54 &   - &   - &  44 &   - &   - &  63 &   - &  55 &   - \\
6439.08 & Ca1 & 2.520 &  0.390 &   - &   - & 106 &   - &   - & 121 &   - & 147 & 146 &   - & 159 & 194 \\
6455.60 & Ca1 & 2.520 & -1.290 &   - &   - &   - &   - &   - &   - &   - &  46 &  28 &   - &  43 &  60 \\
6456.39 & Fe2 & 3.900 & -2.080 &  61 &  81 &   - & 142 &   - &  52 &  54 &  69 &  67 &   - &  66 &  71 \\
6496.91 & Ba2 & 0.600 & -0.380 &  81 & 172 &  69 &   - &   - & 118 & 175 &   - & 187 &   - & 162 &   - \\
6498.94 & Fe1 & 0.960 & -4.690 &  55 & 111 &  51 &   - &   - &  83 &  53 &   - & 108 &   - &   - & 140 \\
6499.65 & Ca1 & 2.520 & -0.820 &   - &   - &  38 &  58 &   - &  33 &   - &  65 &  53 &   - &  79 & 109 \\
6516.08 & Fe2 & 2.890 & -3.450 &  34 &  70 &   - &   - &   - &  48 &  74 &  61 &  43 &   - &  48 &  56 \\
6518.37 & Fe1 & 2.830 & -2.460 &   - &   - &  33 &   - &   - &  31 &   - &  52 &  37 &   - &  63 &  93 \\
6554.24 & Ti1 & 1.440 & -1.218 &   - &  29 &   - &   - &  24 &   - &   - &   - &  23 &   - &  27 &  58 \\
6556.08 & Ti1 & 1.460 & -1.074 &   - &  33 &   - &   - &   - &  27 &   - &   - &   - &   - &   - &   - \\
6574.23 & Fe1 & 0.990 & -5.020 &  39 &  43 &  16 &   - &  59 &  37 &   - &  65 &  51 &  40 &  75 & 124 \\
6581.22 & Fe1 & 1.480 & -4.680 &   - &  32 &   - &   - &   - &  23 &  22 &  33 &  29 &   - &  56 &   - \\
6593.88 & Fe1 & 2.430 & -2.390 &  58 & 115 &  41 & 108 &   - &  84 & 131 &  99 & 110 &  31 &  98 & 136 \\
6608.03 & Fe1 & 2.280 & -3.940 &   - &  30 &   - &   - &   - &   - &   - &  37 &   - &   - &  25 &  49 \\
6609.12 & Fe1 & 2.560 & -2.660 &  45 &  85 &  48 &  74 &  64 &  68 &  78 &  73 &  77 &   - &  86 & 118 \\
6645.13 & Eu2 & 1.370 &  0.200 &   - &   - &   - &   - &   - &   - &   - &   - &   - &   - &   - &   - \\
\hline
\newpage
\caption{List of all spectral lines, their atomic parameters and their measured EWs (in m$\AA$) for all the RGB stars in the Carina dSph. Part 2/3}.\\
\hline
\hline
$\lambda$&Elem&$\chi_{ex}$&$log~gf$& 708 & 729 & 733 & 740 & 743 & 770 & 780 & 812 & 825 & 840 & 842 & 880 \\
  $\AA$  &    &           &        &     &     &     &     &     &     &     &     &     &     &     &     \\
\hline	
5339.93 & Fe1 & 3.270 & -0.680 & 117 & 150 & 106 &   - &   - &   - & 148 & 100 & 151 &   - & 130 & 147 \\
5364.86 & Fe1 & 4.450 &  0.220 &   - & 110 &  74 & 126 & 126 &  72 &  99 &   - &   - &   - & 105 & 120 \\
5367.48 & Fe1 & 4.420 &  0.550 & 116 &  98 &  71 &   - &   - &   - &   - & 151 & 126 & 133 & 131 & 120 \\
5369.96 & Fe1 & 4.370 &  0.540 &   - & 120 & 123 & 163 & 163 &   - & 128 & 138 &   - & 107 & 154 & 116 \\
5371.50 & Fe1 & 0.960 & -1.644 &   - &   - &   - &   - &   - &   - &   - &   - &   - &   - &   - &   - \\
5381.01 & Ti2 & 1.570 & -1.780 &   - & 132 &  93 & 133 & 133 & 101 &  86 & 187 & 124 & 118 & 105 & 112 \\
5383.37 & Fe1 & 4.310 &  0.500 &  97 & 134 & 125 & 174 & 174 & 101 &  94 & 186 & 135 &   - & 133 & 145 \\
5389.48 & Fe1 & 4.420 & -0.400 &   - &  87 &   - &  98 &   98&  56 &  58 &  91 &  97 &  71 &   - &   - \\
5393.17 & Fe1 & 3.240 & -0.920 &   - &   - &   - & 159 & 159 &   - & 107 &   - & 158 & 151 & 156 & 141 \\
5397.14 & Fe1 & 0.910 & -1.992 &   - &   - &   - &   - &   - &   - &   - &   - &   - &   - &   - &   - \\
5400.51 & Fe1 & 4.370 & -0.150 &   - & 108 &  83 & 111 & 111 &  71 &  75 & 121 & 117 &   - &  86 &  99 \\
5402.78 &  Y2 & 1.840 & -0.510 &   - &   - &   - &   - &   - &   - &   - &   - &   - &   - &   - &  25 \\
5405.79 & Fe1 & 0.990 & -1.852 &   - &   - & 199 &   - &   - &   - &   - &   - &   - &   - &   - &   - \\
5407.42 & Mn1 & 2.140 & -1.743 &   - &  53 &  48 &   - &   - &   - &   - &   - &  62 &   - &   - &  50 \\
5409.80 & Cr1 & 1.030 & -0.720 & 133 & 193 & 122 & 157 & 157 & 145 & 110 & 152 & 181 & 170 & 140 &   - \\
5415.19 & Fe1 & 4.390 &  0.510 & 108 & 119 & 112 & 204 &   - & 103 & 106 & 150 & 139 & 110 &   - & 132 \\
5418.77 & Ti2 & 1.580 & -2.110 & 105 &   - &   - & 101 & 101 &  92 &  98 & 114 &  93 & 114 & 100 &  99 \\
5420.36 & Mn1 & 2.140 & -1.460 &   - &   - &   - &  83 &  83 &   - &   - &   - &   - &  53 &   - &  67 \\
5424.07 & Fe1 & 4.320 &  0.520 & 117 & 153 &   - & 177 & 177 & 122 & 112 & 142 & 168 & 131 & 153 & 147 \\
5425.25 & Fe2 & 3.200 & -3.360 &   - &   - &  67 &   - &   - &  52 &  38 &   - &  39 &  63 &   - &   - \\
5432.55 & Mn1 & 0.000 & -3.795 &   - &   - &   - &   - &   - &   - &  65 &   - & 123 &  89 &   - & 116 \\
5476.92 & Ni1 & 1.830 & -0.890 & 174 &   - & 177 &   - &   - &   - & 156 &   - &   - & 190 & 188 &   - \\
5483.34 & Co1 & 1.710 & -1.488 &   - &  77 &   - &  60 &  60 &  45 &   - &   - & 104 &   - &  62 &  97 \\
5485.71 & Nd2 & 1.260 & -0.120 &   - &   - &   - &   - &   - &   - &   - &   - &    -&   - &   - &   - \\
5490.16 & Ti1 & 1.460 & -0.933 &   - &  38 &   - &   - &   - &   - &   - &   - &  38 &   - &   - &  43 \\
5501.48 & Fe1 & 0.960 & -3.050 & 176 & 168 & 154 & 167 & 167 & 166 & 166 & 160 &   - & 196 & 156 &   - \\
5506.79 & Fe1 & 0.990 & -2.790 & 161 & 192 & 168 & 177 & 177 & 176 & 169 & 195 &   - & 186 &   - &   - \\
5516.77 & Mn1 & 2.180 & -1.847 &   - &  27 &   - &   - &   - &   - &   - &   - &   - &   - &   - &  32 \\
5528.41 & Mg1 & 4.350 & -0.357 & 154 & 197 & 131 & 131 & 131 & 154 & 128 &   - & 172 & 184 & 177 & 177 \\
5534.85 & Fe2 & 3.240 & -2.920 &  61 &  79 &  61 & 108 & 108 & 68  & 74  &  67 &  80 &  77 &  96 &  64 \\
6120.26 & Fe1 & 0.910 & -5.940 &   - &   - &   - &  73 &  73 &   - &   - &   - &  20 &   - &   - &   - \\
6122.23 & Ca1 & 1.890 & -0.320 &   - & 173 & 126 &   - &   - &   - & 138 & 161 & 157 & 171 &   - & 168 \\
6126.22 & Ti1 & 1.070 & -1.425 &   - &  52 &   - &   - &   - &   - &   - &   - &  59 &   - &   - &  49 \\
6136.62 & Fe1 & 2.450 & -1.500 & 156 &   - &   - & 143 & 143 &   - &   - & 160 &   - &   - & 124 &   - \\
6137.70 & Fe1 & 2.590 & -1.366 &   - &   - & 124 &   - &   - &   - & 112 & 146 & 168 & 171 &   - & 173 \\
6141.73 & Ba2 & 0.700 & -0.077 &   - & 176 & 141 &   - &   - &   - & 161 & 146 & 170 & 162 &   - & 171 \\
6149.25 & Fe2 & 3.890 & -2.720 &   - &  65 &   - &   - &   - &  27 &  35 &   - &  38 &  37 &  55 &   - \\
6151.62 & Fe1 & 2.180 & -3.370 &   - &  63 &   - &   - &   - &  44 &  60 &   - &  77 &   - &  51 & 101 \\
6157.75 & Fe1 & 4.070 & -1.260 &   - &   - &   - &   - &   - &  38 &  48 &   - &  64 &  56 &  68 &   - \\
6160.75 & Na1 & 2.100 & -1.260 &   - &   - &   - &   - &   - &   - &   - &   - &   - &   - &   - &   - \\
6161.30 & Ca1 & 2.520 & -1.270 &   - &  74 &   - &   - &   - &  34 &   - &   - &   - &  71 &   - &  66 \\
6165.36 & Fe1 & 4.140 & -1.470 &   - &   - &  42 &   - &   - &   - &  40 &   - &  44 &   - &   - &  38 \\
6166.44 & Ca1 & 2.520 & -1.140 &   - & 114 &   - &   - &   - &   - &   - &   - &  64 &  82 &  81 &  58 \\
6169.04 & Ca1 & 2.520 & -0.800 &   - &  81 &   - &   - &   - &  73 &   - &   - &  93 &   - &   - &  91 \\
6169.56 & Ca1 & 2.520 & -0.480 &   - & 113 &  63 &   - &   - &  91 &  73 &   - & 115 & 116 &   - & 128 \\
6173.34 & Fe1 & 2.220 & -2.850 &   - & 113 &  91 &   - &   - &  70 &  71 &  96 & 108 &  94 &   - & 119 \\
6176.82 & Ni1 & 4.090 & -0.430 &   - &  72 &   - &   - &   - &  46 &   - &   - &  40 &   - &   - &  50 \\
6177.25 & Ni1 & 1.830 & -3.500 &   - &   - &   - &   - &   - &   - &   - &   - &   - &   - &   - &  30 \\
6187.99 & Fe1 & 3.940 & -1.580 &   - &   - &   - &   - &   - &   - &   - &   - &  51 &   - &   - &  34 \\
6191.57 & Fe1 & 2.430 & -1.416 &   - &   - & 147 &   - &   - &   - & 131 &   - & 167 & 164 & 129 & 182 \\
6213.43 & Fe1 & 2.220 & -2.660 &  74 &   - &  48 &   - &   - & 103 & 117 & 112 & 123 & 118 &   - & 141 \\
6219.29 & Fe1 & 2.200 & -2.438 &  87 & 105 & 112 &   - &   - & 113 & 113 & 111 & 139 & 139 & 110 & 133 \\
6229.23 & Fe1 & 2.840 & -2.900 &   - &   - &   - &   - &   - &   - &   - &   - &   - &  65 &   - &  57 \\
6230.74 & Fe1 & 2.560 & -1.276 & 148 & 167 & 126 &   - &   - & 149 &   - & 186 &   - & 178 &   - & 197 \\
6238.38 & Fe2 & 3.890 & -2.480 &  50 &   - &  45 &   - &   - &   - &  43 &   - &  49 &  55 &   - &   - \\
6240.66 & Fe1 & 2.220 & -3.230 &   - &   - &   - &   - &   - &  62 &   - & 103 &  87 &  91 &  83 &  83 \\
6243.82 & Si1 & 5.610 & -1.270 &   - &   - &   - &   - &   - &   - &   - &   - &   - &   - &   - &   - \\
\hline
\newpage
\caption{Continued.}\\
\hline
\hline
$\lambda$&Elem&$\chi_{ex}$&$log~gf$& 708 & 729 & 733 & 740 & 743 & 770 & 780 & 812 & 825 & 840 & 842 & 880 \\
  $\AA$  &    &           &        &     &     &     &     &     &     &     &     &     &     &     &     \\
\hline
6244.48 & Si1 & 5.610 & -1.270 &   - &   - &   - &   - &   - &   - &   - &   - &   - &   - &   - &  24 \\
6247.56 & Fe2 & 3.890 & -2.360 &   - &   - &   - &   - &   - &  54 &   - &   - &  54 &  63 &   - &  59 \\
6252.57 & Fe1 & 2.400 & -1.757 &   - & 189 & 163 &   - &   - & 128 & 116 & 163 & 142 & 161 & 123 & 166 \\
6258.10 & Ti1 & 1.443 & -0.355 &   - &  61 &   - &   - &   - &   - &   - &   - & 100 &  55 &   - &  83 \\
6290.97 & Fe1 & 4.730 & -0.760 &   - &   - &   - &   - &   - &   - &   - &   - &   - &   - &   - &  32 \\
6297.80 & Fe1 & 2.220 & -2.740 &   - &   - &   - &   - &   - &   - & 104 &   - &   - &   - &   - & 123 \\
6301.50 & Fe1 & 3.650 & -0.720 &   - &   - &   - &   - &   - &   - & 111 & 120 & 157 &   - & 107 &   - \\
6302.49 & Fe1 & 3.690 & -1.150 &   - &   - &   - &   - &   - &   - &   - &   - &  81 &  99 &  67 &  83 \\
6309.90 & Sc2 & 1.500 & -1.520 &   - &   - &   - &   - &   - &   - &   - &   - &   - &  41 &   - &   - \\
6311.51 & Fe1 & 2.830 & -3.220 &   - &   - &   - &   - &   - &   - &  38 &   - &  26 &   - &   - &  80 \\
6320.43 & La2 & 0.170 & -1.562 &   - &   - &   - &   - &   - &   - &   - &   - &  52 &  45 &   - &   - \\
6330.09 & Cr1 & 0.940 & -2.920 &   - &   - &   - &   - &   - &   - &   - &   - &   - &   - &   - &  52 \\
6355.04 & Fe1 & 2.840 & -2.290 & 102 &  91 &  69 & 111 & 111 &  86 &  61 &  73 &  98 & 102 &  86 &  96 \\
6369.46 & Fe2 & 2.890 & -4.250 &   - &   - &   - &   - &   - &  70 &   - &   - &   - &   - &   - &   - \\
6380.75 & Fe1 & 4.190 & -1.500 &   - &   - &  75 &  74 &  74 &   - &  53 &   - &  49 &  55 &   - &   - \\
6390.46 & La2 & 0.320 & -1.400 &   - &   - &   - &   - &   - &   - &   - &   - &   - &   - &   - &   - \\
6392.54 & Fe1 & 2.280 & -3.950 &  82 &  40 &   - &   - &   - &   - &   - &   - &   - &  58 &  75 &  54 \\
6393.61 & Fe1 & 2.430 & -1.630 & 117 & 172 & 156 & 167 & 167 & 134 & 127 & 193 & 181 &   - &   - & 199 \\
6416.93 & Fe2 & 3.890 & -2.790 &   - &   - &   - & 115 & 115 &  40 &   - &   - &   - &   - &   - &   - \\
6419.96 & Fe1 & 4.730 & -0.240 &   - &   - &   - &   - &   - &   - &   - &   - &  72 &   - &   - &   - \\
6421.36 & Fe1 & 2.280 & -2.014 &   - & 147 &   - & 201 &   - & 109 & 159 &   - & 147 & 125 & 105 & 169 \\
6430.86 & Fe1 & 2.180 & -1.946 &   - & 141 & 107 & 213 &   - & 150 & 141 &   - & 175 & 189 & 163 & 180 \\
6432.68 & Fe2 & 2.890 & -3.710 &   - &   - &  75 &   - &   - &   - &  50 &   - &  61 &   - &   - &  57 \\
6439.08 & Ca1 & 2.520 &  0.390 &   - & 175 & 155 & 149 & 149 &   - &   - &   - &   - &   - &   - & 171 \\
6455.60 & Ca1 & 2.520 & -1.290 &   - &   - &   - &   - &   - &   - &   - &   - &   - &  67 &   - &  60 \\
6456.39 & Fe2 & 3.900 & -2.080 &   - &  55 &   - & 107 & 107 &  86 &  80 & 122 &  54 & 110 & 116 &  73 \\
6496.91 & Ba2 & 0.600 & -0.380 &   - & 147 &   - &   - &   - &   - &   - &   - &   - & 161 & 119 &   - \\
6498.94 & Fe1 & 0.960 & -4.690 &   - &   - & 105 &  80 &  80 &  85 &  55 &  67 &   - & 110 &  74 & 129 \\
6499.65 & Ca1 & 2.520 & -0.820 &   - &   - &   - &   - &   - &  54 &   - &   - &   - &  72 &   - &  95 \\
6516.08 & Fe2 & 2.890 & -3.450 &   - &  72 &   - &  65 &  65 &  59 &  62 &   - &  52 &  56 &  71 &  56 \\
6518.37 & Fe1 & 2.830 & -2.460 &   - &  87 &   - &  60 &  60 &   - &  50 &   - &  50 &  56 &   - &  81 \\
6554.24 & Ti1 & 1.440 & -1.218 &   - &   - &   - &   - &   - &   - &   - &   - &  32 &   - &   - &  39 \\
6556.08 & Ti1 & 1.460 & -1.074 &   - &   - &   - &  69 &  69 &   - &   - &   - &   - &   - &   - &  44 \\
6574.23 & Fe1 & 0.990 & -5.020 &   - &   - &  55 &  73 &  73 &  59 &   - &   - & 108 &  90 &  37 &  85 \\
6581.22 & Fe1 & 1.480 & -4.680 &  43 &   - &   - &  93 &  93 &  42 &   - &   - &  53 &   - &   - &   - \\
6593.88 & Fe1 & 2.430 & -2.390 &  80 & 118 & 100 & 151 & 151 &  74 & 133 &   - & 110 & 113 &  94 & 118 \\
6608.03 & Fe1 & 2.280 & -3.940 &   - &  43 &   - &   - &   - &  35 &   - &   - &  32 &  49 &  48 &  46 \\
6609.12 & Fe1 & 2.560 & -2.660 &   - &  85 &  84 & 113 & 113 &  55 &  44 &  74 & 106 &  89 &  68 & 120 \\
6645.13 & Eu2 & 1.370 &  0.200 &   - &   - &   - &   - &   - &   - &   - &   - &   - &   - &   - &  27 \\
\hline
\newpage
\caption{List of all spectral lines, their atomic parameters and their measured EWs (in m$\AA$) for all the RGB stars in the Carina dSph. Part 3/3.}\\
\hline
\hline
$\lambda$&Elem&$\chi_{ex}$&$log~gf$& 900 & 902 & 914 & 916 & 925 & 948 & 976 & 1007& 1009& 1012& 1061&  \\
  $\AA$  &    &           &        &     &     &     &     &     &     &     &     &     &     &     &  \\
\hline								 
5339.93 & Fe1 & 3.270 & -0.680 &   - & 113 &   - & 128 & 114 & 126 & 126 &  97 & 141 & 128 &   - & \\
5364.86 & Fe1 & 4.450 &  0.220 & 106 &   - &  61 & 124 &   - &  81 &   - & 128 &  81 &  84 &  75 & \\
5367.48 & Fe1 & 4.420 &  0.550 &   - &  96 &  79 & 138 & 119 &   - &   - & 126 & 101 & 110 & 129 & \\
5369.96 & Fe1 & 4.370 &  0.540 &   - &   - &   - &   - &   - &   - & 134 &   - &  90 &   - & 147 & \\
5371.50 & Fe1 & 0.960 & -1.644 &   - &   - &   - & 269 &   - &   - &   - &   - &   - &   - &   - & \\
5381.01 & Ti2 & 1.570 & -1.780 & 117 &  90 &  69 & 185 &   - &  95 &   - &   - &   - &  89 & 108 & \\
5383.37 & Fe1 & 4.310 &  0.500 & 119 & 106 &  64 & 181 & 116 &   - &   - &   - &  97 & 120 &   - & \\
5389.48 & Fe1 & 4.420 & -0.400 &  56 &   - &   - &   - &  70 &   - &   - &   - & 100 &  72 &   - & \\
5393.17 & Fe1 & 3.240 & -0.920 & 129 & 121 &  71 & 139 & 135 & 107 & 135 &   - &  92 & 134 &   - & \\
5397.14 & Fe1 & 0.910 & -1.992 &   - &   - &   - & 248 &   - &   - &   - &   - &   - &   - &   - & \\
5400.51 & Fe1 & 4.370 & -0.150 &  88 &  59 &   - &   - &  88 &   - &  89 &  77 &  74 &  66 &  75 & \\
5402.78 &  Y2 & 1.840 & -0.510 &   - &   - &   - &   - &   - &   - &   - &   - &   - &   - &   - & \\
5405.79 & Fe1 & 0.990 & -1.852 &   - &   - &   - & 221 &   - &   - &   - &   - &   - &   - &   - & \\
5407.42 & Mn1 & 2.140 & -1.743 &  36 &   - &   - &   - &   - &   - &   - &  58 &   - &   - &   - & \\
5409.80 & Cr1 & 1.030 & -0.720 & 170 &   - &  83 & 173 & 118 & 119 & 167 & 178 & 170 & 149 & 153 & \\
5415.19 & Fe1 & 4.390 &  0.510 & 111 & 104 &  73 & 117 & 113 &  81 &   - & 120 &  98 & 129 & 120 & \\
5418.77 & Ti2 & 1.580 & -2.110 &  91 &  65 &   - & 111 & 111 &   - &   - &  90 &  80 &  56 & 104 & \\
5420.36 & Mn1 & 2.140 & -1.460 &   - &   - &   - &   - &   - &   - &   - &  53 &   - &   - &   - & \\
5424.07 & Fe1 & 4.320 &  0.520 & 126 & 113 &  82 & 170 & 124 & 123 & 164 & 121 & 114 & 152 & 112 & \\
5425.25 & Fe2 & 3.200 & -3.360 &   - &   - &   - &   - &   - &   - &   - &   - &  47 &   - &   - & \\
5432.55 & Mn1 & 0.000 & -3.795 & 111 &  54 &   - &   - &   - &   - &   - & 102 &  71 &   - &   - & \\
5476.92 & Ni1 & 1.830 & -0.890 &   - & 114 & 125 & 196 & 191 &   - & 139 &   - &   - & 155 & 194 & \\
5483.34 & Co1 & 1.710 & -1.488 &  71 &   - &   - &   - &   - &   - &   - &  98 &   - &  60 &  60 & \\
5485.71 & Nd2 & 1.260 & -0.120 &  22 &   - &   - &   - &   - &   - &   - &   - &   - &   - &   - & \\
5490.16 & Ti1 & 1.460 & -0.933 &  27 &   - &   - &   - &   - &   - &  52 &  57 &   - &   - &  34 & \\
5501.48 & Fe1 & 0.960 & -3.050 & 183 & 170 & 137 &   - & 153 & 170 & 159 & 189 & 150 & 186 & 179 & \\
5506.79 & Fe1 & 0.990 & -2.790 & 194 & 182 & 146 & 211 & 140 & 161 & 140 &   - & 188 & 180 & 188 & \\
5516.77 & Mn1 & 2.180 & -1.847 &  29 &   - &   - &   - &   - &   - &   - &  37 &   - &   - &   - & \\
5528.41 & Mg1 & 4.350 & -0.357 & 164 &  98 & 124 & 170 & 178 &   - & 155 & 174 & 148 & 167 & 189 & \\
5534.85 & Fe2 & 3.240 & -2.920 &  55 &  74 &  37 & 103 &  58 &   - &  74 &  57 &  71 &  79 &  70 & \\
6120.26 & Fe1 & 0.910 & -5.940 &  21 &   - &   - &   - &   - &   - &   - &   - &   - &   - &   - & \\
6122.23 & Ca1 & 1.890 & -0.320 & 181 &   - & 133 &   - &   - & 117 &   - &   - & 156 & 148 &   - & \\
6126.22 & Ti1 & 1.070 & -1.425 &  52 &   - &   - &   - &   - &   - &   - &  76 &   - &   - &   - & \\
6136.62 & Fe1 & 2.450 & -1.500 &   - &   - &  97 &   - &   - &   - & 143 &   - & 162 & 169 & 158 & \\
6137.70 & Fe1 & 2.590 & -1.366 & 157 & 120 & 106 &   - &   - & 138 & 142 &   - & 134 & 121 & 138 & \\
6141.73 & Ba2 & 0.700 & -0.077 & 178 &   - & 104 & 211 & 143 & 182 & 163 &   - &   - & 161 &   - & \\
6149.25 & Fe2 & 3.890 & -2.720 &   - &   - &   - &   - &   - &   - &   - &   - &   - &  36 &   - & \\
6151.62 & Fe1 & 2.180 & -3.370 &  67 &   - &   - &   - &  58 &  62 &   - &   - &  46 &  74 &  82 & \\
6157.75 & Fe1 & 4.070 & -1.260 &  48 &   - &   - &  66 &   - &   - &   - &   - &  41 &   - &   - & \\
6160.75 & Na1 & 2.100 & -1.260 &   - &   - &   - &   - &   - &   - &   - &   - &   - &   - &  37 & \\
6161.30 & Ca1 & 2.520 & -1.270 &  52 &   - &   - &  83 &  73 &   - &   - &   - &  52 &  37 &   - & \\
6165.36 & Fe1 & 4.140 & -1.470 &   - &   - &   - &   - &  58 &   - &  49 &   - &   - &  31 &  48 & \\
6166.44 & Ca1 & 2.520 & -1.140 &  61 &   - &   - &   - &   - &   - &   - &   - &  59 &   - &   - & \\
6169.04 & Ca1 & 2.520 & -0.800 &  71 &   - &   - &   - &   - &   - &   - &   - &  74 &   - &   - & \\
6169.56 & Ca1 & 2.520 & -0.480 & 102 &  63 &   - &   - & 111 &   - &  86 & 113 &  92 & 106 &   - & \\
6173.34 & Fe1 & 2.220 & -2.850 & 102 &  57 &   - & 114 &  75 &  63 &  70 &   - &  93 &  72 &   - & \\
6176.82 & Ni1 & 4.090 & -0.430 &  32 &   - &   - &   - &   - &   - &   - &   - &   - &  29 &   - & \\
6177.25 & Ni1 & 1.830 & -3.500 &   - &   - &   - &   - &   - &   - &   - &   - &   - &   - &   - & \\
6187.99 & Fe1 & 3.940 & -1.580 &   - &  45 &   - &   - &   - &   - &   - &   - &   - &   - &   - & \\
6191.57 & Fe1 & 2.430 & -1.416 &   - &   - &   - &   - &   - & 145 & 166 & 147 & 146 & 130 &   - & \\
6213.43 & Fe1 & 2.220 & -2.660 & 120 &  97 &  59 &   - & 104 &  77 &  84 & 134 & 104 & 107 & 100 & \\
6219.29 & Fe1 & 2.200 & -2.438 & 124 & 101 &  77 &   - &  89 &  94 &   - & 135 & 123 & 108 & 136 & \\
6229.23 & Fe1 & 2.840 & -2.900 &   - &   - &   - &   - &   - &   - &   - &   - &  49 &   - &   - & \\
6230.74 & Fe1 & 2.560 & -1.276 & 172 &   - &   - & 169 &   - & 136 & 155 &   - & 131 & 152 & 157 & \\
6238.38 & Fe2 & 3.890 & -2.480 &   - &   - &   - &   - &   - &   - &   - &   - &  51 &  39 &  55 & \\
6240.66 & Fe1 & 2.220 & -3.230 &  74 &   - &   - &   - &   - &  42 &  51 &   - &  44 &   - &   - & \\
6243.82 & Si1 & 5.610 & -1.270 &   - &   - &   - &   - &   - &   - &   - &   - &   - &   - &  57 & \\
\hline
\newpage
\caption{Continued.}\\
\hline
\hline
$\lambda$&Elem&$\chi_{ex}$&$log~gf$& 900 & 902 & 914 & 916 & 925 & 948 & 976 & 1007& 1009& 1012& 1061&  \\
  $\AA$  &    &           &        &     &     &     &     &     &     &     &     &     &     &     &  \\
\hline	
6244.48 & Si1 & 5.610 & -1.270 &   - &   - &   - &   - &   - &   - &   - &   - &   - &   - &   - & \\
6247.56 & Fe2 & 3.890 & -2.360 &   - &  47 &   - &   - &   - &  49 &   - &   - &   - &  52 &  61 & \\
6252.57 & Fe1 & 2.400 & -1.757 & 156 & 132 &   - & 138 & 153 & 122 & 139 & 151 &   - & 144 & 162 & \\
6258.10 & Ti1 & 1.443 & -0.355 &  70 &   - &   - &   - &   - &   - &   - &  75 &   - &   - &   - & \\
6290.97 & Fe1 & 4.730 & -0.760 &   - &   - &   - &   - &   - &   - &   - &   - &   - &   - &   - & \\
6297.80 & Fe1 & 2.220 & -2.740 &   - &  75 &   - &   - &  77 &   - &   - &   - &  97 &  96 & 119 & \\
6301.50 & Fe1 & 3.650 & -0.720 &   - &   - &   - &   - &   - &   - &   - &   - & 100 &  94 & 137 & \\
6302.49 & Fe1 & 3.690 & -1.150 &  77 &   - &   - &   - &   - &   - &   - &  86 &  80 &   - &   - & \\
6309.90 & Sc2 & 1.500 & -1.520 &   - &   - &   - &   - &   - &   - &   - &   - &   - &   - &   - & \\
6311.51 & Fe1 & 2.830 & -3.220 &  41 &   - &   - &  94 &   - &   - &   - &  53 &   - &  56 &   - & \\
6320.43 & La2 & 0.170 & -1.562 &   - &   - &   - &   - &   - &   - &   - &  47 &   - &   - &   - & \\
6330.09 & Cr1 & 0.940 & -2.920 &   - &   - &   - &   - &   - &   - &   - &   - &   - &   - &   - & \\
6355.04 & Fe1 & 2.840 & -2.290 & 101 &  69 &  77 & 139 &   - &  67 &   - &   - &   - &  81 &  93 & \\
6369.46 & Fe2 & 2.890 & -4.250 &   - &   - &   - &   - &   - &   - &   - &   - &   - &  50 &   - & \\
6380.75 & Fe1 & 4.190 & -1.500 &  33 &  47 &   - &   - &   - &   - &   - &   - &  41 &  56 &   - & \\
6390.46 & La2 & 0.320 & -1.400 &   - &   - &   - &   - &   - &   - &   - &   - &   - &   - &   - & \\
6392.54 & Fe1 & 2.280 & -3.950 &  26 &  41 &   - &   - &   - &   - &   - &   - &   - &   - &  38 & \\
6393.61 & Fe1 & 2.430 & -1.630 &   - &   - &   - & 139 & 125 & 123 & 170 & 169 & 138 &   - & 145 & \\
6416.93 & Fe2 & 3.890 & -2.790 &   - &   - &   - &  67 &   - &   - &   - &   - &   - &   - &   - & \\
6419.96 & Fe1 & 4.730 & -0.240 &   - &   - &   - &   - &   - &   - &   - &   - &  69 &   - &   - & \\
6421.36 & Fe1 & 2.280 & -2.014 & 151 &  86 &  73 & 135 & 158 & 126 &   - & 127 & 115 & 130 & 152 & \\
6430.86 & Fe1 & 2.180 & -1.946 &   - & 130 & 102 & 212 & 123 & 127 & 121 & 159 &  80 &   - & 161 & \\
6432.68 & Fe2 & 2.890 & -3.710 &  45 &   - &   - &  96 &   - &   - &   - &  43 &   - &   - &  77 & \\
6439.08 & Ca1 & 2.520 &  0.390 & 159 &   - &  91 &   - &   - &   - &   - & 161 &   - & 133 & 159 & \\
6455.60 & Ca1 & 2.520 & -1.290 &  40 &   - &   - &   - &   - &   - &   - &   - &   - &   - &   - & \\
6456.39 & Fe2 & 3.900 & -2.080 &  88 &  46 &  29 &   - &  73 &   - &   - &   - &   - &   - &  89 & \\
6496.91 & Ba2 & 0.600 & -0.380 &   - & 113 &   - &   - & 132 & 180 &   - & 168 & 172 & 148 &   - & \\
6498.94 & Fe1 & 0.960 & -4.690 &   - &   - &  53 &  96 &  57 &   - &   - & 117 &  77 &  64 &  67 & \\
6499.65 & Ca1 & 2.520 & -0.820 &  68 &  33 &   - &   - &   - &   - &   - &   - &  75 &  76 &   - & \\
6516.08 & Fe2 & 2.890 & -3.450 &   - &  40 &   - &  52 &  54 &  36 &   - &  52 &   - &   - &  48 & \\
6518.37 & Fe1 & 2.830 & -2.460 &  77 &  47 &   - &  85 &   - &   - &   - &  55 &  60 &  60 &  49 & \\
6554.24 & Ti1 & 1.440 & -1.218 &   - &   - &   - &   - &   - &   - &   - &  35 &   - &  18 &   - & \\
6556.08 & Ti1 & 1.460 & -1.074 &   - &   - &   - &   - &   - &   - &   - &  49 &   - &   - &  32 & \\
6574.23 & Fe1 & 0.990 & -5.020 &  84 &  37 &  40 &  52 &  57 &  40 &   - &  64 &  55 &  62 &  53 & \\
6581.22 & Fe1 & 1.480 & -4.680 &  55 &  30 &   - &  65 &   - &   - &   - &  68 &   - &   - &  48 & \\
6593.88 & Fe1 & 2.430 & -2.390 & 126 &  75 &  42 & 101 &  96 &  76 &   - & 112 &  73 &  73 &  93 & \\
6608.03 & Fe1 & 2.280 & -3.940 &  39 &   - &   - &   - &   - &   - &   - &  56 &   - &   - &   - & \\
6609.12 & Fe1 & 2.560 & -2.660 &  98 &   - &   - & 101 &  62 &  57 &   - & 102 &  93 &  62 &  74 & \\
6645.13 & Eu2 & 1.370 &  0.200 &   - &   - &   - &   - &   - &   - &   - &   - &   - &   - &   - & \\
\hline
\end{longtable}
}

\end{document}